\documentclass[aps,twocolumn,superscriptaddress,preprintnumbers]{revtex4-1}

\usepackage[T1]{fontenc}
\usepackage{amsmath}
\usepackage{float}
\usepackage{amstext}
\usepackage{amssymb}
\usepackage{graphicx}
\usepackage{esint}
\usepackage{color}
\usepackage{hyperref}
\usepackage[title]{appendix}
\definecolor{colorref}{rgb}{0.0, 0.408, 0.647}
\definecolor{grey}{rgb}{0.95, 0.95, 0.95}
\hypersetup{
colorlinks = true,
allcolors = colorref
}
\usepackage{dcolumn}
\usepackage{bm}

\begin{document}

\title{Quantum many-body dynamics for fermionic $t$-$J$ model simulated with atom arrays}

 \author{Ye-Bing Zhang}
 \affiliation{International Center for Quantum Materials and School of Physics, Peking University, Beijing 100871, China}
 \affiliation{Hefei National Laboratory, Hefei 230088, China}

 \author{Xin-Chi Zhou}
 \affiliation{International Center for Quantum Materials and School of Physics, Peking University, Beijing 100871, China}
 \affiliation{Hefei National Laboratory, Hefei 230088, China}

 \author{Bao-Zong Wang}
 \affiliation{International Center for Quantum Materials and School of Physics, Peking University, Beijing 100871, China}
 \affiliation{Hefei National Laboratory, Hefei 230088, China}

 \author{Xiong-Jun Liu}
 \email{Corresponding author:xiongjunliu@pku.edu.cn}
 \affiliation{International Center for Quantum Materials and School of Physics, Peking University, Beijing 100871, China}
 \affiliation{Hefei National Laboratory, Hefei 230088, China}
 \affiliation{International Quantum Academy, Shenzhen 518048, China}

\begin{abstract}
The fermionic $t$-$J$ model has been widely recognized as a canonical model for broad range of strongly correlated phases, particularly the high-$T_{\mathrm{c}}$ superconductor. Simulating this model with controllable quantum platforms offers new possibilities to probe high-$T_{\mathrm{c}}$ physics, yet suffering challenges.
Here we propose a novel scheme to realize a highly-tunable extended $t$-$J$ model in a programmable Rydberg-dressed tweezer array.
Through engineering the Rydberg-dressed dipole-dipole interaction and inter-tweezer couplings, the fermionic $t$-$J$ model with independently tunable exchange and hopping couplings is achieved.
With the high tunability, we explore quantum many-body dynamics
in the large $J/t$ limit, a regime well beyond the conventional optical lattices and cuprates, and predict an unprecedented many-body self-pinning effect enforced by local quantum entanglement with emergent conserved quantities. The self-pinning effect leads to novel nonthermal quantum many-body dynamics, which violates eigenstate thermalization hypothesis in Krylov subspace. Our prediction opens a new horizon in exploring exotic quantum many-body physics with $t$-$J$ model, and shall also make a step towards simulating the high-$T_{\mathrm{c}}$ physics in neutral atom systems.
\end{abstract}

\maketitle

\textcolor{blue}{\em Introduction.}-- The high-$T_{\mathrm{c}}$ superconductivity (SC) in doped cuprates has been discovered for several decades \cite{Bednorz1986}, yet a comprehensive understanding of their phase diagram remains elusive \cite{Keimer2015,Zhou2021} due to the limitations of analytical and numerical methods. The quantum simulation based on controllable artificial systems \cite{Cirac2012,RevModPhys.86.153,Daley2022} has emerged as a promising tool to probe its underlying physics. The Fermi-Hubbard model and related $t$-$J$ model are recognized as canonical models \cite{PhysRevB.18.3453,doi:10.1126/science.235.4793.1196,PhysRevB.37.3759,YuriiAIzyumov_1997,RevModPhys.78.17,10.12693/aphyspola.111.409,Ogata_2008} for high-$T_{\mathrm{c}}$ SC. While experimental realizations of the single-band Hubbard model in optical lattices have been widely reported \cite{annurev-conmatphys-070909-104059,PhysRevLett.104.080401,Hart2015,doi:10.1126/science.aag3349,Mazurenko2017,TARRUELL2018365,Koepsell2019,BOHRDT2021168651,Xu2023,Shao2024}, these studies face fundamental, non-technical limitations.
First, in the Mott regime of single-band Hubbard model, $t$ and $J=4t^2/U$ are intrinsically linked, constraining the tunability of $J/t$. In real cuprate materials, however, these parameters can vary more freely, as originated from the three-band model \cite{PhysRevB.37.3759,RevModPhys.78.17,Ogata_2008}. Second, recent numerical studies suggest that the high-$T_{\mathrm{c}}$ SC crucially depends on the next-nearest-neighbour (NNN)
hopping \cite{doi:10.1126/science.aal5304,PhysRevLett.127.097003,doi:10.1073/pnas.2109978118,PhysRevLett.127.097002,PhysRevLett.132.066002,doi:10.1126/science.adh7691,PhysRevLett.133.256003,PhysRevResearch.2.033073,Chen2025}, which is hard to engineer in optical lattices. Recently a new simulation scheme based on polar molecules \cite{Miller2024,Carroll2025} have been proposed, while realizing a programmable fermionic $t$-$J$ model with highly tunable parameters remains an outstanding issue.

Rydberg atoms in tweezer arrays \cite{doi:10.1126/science.aah3752,doi:10.1126/science.aah3778,Barredo2018} may present a promising solution to address the problem. The intrinsic long-range interaction \cite{RevModPhys.82.2313,Adams_2020,Wu_2021} render Rydberg atoms powerful quantum simulators for many-body physics \cite{Browaeys2020}, especially for quantum spin models \cite{doi:10.1126/science.aav9105,PhysRevA.106.L021101,Chen2023,PhysRevResearch.6.L042054,https://doi.org/10.1002/qute.202300356,Labuhn2016,Scholl2021,Bernien2017,Ebadi2021}. Moreover, ``Rydberg dressing'' \cite{PhysRevA.82.033412,Balewski_2014} enables further control over
interaction and lifetime of the system. Applications of Rydberg-dressed atoms have included simulations of supersolid states \cite{PhysRevLett.104.195302,PhysRevLett.108.265301}, XYZ spin model \cite{PhysRevLett.114.243002,PhysRevLett.114.173002,PhysRevLett.130.243001}, long-range Ising chain \cite{PhysRevX.7.041063} and extended Fermi(Bose)-Hubbard model \cite{PhysRevX.11.021036,Weckesser2024}. However, most of these studies focus on quantum magnets without charge degree of freedom \cite{Chen2023,PhysRevResearch.6.L042054,PhysRevLett.114.243002,PhysRevLett.114.173002,PhysRevLett.130.243001,PhysRevX.7.041063,PhysRevX.11.021036}. Recent advancements in coherent transport of tweezers \cite{Beugnon2007,Bluvstein2022,Bluvstein2024} and tunnel-coupled tweezers \cite{https://doi.org/10.1002/qute.202300176,doi:10.1126/science.1250057,Bergschneider2019} realized atom hopping between adjacent tweezers,
enabling the realization of Fermi-Hubbard model in tweezer arrays \cite{PhysRevLett.128.223202,PhysRevLett.129.123201} and the fermion tunneling gates \cite{doi:10.1073/pnas.2304294120}. This capability opens up the further simulation of intricate doped quantum magnet in tweezer arrays.

In this letter, we propose a novel scheme to realize the highly tunable fermionic $t$-$J$ model in a programmable Rydberg-dressed tweezer array, with which we uncover exotic nonthermal quantum many-body dynamics in the large $J/t$ regime, a regime far beyond conventional optical lattices and cuprates, but well achievable in the current realization. The Hamiltonian is given by
\begin{eqnarray}
H&=&H_{t,\mathrm{N}}+H_{t',\mathrm{NN}}+H_{J},\\
 \label{HtN} H_{t,\mathrm{N}}&=&  \sum_{\langle i,j\rangle,\mathrm{\sigma}} (-t P_{i}c_{i,\mathrm{\sigma}}^{\dagger}c_{j,\mathrm{\sigma}}P_{j}+\mathrm{H.c.}),\nonumber\\
 \label{HtNN} H_{t',\mathrm{NN}}&=&\sum_{\langle \langle i,j\rangle \rangle, \mathrm{\sigma}} (-t^{\prime} P_i c_{i,\mathrm{\sigma}}^\dagger c_{j, \mathrm{\sigma}} P_j+\mathrm{H.c.}),\nonumber\\
 \label{Hj} H_{J}&=&  \sum_{i\neq j} [ J_{\mathrm{ex},ij}(S_{i,x}S_{j,x}+S_{i,y}S_{j,y})+ \notag\nonumber\\
  &&+ J_{z,ij}(S_{i,z}S_{j,z}-\frac{1}{4}n_{i}n_{j}) ],\nonumber
\end{eqnarray}
where $P_i=1-n_{i,\uparrow}n_{i,\downarrow}$ excludes double occupancy on site $i$, with $n_{i,\sigma}=c_{i,\sigma}^\dag c_{i,\sigma}$ ($\sigma=\uparrow,\downarrow$) the fermion number operator. The kinetic terms $H_{t,\mathrm{N}}$ and $H_{t',\mathrm{NN}}$ describe hopping processes between nearest-neighbor (NN) sites with amplitude $t$ and NNN sites with amplitude $t^{\prime}$, respectively. Our scheme allows a free tuning of 
spin interaction strength $J/t$ and NNN hopping strength $t^{\prime}/t$, a key feature to simulate the high-$T_{\mathrm{c}}$ SC~\cite{MOREIRA2001183}. In the large $J/t$ limit, we show the existence of Hilbert space fragmentation (HSF) \cite{PhysRevX.10.011047,PhysRevB.101.174204,PhysRevLett.130.010201,Adler2024} characterized by the emergent conserved quantities of the local quantum entangled states. Most nontrivially, we predict an unprecedented many-body phenomena, dubbed {\em many-body self-pinning effect}, in which the local quantum entanglement enforces the many-body state to preserve its initial configuration. The self-pinning effect leads to the breakdown of Krylov restricted thermalization, a new mechanism in contrast to previous theory relying on integrability~\cite{Hahn2021InformationDI,Moudgalya_2022,PhysRevX.15.011035,PhysRevB.111.144313,doi:10.1142/9789811231711_0009,PhysRevLett.124.207602,doi:10.1142/9789811231711_0009, PhysRevLett.124.207602,PhysRevB.100.214313,PhysRevB.103.134207}.

\begin{figure}[t]
	\begin{centering}
		\includegraphics[scale=1.15]{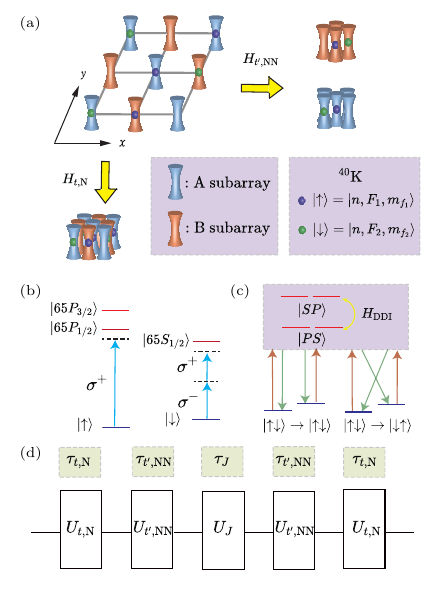}
\caption{Scheme for simulating the $t$-$J$ model. (a) Realization of the NN and NNN hopping terms $H_{t,\mathrm{N}}$ and $H_{t^{\prime},\mathrm{NN}}$ in a 2D square tweezer array, which is  composed of A and B subarrays. The spin up $\lvert\uparrow\rangle$  and spin down $\lvert\downarrow\rangle$ states are encoded in two hyperfine states of fermionic (e.g. $^{40}\mathrm{K}$) atoms. (b) Ground states are dressed to the Rydberg states with principle quantum number $n=65$. $\lvert\uparrow\rangle$ is dressed to states $\lvert 65P\rangle$ via a single-photon process using $\sigma^+$ polarized light, and $\lvert\downarrow\rangle$ is dressed to state $\lvert 65S\rangle$ through a two-photon process involving $\sigma^{+}$ and $\sigma^{-}$ polarized light \cite{supplement}. (c) Realization of the interaction term $H_J$. The diagonal channels (left) generate the Ising interaction $J_{z,ij}$, while the off-diagonal channels (right) generate the spin exchange interaction $J_{\mathrm{ex},ij}$. (d) A digital combination of the hopping and interaction terms is employed to construct the full $t$-$J$ Hamiltonian.} 
		\label{fig: sketch}
		\par\end{centering}
\end{figure}

\textcolor{blue}{\em Model realization.}--We propose to realize the extended $t$-$J$ model with a digital combination of  inter-tweezer couplings and Rydberg-dressed dipole-dipole interaction (DDI), enabling the realization of both one-dimensional (1D) chain and 2D square array. The fermionic (e.g. $^{40}\mathrm{K}$) atoms are trapped in optical tweezers, with two ground ($4^{2}S_{\frac{1}{2}}$) hyperfine states defining spin states:
$\lvert \uparrow\rangle=\lvert F=\frac{9}{2},m_{f}=-\frac{9}{2}\rangle$ and $\lvert \downarrow\rangle=\lvert F=\frac{7}{2},m_{f}=-\frac{7}{2}\rangle$.
The kinetic terms $H_{t,\mathrm{N}}$ and $H_{t',\mathrm{NN}}$ are realized through controlled atomic tunneling between adjacent tweezers \cite{https://doi.org/10.1002/qute.202300176,doi:10.1126/science.1250057,Bergschneider2019,PhysRevLett.128.223202,PhysRevLett.129.123201,doi:10.1073/pnas.2304294120}. 

Fig.~\ref{fig: sketch}(a) illustrates the realization for the 2D configuration, in which 
$H_{t,\mathrm{N}}$ is realized by coherently coupling the nearest (AB) tweezers, and $H_{t',\mathrm{NN}}$ is achieved by coupling only those tweezers within the same AA or BB subarray. The similar subarray-based coupling scheme applies in the 1D case. To eliminate the double occupancy of each tweezer during the tunneling process, one can introduce a strong intra-tweezer repulsion generated by the mature Feshbach resonance technique~\cite{RevModPhys.82.1225,footnote2}.

The interaction term $H_J$ is realized using dressed DDI, as depicted in Fig.~\ref{fig: sketch}(b). The total Hamiltonian is
\begin{equation}
H^{\mathrm{tot}}_{J}=H_{\Omega}+H_{\mathrm{DDI}}+H_{\Delta},\label{full_H}
\end{equation}
where $H_{\Omega}$ denotes the coupling between ground spin state $\lvert \uparrow \rangle$ (or $\lvert \downarrow\rangle$) and Rydberg $p$-(or $s$-)orbital states $\lvert nP\rangle$ (or $\lvert nS\rangle$), with $n$ taken as $n=65$ below for concrete calculation. $H_{\mathrm{DDI}}$ describes dipolar interactions between relevant Rydberg states and $H_{\Delta}$ represents the residual detuning for the coupling between ground and relevant Rydberg states~\cite{supplement}. The spin exchange $J_{\mathrm{ex},ij}$ and Ising interaction $J_{z,ij}$ are derived from the off-diagonal and diagonal contributions of the effective Hamiltonian
\begin{equation}
    H_J^{\mathrm{eff}}=\sum_{n={4,6,8,\cdots}}H_{\Omega}(G H_{\Omega})^{n-1}.
    \label{HJeff}
\end{equation}
Here, $G=(1-P_0)(E_{0}-H_{\Delta}-H_{\mathrm{DDI}})^{-1}(1-P_0)$, with $P_0$ the projection operator in the low-energy subspace spanned by $H_{\mathrm{DDI}}+H_{\Delta}$, and $E_{0}$ is the unperturbed energy. The leading order contribution is illustrted in Fig.~\ref{fig: sketch}(c). Specifically, $J_{\mathrm{ex},ij}$ is given by
$ J_{\mathrm{ex},ij}/2=\langle\cdots\uparrow_i\cdots\downarrow_j\cdots\lvert H_J^{\mathrm{eff}}\rvert\cdots\downarrow_i\cdots\uparrow_j\cdots\rangle$ and $J_{z,ij}$ is determined by $-(J_{z,ij}/2)= \langle\cdots\uparrow_i\cdots\downarrow_j\cdots\lvert H_J^{\mathrm{eff}}\rvert\cdots\uparrow_i\cdots\downarrow_j\cdots\rangle
		-\langle\cdots\uparrow_i\cdots\downarrow_j\cdots\lvert H_J^{\mathrm{eff}}(\mathrm{DDI}=0)\rvert\cdots\uparrow_i\cdots\downarrow_j\cdots\rangle$~\cite{supplement}.

The extended $t$-$J$ model is realized by the digital combination of $H_{t,\mathrm{N}}$, $H_{t',\mathrm{NN}}$ and $H_J$ through a sequence illustrated in Fig.~\ref{fig: sketch}(d) \cite{footnote3}, with the coupling times being $\tau_{t,\mathrm{N}}$, $\tau_{t',{\mathrm{NN}}}$, and $\tau_J$, respectively. The time evolution over one cycle is approximated using the standard second-order Suzuki-Trotter decomposition \cite{10.1063/1.529425,Berry2007,PRXQuantum.4.030319}, yielding the full effective model
\begin{equation}
 H_{t\text{-}J}=\frac{2\tau_{t,\mathrm{N}}}{\tau_J}H_{t_0,\mathrm{N}}+\frac{2\tau_{t',\mathrm{NN}}}{\tau_J}H_{t_0^{\prime},\mathrm{NN}}+H_J+o(\tau^2H^3),
\end{equation}
where $H_{t_0,\mathrm{N}}$ and $H_{t_0^{\prime},\mathrm{NN}}$ are the bare kinetic terms.

The present scheme directly simulates the extended $t$-$J$ model, rather than being derived from Hubbard model. This enables a free and independent control of each parameter in $H_{t\text{-}J}$ by manipulating $\tau_{t,\mathrm{N}}$, $\tau_{t',{\mathrm{NN}}}$, and $\tau_J$. The maximum hopping $t$ and $t^\prime$ can be tuned 
to be over $0.3$ kHz~\cite{doi:10.1126/science.1250057}, and be further effectively adjusted by the ratio $2\tau_{t,\mathrm{N}}(2\tau_{t',\mathrm{NN}})/\tau_J$ 
    to be several kHz. The interaction strength $J_{\mathrm{ex}} \approx J_z \approx 2$ kHz is achieved at a distance of 8.2 $\mu\mathrm{m}$ for $^{40}$K atoms 
    (see numerical details in Supplementary Material~\cite{supplement}). This allow to realize the broad regimes ranging from $J/t\ll1$ to $J/t \gg 1$, well surpassing the limitations of simulating Fermi-Hubbard model in optical lattices and beyond the regimes in cuprates.

In experiment the coupling times are restricted by the lifetimes of Rydberg-dressed states (denoted as $\tau_{\mathrm{dress}}$) and tweezers (denoted as $\tau_{\mathrm{tweezer}}$). 
In a cryogenic environment at approximately 4 K \cite{PhysRevApplied.16.034013,PhysRevApplied.22.024073}, the lifetime of Rydberg-dressed state is $\tau_{\mathrm{dress}} \approx 60$ ms, which gives $J_{{\rm ex},z}\tau_{\mathrm{dress}}\sim120\hbar$ and is sufficient for the simulation.
The lifetime of tweezers $\tau_{\mathrm{tweezer}}$ can be up to $10^3$ seconds \cite{PhysRevApplied.16.034013,PhysRevApplied.22.024073}, far exceeding the total operation time of the present simulation, which is several seconds \cite{supplement}. 

\begin{figure}[t]
	\begin{centering}
		\includegraphics[scale=1.1]{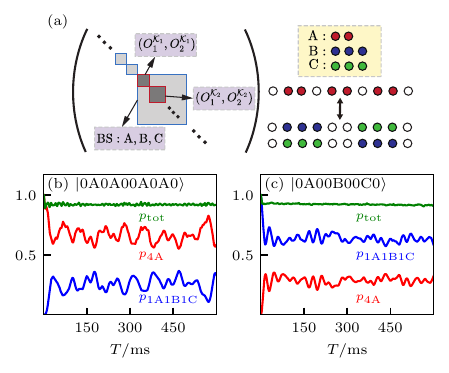}
		\caption{Hilbert space fragmentation (HSF) and self-pinning effect of 1D $t$-$J$ model in the $J/t\gg 1$ regime. (a) Matrix representation of the effective Hamiltonian showing dynamically disconnected Krylov subspaces $\mathcal{K}_i$. Subspaces containing characteristic bound states $\lvert \mathrm{A} \rangle$, $\lvert \mathrm{B} \rangle$ and $\lvert \mathrm{C} \rangle$ are labeled by emergent conserved quantities $(O_{1}^{\mathcal{K}_i},O_{2}^{\mathcal{K}_i})$. Resonant transitions occur between two elementary configurations: one with three $\lvert \mathrm{A} \rangle$ states and another with one $\lvert \mathrm{B} \rangle$ and one $\lvert \mathrm{C} \rangle$ state. (b)-(c) Time evolution of probabilities $p_c$ for observing the configurations $c=\{\mathrm{4A}, \mathrm{1A1B1C}\}$, starting from different initial states in a 14-site chain within the $(O_{1}^{\mathcal{K}_i},O_{2}^{\mathcal{K}_i})=(4,1)$ Krylov subpace. $p_{\mathrm{tot}}=p_{4\mathrm{A}}+p_{1\mathrm{A}1\mathrm{B}1\mathrm{C}}$. TEBD simulations use $t$=0.1 kHz, $J$=2.0 kHz, with a bond dimension up to 800.}
		\label{fig: hsf}
		\par\end{centering}
\end{figure}

\textcolor{blue}{\em HSF with entangled bases.}--The high tunability of our scheme enables access to an underexplored parameter regime with $J/t \gg 1$. We focus on the 1D $t$-$J$ chain with NNN hopping $t^\prime=0$ and isotropic interaction $J_{\mathrm{ex}}=J_z=J$, with
the unimportant anisotropy being neglected~\cite{supplement}. In this regime, the dominant energy scale from $H_J$ induces a novel type of HSF characterized by the quantum entangled states and associated emergent conserved quantities. This contrasts with the previous studies of HSF which is built on product states~\cite{PhysRevX.10.011047,PhysRevB.101.174204,PhysRevLett.130.010201,Adler2024,PhysRevX.15.011035,PhysRevB.111.144313,doi:10.1142/9789811231711_0009,PhysRevLett.124.207602,doi:10.1142/9789811231711_0009, PhysRevLett.124.207602,PhysRevB.100.214313,PhysRevB.103.134207}. The simplest nontrivial fragmented Krylov subspaces are formed by the three entangled spin bound states as
\begin{equation}
	\begin{array}{ll}
		\lvert\mathrm{A}\rangle=\frac{1}{\sqrt{2}}(\lvert\uparrow\downarrow\rangle-\lvert\downarrow\uparrow\rangle), & E_{\mathrm{A}}=-J,  \\
		\lvert\mathrm{B}\rangle=\frac{1}{\sqrt{6}}(\lvert\uparrow\downarrow\downarrow\rangle-2\lvert\downarrow\uparrow\downarrow\rangle+\lvert\downarrow\downarrow\uparrow\rangle), &  E_{\mathrm{B}}=-\frac{3}{2}J,\\
		\lvert\mathrm{C}\rangle=\frac{1}{\sqrt{6}}(\lvert\downarrow\uparrow\uparrow\rangle-2\lvert\uparrow\downarrow\uparrow\rangle+\lvert\uparrow\uparrow\downarrow\rangle), & E_{\mathrm{C}}=-\frac{3}{2}J.
	\end{array}\label{spin_bound state}
\end{equation}
The energies $E_i$ for the state $\lvert i\rangle$ satisfy the relation $3E_{\mathrm{A}}=E_{\mathrm{B}}+E_{\mathrm{C}}$. Then within the Krylov subspace any cluster characterized by three $\lvert\mathrm{A}\rangle$ states can be mixed with that by one $\lvert\mathrm{B}\rangle$ and one $\lvert\mathrm{C}\rangle$ state~\cite{footnote1}. Therefore, the three states span dynamically disconnected Krylov subspaces $\mathcal{K}_i$ [see Fig. \ref{fig: hsf}(a)], with each subspace labeled by two emergent conserved quantities $(O^{\mathcal{K}_i}_{1},O^{\mathcal{K}_i}_{2})$:
\begin{equation}\label{eq:ConservedNumber}	(O^{\mathcal{K}_i}_{1},O^{\mathcal{K}_i}_{2})=(N_{\mathrm{A}}^{\mathcal{K}_i}+\frac{3}{2}N_{\mathrm{B}}^{\mathcal{K}_i}+\frac{3}{2}N_{\mathrm{C}}^{\mathcal{K}_i},N_{\mathrm{B}}^{\mathcal{K}_i}-N_{\mathrm{C}}^{\mathcal{K}_i}),
\end{equation}
where $N_{\mathrm{A}}^{\mathcal{K}_i}$, $N_{\mathrm{B}}^{\mathcal{K}_i}$ and $N_{\mathrm{C}}^{\mathcal{K}_i}$ denote the numbers of the three types of spin bound states. 
The two quantities remain conserved under allowed transitions between configurations of spin bound states, thereby characterizing the quantum many-body dynamics within the Krylov subspace $\mathcal{K}_i$. A detailed analysis involving multiple spin states establishes a refined condition, $t/J\ll0.3714$ and $(t/J)^2\ll0.01474$, to fully suppress the transitions between Krylov subspaces and ensure robust HSF \cite{supplement}.

\begin{figure*}[t]
	\includegraphics[scale=1.1]{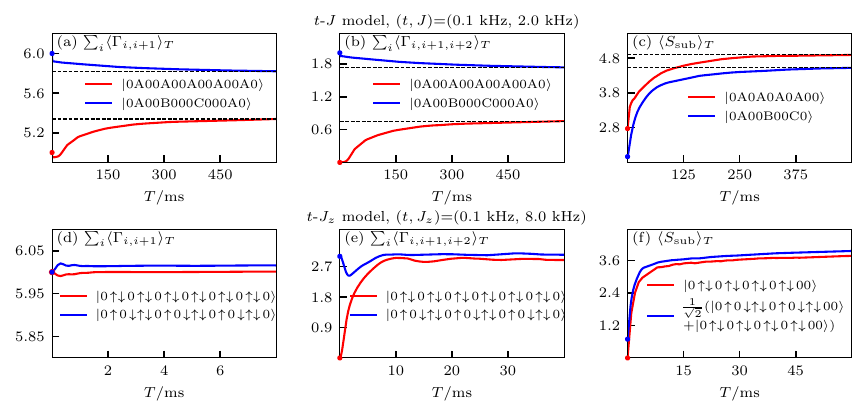}
\caption{\label{fig: nonthermal}
Nonthermal and thermal dynamics for the $t$-$J$ model and the $t$-$J_z$ model. Long-time averages of density correlation operators ($\Gamma_{i,i+1}=n_i n_{i+1}$, $\Gamma_{i,i+1,i+2}=n_i n_{i+1}n_{i+2}$) and sublattice entanglement entropy (EE) are calculated across various initial states for $t$-$J$ and $t$-$J_z$ models. (a-c) show the long-time averages of density correlation operators for a 20-site $t$-$J$ chain and averages of sublattice EE for a 14-site $t$-$J$ chain, with parameters $t=0.1$ kHz and $J=2.0$ kHz. (d-f) show the long-time averages of the density operator for a 19-site $t$-$J_z$ chain and sublattice EE for a 14-site $t$-$J_z$ chain, with parameters $t=0.1$ kHz and $J_z=8.0$ kHz. Compared to $t$-$J_z$ model, the long-time averages converge to significantly different values for $t$-$J$ model.  TEBD calculation used the bond dimensions up to 800 for (a-b), 400 for (c), 306 for (d-e), and 100 for (f). The blue and red solid lines in (f) corresponds to two different initial states with the same energy but different $S_{\mathrm{sub}}$. }
\end{figure*}
\textcolor{blue}{\em Many-body self-pinning effect.}--We predict
a highly nontrivial effect, dubbed many-body self-pinning, which emerges in conjunction with robust HSF in the $t$-$J$ model. This effect, enforced by the entangled nature of spin bound states, hinders transitions between distinct many-body configurations even within the same fragmented subspace. Using the time-evolving block decimation (TEBD) method~\cite{PhysRevLett.93.040502,SCHOLLWOCK201196}, we first numerically simulate a 14-site chain within the $(O^{\mathcal{K}_i}_{1},O^{\mathcal{K}_i}_{2})=(4.1)$ Krylov subspace. The basis of this subspace consists of two elementary configurations: one with any valid arrangement of four $\lvert \mathrm{A}\rangle$ states, and another with one $\lvert \mathrm{A}\rangle$, one $\lvert \mathrm{B}\rangle$, and one $\lvert \mathrm{C}\rangle$. The probability $p_c$ of observing these configurations, $c=\{\mathrm{4A}, \mathrm{1A1B1C}\}$, shows that the system evolves within the Krylov subspace labeled by the conserved quantities in Eq.~(\ref{eq:ConservedNumber}) of the initial states. From Fig. \ref{fig: hsf}(b)-(c), the total probability $p_{\mathrm{tot}}=p_{4\mathrm{A}}+p_{1\mathrm{A}1\mathrm{B}1\mathrm{C}}\approx1$, confirming the HSF of the system. Moreover, the dynamics strongly depend on the initial configuration, retaining a significant population in the initial pattern, manifesting the many-body self-pinning effect.

To elucidate the self-pinning mechanism, we provide a general framework and show how entanglement enforces this effect. Consider a general interacting Hamiltonian $H=\alpha_{\mathrm{kin}} H_{\mathrm{kin}}+\alpha_{\mathrm{int}} H_{\mathrm{int}}$ with kinetic part $H_{\mathrm{kin}}$ and interaction part $H_{\mathrm{int}}$ in the strong coupling regime $|\alpha_{\mathrm{int}}/\alpha_{\mathrm{kin}}|\gg 1$. For two resonant eigenstates $|\phi_{\mathrm{I}1}\rangle$ and $|\phi_{\mathrm{I}2}\rangle$ of $H_{\mathrm{int}}$, the ratio of the effective energy splitting $\delta_{\mathrm{eff}}$ to the transition amplitude $\lambda_{\mathrm{eff}}$ induced by $H_{\mathrm{kin}}$ is \cite{supplement}
\begin{equation}
	\frac{\delta_{\mathrm{eff}}}{\lambda_{\mathrm{eff}}}=\frac{\sum_{n=n^{\delta}_0}^{\infty} g_n^{\delta} (\alpha_{\mathrm{kin}}^{n}/\alpha_{\mathrm{int}}^{n-1})}{\sum_{n=n^{\lambda}_0}^{\infty} g_n^{\lambda} (\alpha_{\mathrm{kin}}^{n}/\alpha_{\mathrm{int}}^{n-1}) }\approx \left\{
	\begin{array}{c c}
		\infty, &  \ n_0^{\lambda} > n_0^{\delta},\\
		\mathrm{const.},  &  \ n_0^{\lambda} = n_0^{\delta},\\
		0, & \ n_0^{\lambda} < n_0^{\delta},
	\end{array}
	\right. \label{self-pin}
\end{equation}
where $n_0^{\lambda}$ and $n_0^{\delta}$ are the leading perturbative orders of $\lambda_{\mathrm{eff}}$ and $\delta_{\mathrm{eff}}$, respectively. When $n_0^{\delta}\leq n_0^{\lambda}$, transitions between $\lvert\phi_{\mathrm{I}1}\rangle$ and $\lvert\phi_{\mathrm{I}2}\rangle$ are suppressed, leaving the quantum state pinned to its initial configuration. Applying this framework to the $(O^{\mathcal{K}_i}_{1},O^{\mathcal{K}_i}_{2})=(4,1)$ subspace, detailed calculations for two fundamental configurations reveal  $n^{\lambda}_0 = n^{\delta}_0 = 2$ \cite{supplement}, explaining the self-pinning effect.

The self-pinning effect of the 1D $t$-$J$ model is a consequence of quantum entanglement of the spin bound states spanning the Krylov subspace. The transition between two many-body states with distinct entangled patterns, e.g. the minimal case being that from three $A$ states to one $B$ plus one $C$ states [Fig. \ref{fig: hsf}(a)], necessitates simultaneous tunneling processes of two or more particles, corresponding to second-order or higher-order perturbations. Then the diagonal splitting, as dominated by second-order perturbation, is always relevant and results in the self-pinning effect. In comparison, we examine the $t$-$J_z$ model \cite{footnotetjz} with $J_z/t \gg 1$, where the HSF is characterized by product spin bonding states $\uparrow\downarrow\cdots\uparrow\downarrow\cdots$ \cite{PhysRevB.101.125126}. The transition can be driven by single-particle tunneling that preserves total number of two-body spin bonds, hence $n_0^{\lambda}=1$ while $n_0^{\delta}\geq2$, indicating no self-pinning~\cite{supplement}.

\textcolor{blue}{\em Nonthermalization in the Krylov subspace.}--HSF leads to non-ergodicity and violations of eigenstate thermalization hypothesis (ETH) \cite{PhysRevA.43.2046,PhysRevE.50.888,Rigol2008,Deutsch_2018} across different subspaces, but thermalization occurs within one non-integrable Krylov subspace \cite{Hahn2021InformationDI,Moudgalya_2022,PhysRevX.15.011035,PhysRevB.111.144313,doi:10.1142/9789811231711_0009,PhysRevLett.124.207602}. The self-pinning effect predicted here renders a new mechanism breaking ETH within Krylov subspaces of $t$-$J$ model, 
which distinguishes it from previous works where non-ergodicity arises from integrability within each subspace~\cite{doi:10.1142/9789811231711_0009, PhysRevLett.124.207602,PhysRevB.100.214313,PhysRevB.103.134207}.
We demonstrate the nonthermal (thermal) dynamics for the $t$-$J$ model ($t$-$J_z$ model) in the strong coupling limits ($J/t \gg 1$ and $J_z/t \gg 1$). The key distinction lies in the quantum entanglement of few-body states in the $t$-$J$ model, which drives nonthermalization. Specifically, we analyze the saturated long-time average values of multi-site density correlations $\langle n_{i}n_{i+1} \rangle_T$ and $\langle n_{i}n_{i+1}n_{i+2} \rangle_T$, and the sublattice entanglement entropy defined by
\begin{equation}
    S_{\mathrm{sub}}=-\mathrm{Tr}[\rho_{\mathrm{sub}}\log(\rho_{\mathrm{sub}})],
\end{equation}
where $\rho_{\mathrm{sub}}=\mathrm{Tr}_{\mathrm{even\ sites}}(\rho)$, is the reduced density matrix obtained by tracing out the even sites.
The average of an observable $O$ over time $T$ reads $\langle O \rangle_T=(1/T)\int_0^{T} dt \langle O \rangle$. For $t$-$J$ model, $\sum_i\langle n_{i}n_{i+1} \rangle_T$ and $\sum_i\langle n_{i}n_{i+1}n_{i+2} \rangle_T$ show a strong dependence on the initial states within the same Krylov subspace [Fig.~\ref{fig: nonthermal}(a)-(b)]. These averages converge to distinct values, signaling nonthermalization. For the $t$-$J_z$ model, the averaged two body-correlation $\sum_i\langle n_{i}n_{i+1} \rangle_T$ relates to the total number of two-body spin pairs, and thus is constant for any state in the same Krylov subspace, as shown by the coincidence during evolution [Fig.~\ref{fig: nonthermal}(d)]. Further, the average of three-body correlation $\sum_i\langle n_{i}n_{i+1}n_{i+2} \rangle_T$ saturates to nearly identical values, implying the Krylov-restricted thermalization in $t$-$J_z$ model [Fig.~\ref{fig: nonthermal}(e)]. The normalized discrepancies approaches zero as the system size increases~\cite{supplement}.

The long-time average of the sublattice entanglement entropy $\langle S_{\mathrm{sub}} \rangle_T$ is shown in Fig.~\ref{fig: nonthermal}(c) and (f). For the $t$-$J$ model, $\langle S_{\mathrm{sub}} \rangle_T$ saturate to distinct values depending on the initial states, whereas for $t$-$J_z$ model, the values converge. The results demonstrate that within the Krylov subspaces of $t$-$J$ model, the long-time evolution cannot be described by a single statistical ensemble, while for $t$-$J_z$ model,  Krylov-restricted thermalization occurs. This contrast demonstrates that local entanglement proliferated self-pinning effect is a novel mechanism driving the breakdown of Krylov-restricted thermalization.

\textcolor{blue}{\em Conclusion and outlook.}-- We have proposed the realization of a highly tunable fermionic extended $t$-$J$ model through a programmable Rydberg-dressed tweezer array, and predicted an entanglement-enforced self-pinning effect in the $J/t\gg1$ regime, which renders a new mechanism of violating ETH and Krylov restricted thermalization. The quantum entanglement of the few-body states in the $t$-$J$ model characterizes a novel type of Hilbert space fragmentation, and is essential for the predicted non-thermal quantum many-body dynamics. Many interesting issues, such as the dissipative effects which can be engineered in experiment, deserve further investigation. Given the fundamental importance and broad interests of the $t$-$J$ model, this work opens up an intriguing avenue to explore the exotic quantum many-body physics beyond the conventional optical lattices and cuprates, and provides a controllable quantum platform to investigate the high-$T_\mathrm{c}$ SC physics.

\textcolor{blue}{\em Acknowledgments.}--This work was supported by National Key Research and Development Program of China (2021YFA1400900), the National Natural Science Foundation of China (Grants No. 12425401 and No. 12261160368), and the Innovation Program for Quantum Science and Technology (Grant No. 2021ZD0302000), and the Shanghai Municipal Science and Technology Major Project (Grant No.~2019SHZDZX01). The TEBD is implemented by Tenpy library \cite{Tenpy}.


%

\renewcommand{\thesection}{S-\arabic{section}}
\setcounter{section}{0}  
\renewcommand{\theequation}{S\arabic{equation}}
\setcounter{equation}{0}  
\renewcommand{\thefigure}{S\arabic{figure}}
\setcounter{figure}{0}  
\renewcommand{\thetable}{S-\Roman{table}}
\setcounter{table}{0}  
\onecolumngrid \flushbottom 
\newpage

\begin{center}
	\large \textbf{\large Supplemental Material: Quantum many-body dynamics for fermionic $t$-$J$ model simulated with atom arrays}
\end{center}

\section{The Rydberg-dressed states}
\label{app:oH}
In this section, we detail the dressing of the two ground states with Rydberg states through single-photon and two-photon transitions. To begin, we define the notation for the atomic states employed throughout the context. The ground states are denoted as $\lvert n ^{2s+1}L_{J},F,m_f\rangle$, where $n$ is the principal quantum number, $s$ is the spin angular momentum quantum number, $L$ refers to the orbital angular momentum quantum number, $J$ is the total electronic angular momentum quantum number, and $F$ and $m_f$ are the total angular momentum quantum number and the associated magnetic quantum number. The Rydberg states are denoted as $\lvert n^{2s+1}L_{J},m_j\rangle\otimes\lvert I, m_i\rangle$, where $m_j$ is the magnetic quantum number associated with $J$, and $I$ and $m_i$ is the nuclear angular momentum quantum number and the associated magnetic quantum number.  Due to the weak hyperfine coupling in highly excited Rydberg states, the nuclear spin $I$ remains effectively decoupled from the total electronic angular momentum $J$.

\begin{figure}[H]
	\begin{centering}
		\includegraphics[scale=1.2]{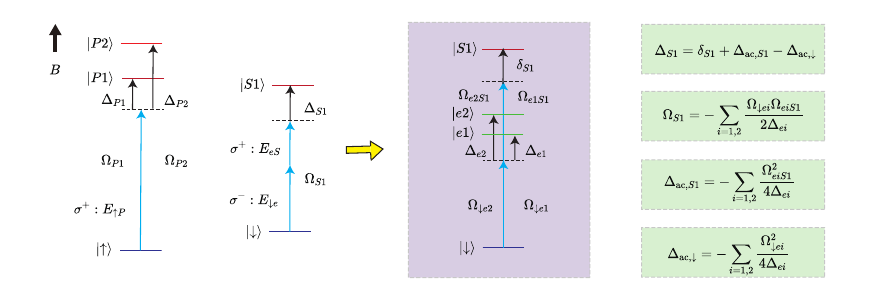}
		\caption{Detailed sketch of the single-photon and two-photon process. $\Omega$ characterizes the transition amplitude between two atom states, and $\Delta$ and $\delta$ characterize the relative detuning between two atom states.  Left: The magnetic field $B$, perpendicular to the lattice, determine the quantization axis of the $^{40}\mathrm{K}$ atoms. $\lvert\uparrow\rangle$ is dressed to Rydberg p-oribital states $\lvert P1(2)\rangle$ via a single-photon process using $\sigma^{+}$ polarized light with the electric field strength $E_{\uparrow P}$. $\lvert\downarrow\rangle$ is dressed to Rydberg $s$-orbital states $\lvert S1\rangle$ via a two-photon process using $\sigma^{+}$ polarized light with the electric field strength $E_{\downarrow e}$ and $\sigma^{-}$ polarized light with the electric field strength $E_{eS}$. Purple box: detailed sketch of the complete two-photon transition process including the intermediate state $\lvert e1(2)\rangle$. Green box: derivation of parameters of two-photon transitions after the elimination of the intermediate states. $\Delta_{\mathrm{ac},S1}$ and $\Delta_{\mathrm{ac},\downarrow}$ denote the AC Stark shifts for $\lvert S1\rangle$ and $\lvert\downarrow\rangle$. 
		}
		\label{fig: parameter_detail}
		\par\end{centering}
\end{figure}

As illustrated in Fig.~\ref{fig: parameter_detail}, the spin-up state $\lvert\uparrow\rangle=\lvert4^2S_{\frac{1}{2}},F=\frac{9}{2},m_f=-\frac{9}{2}\rangle$ in the hyperfine manifold of $^{40}\mathrm{K}$ is dressed to Rydberg $p$-orbital states $\lvert P1(2)\rangle=\lvert65^2P_{\frac{1}{2}(\frac{3}{2})},m_j=\frac{1}{2}\rangle \otimes\lvert I=4,m_i=-4\rangle$ via a single-photon process using $\sigma^{+}$ polarized light. This process is described by the effective dressing Hamiltonian $H_{\uparrow,\mathrm{dress}} = H_{\Omega,\mathrm{\uparrow}}+H_{\Delta,\uparrow}$, where $H_{\Delta,\uparrow}$ represents the on-site detuning and  $H_{\Omega,\mathrm{\uparrow}}$ describes the coupling between the bare state $\lvert\uparrow\rangle$ and Rydberg $p$-orbit states, which are given by 
\begin{equation}
	\begin{aligned}
		H_{\Omega,\uparrow} &=\sum_{i}\left(\frac{1}{2}\Omega_{P1}\, c^{\dagger}_{i,P1}c_{i,\uparrow}+\frac{1}{2}\Omega_{P2} \, c^{\dagger}_{i,P2}c_{i,\uparrow}+\mathrm{H.c.}\right), \\
		H_{\Delta,\uparrow}&=\sum_{i}\left(\Delta_{P1}\, n_{i,P1}+\Delta_{P2} \, n_{i,P2}\right).
	\end{aligned}\label{HSP}
\end{equation} 
Here, $\Omega_{P1(2)}$ and $\Delta_{P1(2)}$  denote the transition amplitude and relative detuning between $\lvert P1(2)\rangle$ and $\lvert\uparrow\rangle$. The spin-down state, $\lvert\downarrow\rangle=\lvert4^2S_{\frac{1}{2}},F=\frac{7}{2},m_f=-\frac{7}{2}\rangle$, is dressed to Rydberg $s$-orbital state $\lvert S1\rangle=\lvert65^2S_{\frac{1}{2}},m_j=\frac{1}{2}\rangle
\otimes \lvert I=4,m_i=-4\rangle$ via a two-photon process involving $\sigma^{+}$ and $\sigma^{-}$ polarized light, as shown in Fig.~\ref{fig: parameter_detail}. This process incorporates intermediate states $\lvert e1(2)\rangle=\lvert {30}^{2}P_{\frac{1}{2}(\frac{3}{2})},m_{j}=-\frac{1}{2}\rangle \otimes\lvert I=4,m_{i}=-4\rangle$ (highlighted in the purple box in Fig.~\ref{fig: parameter_detail}). The effective dressing Hamiltonian including intermediate states for this transition is similarly given by $H_{\downarrow,\mathrm{dress}}^{\mathrm{tot}}=H_{\Omega,\downarrow}^{\mathrm{tot}}+H_{\Delta,\downarrow}^{\mathrm{tot}}$, with $H_{\Delta,\downarrow}^{\mathrm{tot}}$ labeling the on-site detuning and  $H_{\Omega,\mathrm{\downarrow}}^{\mathrm{tot}}$ represents the coupling between the bare state $\lvert\downarrow\rangle$ and Rydberg $s$-orbital state,
\begin{equation}
	\begin{aligned}
		H_{\Omega,\downarrow}^{\mathrm{tot}}=&\sum_{i}\left(\frac{1}{2}\Omega_{\downarrow e1}\, c^{\dagger}_{i,e1}c_{i,\downarrow}+\frac{1}{2}\Omega_{\downarrow e2}\, c^{\dagger}_{i,e2}c_{i,\downarrow}+\right.\\
		&\left. +\frac{1}{2}\Omega_{e1S1} \, c^{\dagger}_{i,S1}c_{i,e1}+\frac{1}{2}\Omega_{e2S1} \, c^{\dagger}_{i,S1}c_{i,e2}+\mathrm{H.c.}\right), \\
		H_{\Delta,\downarrow}^{\mathrm{tot}}=&\sum_{i}\left(\Delta_{e1}\, n_{i,e1}+\Delta_{e2} \, n_{i,e2}+\delta_{S1} \, n_{i,S1}\right).
	\end{aligned}\label{HTP_tot}
\end{equation}
Here, $\Omega_{\downarrow e1(2)}$ and $\Omega_{e1(2)S1}$ represent the transition amplitudes between $\lvert\downarrow\rangle$, $\lvert e1(2)\rangle$ and $\lvert S1\rangle$, while  $\Delta_{e1(2)}$ and $\delta_{S1}$ denote the relative detunings of  $\lvert e1(2)\rangle$ and $\lvert S1\rangle$ to $\lvert\downarrow\rangle$. By eliminating the intermediate states $\lvert e1(2)\rangle$, the effective dressing Hamiltonian $H_{\downarrow,\mathrm{dress}}^{\mathrm{tot}}$ is significantly simplified into $H_{\downarrow,\mathrm{dress}}=H_{\Omega,\downarrow}+H_{\Delta,\downarrow}$ as
\begin{equation}
	\begin{aligned}
		H_{\Omega,\downarrow} &=\sum_{i}\left(\frac{1}{2}\Omega_{S1}\, c^{\dagger}_{i,S1}c_{i,\downarrow} +\mathrm{H.c.}\right), \\
		H_{\Delta,\downarrow}&=\sum_{i}\left(\Delta_{S1}\, n_{i,S1}\right).
	\end{aligned}\label{HTP}
\end{equation} 
Here, $\Omega_{S1}$ represents the effective two-photon transition amplitude between  $\lvert \downarrow\rangle$ and $\lvert S1\rangle$, while $\Delta_{S1}$ denotes the relative detuning between $\lvert S1\rangle$ and $\lvert\downarrow\rangle$ after including corrections from the AC Stark shifts (green box in Fig. \ref{fig: parameter_detail}). Notably, the single-photon (two-photon) process does not couple the spin state $\lvert \downarrow\rangle$ ($\lvert \uparrow\rangle$) to the Rydberg $p$-orbital ($s$-orbital) states due to the large single-particle energy difference of approximately 1.286 GHz between $\lvert \uparrow\rangle$ and $\lvert\downarrow\rangle$. This energy difference does not influence $H_{J}^{\mathrm{tot}}$ owing to the particle number conservation of both $\lvert \uparrow\rangle$ and $\lvert \downarrow \rangle$.

The two-photon process introduces an additional coupling channel, connecting  $\lvert\downarrow\rangle$ to $\lvert S3\rangle=\lvert {65}^{2}S_{\frac{1}{2}},m_{j}=-\frac{1}{2}\rangle \otimes \lvert I=4,m_{i}=-3\rangle$ through the intermediate state $\lvert e3\rangle=\lvert{30}^{2}P_{\frac{3}{2}},m_{j}=-\frac{3}{2}\rangle \otimes \lvert I=4,m_{i}=-3\rangle$. In the presence of DDI, this channel could in principle induce a resonant transition from $\lvert\uparrow\downarrow\rangle$ to another pair of hyperfine states of $^{40}\mathrm{K}$, specifically $\lvert4^2 S_{\frac{1}{2}},F=\frac{9}{2},m_f=-\frac{7}{2}\rangle$ and $\lvert4^2 S_{\frac{1}{2}},F=\frac{7}{2},m_f=-\frac{5}{2}\rangle$. However, numerical simulations in Sec.~\ref{app:ED} indicate that, under the experimental parameters described in Sec.~\ref{app:parameters}, this coupling channel is too weak to produce any physically significant effects and can therefore be safely neglected.

\section{dipole-dipole and VdW interaction}
\label{app:ddi and vdw}
In this section, we detail the relevant Rydberg dipole-dipole and van der Waals (vdW) interaction in our scheme. The Rydberg dipole-dipole interaction (DDI) couples different Rydberg pairs with one $p$-oribital and one $s$-orbital. Assuming the quantization axis of atoms determined by the magnetic field is always perpendicular to the lattice, the Rydberg DDI operator is ($\vec{d}=e\vec{r}$)
\begin{equation}
	V_{\mathrm{DDI}}=\frac{1}{4\pi\epsilon_{0}}\frac{-0.5d_{1+}d_{2-}-0.5d_{1-}d_{2+}+d_{1z}d_{2z}-1.5d_{1+}d_{2+}-1.5d_{1-}d_{2-}}{x^{3}},\label{eq:ddi}
\end{equation}
where $d_{\pm}=(d_x\pm id_y)/\sqrt{2}$, and $x$ is the distance between two atoms. In our scheme, the operator $V_{\mathrm{DDI}}$ ultimately couples a total of 12 Rydberg pairs exhibiting energy differences on the order of DDI, denoted as $\lvert P1S1\rangle$, $\lvert S1P1\rangle$, $ \lvert P3S2\rangle$, $
\lvert S2P3\rangle$, $\lvert P2S1\rangle$, $\lvert S1P2\rangle$, $ \lvert P4S2\rangle$, $\lvert S2P4\rangle$, $\lvert P5S2\rangle$, $\lvert S2P5\rangle$, $\lvert P6S1\rangle$, $\lvert S1P6\rangle
$. The definition of $\lvert P1(2)\rangle$ and $\lvert S1\rangle$ is given in Sec.~\ref{app:oH}, and other states are defined as $\lvert P3\rangle=\lvert {65}^{2}P_{\frac{1}{2}},m_{j}=-\frac{1}{2}\rangle\otimes\lvert I=4,m_{i}=-4\rangle$, $\lvert P4\rangle=\lvert {65}^{2}P_{\frac{3}{2}},m_{j}=-\frac{1}{2}\rangle\otimes\lvert I=4,m_{i}=-4\rangle$, $\lvert P5\rangle=\lvert {65}^{2}P_{\frac{3}{2}},m_{j}=\frac{3}{2}\rangle\otimes\lvert I=4,m_{i}=-4\rangle$, $\lvert P6\rangle=\lvert {65}^{2}P_{\frac{3}{2}},m_{j}=-\frac{3}{2}\rangle\otimes\lvert I=4,m_{i}=-4\rangle$, $\lvert S2\rangle=\lvert {65}^{2}S_{\frac{1}{2}},m_{j}=-\frac{1}{2}\rangle\otimes\lvert I=4,m_{i}=-4\rangle$. The relevant Hamiltonian $H_{\mathrm{DDI}}$ is formulated by the representation of $V_{\mathrm{DDI}}$ in the Hilbert space spanned by 12 Rydberg pairs as
\begin{equation}
	H_{\mathrm{DDI}}= \sum_{i\neq j,m,n,k,l} \left( \frac{\mathrm{C}_{3,(m,n,k,l)}}{|\vec{x}_i-\vec{x}_j|^3}\ c^{\dagger}_{i,Pm}c^{\dagger}_{j,Sn} c_{j,Pk} c_{i,Sl}+\mathrm{H.c.} \right), \label{Hddi}
\end{equation}
where $\mathrm{C}_{3,(m,n,k,l)}$ denotes the coefficients of Rydberg DDI and $\vec{x}_{i(j)}$ represents the position vector of atom $i(j)$. Besides, the relative detuning of additional Rydberg $p$-orbital states to $\lvert\uparrow\rangle$ (denoted as $\Delta_{P3(4,5,6)}$) and additional $s$-orbital state to $\lvert\downarrow\rangle$ (denoted as $\Delta_{S2}$) is incorporated as
\begin{equation}
	H_{\Delta,\mathrm{DDI}}=\sum_i \left(\Delta_{P3}\, n_{i,P3}+\Delta_{P4} \, n_{i,P4}+\Delta_{P5} \, n_{i,P5}+\Delta_{P6} \, n_{i,P6}+\Delta_{S2} \, n_{i,S2}\right).\label{hddi_delta}
\end{equation}
As outlined in the main text, the total Hamiltonian $H_J^\mathrm{tot}$ can be expressed as $H_{\Omega}+H_{\mathrm{DDI}}+H_{\Delta}$, with $H_{\Omega}=H_{\Omega,\uparrow}+H_{\Omega,\downarrow}$ and $H_{\Delta}=H_{\Delta,\uparrow}+H_{\Delta,\downarrow}+H_{\Delta,\mathrm{DDI}}$.

The Rydberg van der Waals (vdW) interaction $H_{\mathrm{vdW}}$ exists between the relevant Rydberg states with the same parity and is formulated as
\begin{equation}
	\begin{aligned}
		H_{\mathrm{vdW}}= &\sum_{i\neq j} \left[\frac{\mathrm{C}_{6,P1P1}}{|\vec{x}_i-\vec{x}_j|^6}\, n_{i,P1}n_{j,P1}+\frac{\mathrm{C}_{6,P2P2}}{|\vec{x}_i-\vec{x}_j|^6}\, n_{i,P2}n_{j,P2}+\frac{\mathrm{C}_{6,S1S1}}{|\vec{x}_i-\vec{x}_j|^6}\, n_{i,S1}n_{j,S1}+\right.\\  &\left.+\frac{\mathrm{C}_{6,P1P2}}{|\vec{x}_i-\vec{x}_j|^6} \, (n_{i,P1}n_{j,P2}+n_{i,P2}n_{j,P1})\right],
	\end{aligned}\label{HVdw}
\end{equation}
where $\mathrm{C}_6$ represents the coefficients of Rydberg vdW interaction, and $\vec{x}_{i(j)}$ denotes the position vector of atom $i$($j$).  The term $H_{\mathrm{vdW}}$ generates the Rydberg-dressed density-density interaction for $\lvert\uparrow\uparrow\rangle$ and $\lvert\downarrow\downarrow\rangle$ \cite{PhysRevA.82.033412}, expressed as $V_{\uparrow\uparrow,ij}n_{i,\uparrow}n_{j,\uparrow}+V_{\downarrow\downarrow,ij}n_{i,\downarrow}n_{j,\downarrow}$. Numerical simulations in Sec.~\ref{app:ED} show that, under the experimental parameters described in Sec.~\ref{app:parameters}, $V_{\uparrow\uparrow}$ and $V_{\downarrow\downarrow}$ are negligible compared to $J_{\mathrm{ex}}$ and $J_z$, and are thus omitted from $H_J^{\mathrm{tot}}$.

\section{Experimental parameters}
\label{app:parameters}
In this section, we present the relevant experimental parameters in $H_J^{\mathrm{tot}}$. All the data and calculations are based on the Alkali Rydberg Calculator (ARC) Toolbox \cite{SIBALIC2017319}. Firstly, we list the typical strengths of the electric fields $E$ of the polarized light and the magnetic field $B$, as depicted in Fig. \ref{fig: parameter_detail}, in Table \ref{tab:em_field}.

\begin{table}[h]
	\centering
	\begin{tabular}{|c|c|c|c|c|c|c|c|}
		\hline $B$ & $\mu_{B}B$ & $E_{\uparrow P}$ & $E_{\downarrow e}$ & $E_{e S}$\\
		\hline -6.25 $\text{G}$ & -8.75 $\text{MHz}$ & $1.658\times10^{5}$ $\text{V/m}$ & $8.062\times10^{5}$ $\text{V/m}$ & $1.855\times10^{4}$ $\text{V/m}$ \\
		\hline
	\end{tabular} 
	\caption{The typical strengths of the electric fields $E$ and the magnetic field $B$, as depicted in Fig. \ref{fig: parameter_detail}.} \label{tab:em_field}
\end{table}

Secondly, we list the typical experimental parameters for $H_{\uparrow_,\mathrm{dress}}$ [Eq.~(\ref{HSP})] in Table \ref{tab:sp}, $H_{\downarrow,\mathrm{dress}}$ [Eq.~(\ref{HTP})] in Table \ref{tab:tp1}, and $H_{\downarrow,\mathrm{dress}}^{\mathrm{tot}}$ [(Eq.~\ref{HTP_tot})] in Table \ref{tab: tp2}. The lifetime $\tau$ is calculated at the cryogenic temperature of 4 K, which is feasible with the current experimental techniques \cite{PhysRevApplied.16.034013,PhysRevApplied.22.024073}. 
\begin{table}[h]
	\centering
	\begin{tabular}{|c|c|c|c|c|}
		\hline $\text{Ryd.}$ & $\tau_{\text{Ryd}}$ $(4$ $\mathrm{K})$  & $\Delta_{P1(2)}/(2\pi)$ & $\Omega_{P1(2)}/(2\pi)$ & $\tau_{\uparrow}$ $(4$ $\mathrm{K})$\\
		\hline $65^{2}P_{\frac{1}{2}}$ & $0.956$ $\text{ms}$  & $5.00$ $\text{MHz}$ & $1.29$ $\text{MHz}$ & $58.47$ $\text{ms}$\\
		\hline $65^{2}P_{\frac{3}{2}}$ & $0.931$ $\text{ms}$ & $81.08$ $\text{MHz}$ & $1.02$ $\text{MHz}$ & $24.6$ $\text{s}$
		\\\hline
	\end{tabular} 
	\caption{The typical experimental parameters for $H_{\uparrow,\mathrm{dress}}$ in Eq.~(\ref{HSP}), as depicted in Fig. \ref{fig: parameter_detail}. $\tau_{\mathrm{Ryd}}$ denotes the lifetime of $\lvert P1(2)\rangle$, and $\tau_{\uparrow}$ denotes the lifetime of $\lvert\uparrow\rangle$ dressed to the Rydberg $p$-orbital states $\lvert P1(2)\rangle$. } \label{tab:sp}
\end{table}
\begin{table}[h]
	\centering
	\begin{tabular}{|c|c|c|c|c|c|c|c|}
		\hline $\text{Ryd.}$ & $\tau_{\text{Ryd}}$ $(4$ $\mathrm{K})$ & $\Delta_{S1}/(2\pi)$ & $\Omega_{S1}/(2\pi)$ & $\Delta_{ac,S1}/(2\pi)$ & $\Delta_{ac,\downarrow}/(2\pi)$ & $\delta_{S1}/(2\pi)$ & $\tau_{\downarrow}$ $(4$ $\mathrm{K})$\\
		\hline $65^{2}S_{\frac{1}{2}}$ & $0.287$ $\text{ms}$ & $10.078$ $\text{MHz}$ & $-1.406$ $\text{MHz}$ & $-6.653$ $\text{MHz}$ & $-0.074$ $\text{MHz}$& $16.657$ $\text{MHz}$& $58.935$ $\text{ms}$
		\\\hline
	\end{tabular} 
	\caption{The typical experimental parameters for $H_{\downarrow,\mathrm{dress}}$ in Eq.~(\ref{HTP}), as depicted in Fig. \ref{fig: parameter_detail}. $\tau_{\mathrm{Ryd}}$ denotes the lifetime of $\lvert S1\rangle$, and $\tau_{\downarrow}$ denotes the lifetime of $\lvert\downarrow\rangle$ dressed to the Rydberg $s$-orbital state $\lvert S1\rangle$. } \label{tab:tp1}
\end{table}
\begin{table}[!htb]
	\centering
	\begin{tabular}{|c|c|c|c|c|c|}
		\hline $\text{Ryd.}$ & $\tau_{\text{Ryd}}$ $(4$ $\mathrm{K})$ & $\Delta_{e1(2)}/(2\pi)$ & $\Omega_{\downarrow e1(2)}/(2\pi)$ & $\tau_{\downarrow}$ $(4$ $\mathrm{K})$ & $\Omega_{e1(2)S1}/(2\pi)$\\
		\hline $30^{2}P_{\frac{1}{2}}$ & $0.0886$ $\text{ms}$ & $2000$ $\text{MHz}$ & $20.40$ $\text{MHz}$ & $3.406$ $\text{s}$ & $199.4$ $\text{MHz}$\\
		\hline $30^{2}P_{\frac{3}{2}}$ & $0.0862$ $\text{ms}$ & $2887.02$ $\text{MHz}$ & $16.097$ $\text{MHz}$ & $11.09$ $\text{s}$ & $139.5$ $\text{MHz}$
		\\\hline
	\end{tabular} 
	\caption{The typical experimental parameters for $H_{\downarrow,\mathrm{dress}}^{\mathrm{tot}}$ in Eq.~(\ref{HTP_tot}), as depicted in Fig. \ref{fig: parameter_detail}. $\tau_{\mathrm{Ryd}}$ denotes the lifetime of $\lvert e1(2)\rangle$, and $\tau_{\downarrow}$ denotes the lifetime of $\lvert\downarrow\rangle$ dressed to the intermediate states $\lvert e1(2)\rangle$. } \label{tab: tp2}
\end{table}

Thirdly, we list the $
\mathrm{C}_3$ coefficients of the Rydberg DDI for $H_{\mathrm{DDI}}$ [Eq.~(\ref{Hddi})] in Table \ref{tab:C3} (A subtle but important point regarding the use of the ARC toolbox is the definition of $d_{\pm
}$: in ARC, $d_\pm$ is defined as $\mp(d_x \pm i d_y)/\sqrt{2}$, whereas in our convention it is taken as $(d_x \pm i d_y)/\sqrt{2}$.). The symmetry of $
\mathrm{C}_3$ helps to reduce the number of independent coefficients. For instance, ${\mathrm{C}}_{3,(m,n,k,l)}$ satisfies ${\mathrm{C}}_{3,(m,n,k,l)}={\mathrm{C}}_{3,(n,m,l,k)}$ by definition, and spin-inversion symmetry further leads to additional constraints. So we only list the non-zero independent $\mathrm{C}_3$ coefficients for $H_{\mathrm{DDI}}$ in Table \ref{tab:C3}. Besides, we also list the detuning of other relevant Rydberg states for $H_{\Delta,\mathrm{DDI}}$ [Eq.~(\ref{hddi_delta})] in Table \ref{tab:ryd_detuning}.

\begin{table}[!htp]
	\centering
	\begin{tabular}{|c|c|c|c|c|}
		\hline  ${\mathrm{C}}_{3,(1,1,1,1)}$  & ${\mathrm{C}}_{3,(1,1,3,2)}$ & ${\mathrm{C}}_{3,(2,1,2,1)}$ & ${\mathrm{C}}_{3,(2,1,4,2)}$ & ${\mathrm{C}}_{3,(5,2,6,1)}$ \\
		\hline $2.306$ $\mathrm{GHz/\mu m^3}$ & $-6.918$ $\mathrm{GHz/\mu m^3}$ & $4.580$ $\mathrm{GHz/\mu m^3}$ & $3.435$ $\mathrm{GHz/\mu m^3}$ & $10.305$ $\mathrm{GHz/\mu m^3}$ \\
		\hline ${\mathrm{C}}_{3,(1,1,2,1)}$ & ${\mathrm{C}}_{3,(1,1,4,2)}$ & ${\mathrm{C}}_{3,(1,1,5,2)}$ & ${\mathrm{C}}_{3,(2,1,5,2)}$ & $\sim$\\
		\hline $-3.250$ $\mathrm{GHz/\mu m^3}$ & $4.875$ $\mathrm{GHz/\mu m^3}$ & $-2.8145$ $\mathrm{GHz/\mu m^3}$ & $-1.983$ $\mathrm{GHz/\mu m^3}$ & $\sim$\\
		\hline
	\end{tabular} 
	\caption{The non-zero independent $\mathrm{C}_3$ coefficients for $H_{\mathrm{DDI}}$ in Eq.~(\ref{Hddi}).} \label{tab:C3}
\end{table}

\begin{table}[H]
	\centering
	\begin{tabular}{|c|c|c|c|c|}
		\hline  $\Delta_{P3}/(2\pi)$  & $\Delta_{P4}/(2\pi)$ & $\Delta_{P5}/(2\pi)$ & $\Delta_{P6}/(2\pi)$ & $\Delta_{S2}/(2\pi)$\\
		\hline  $10.83$ $\text{MHz}$ & $92.75$ $\text{MHz}$ & $69.42$ $\text{MHz}$ & $104.41$ $\text{MHz}$& $34.16$ $\text{MHz}$\\
		\hline
	\end{tabular} 
	\caption{The detuning of other relevant Rydberg states for $H_{\Delta,\mathrm{DDI}}$ in Eq.~(\ref{hddi_delta}).} \label{tab:ryd_detuning}
\end{table}

Finally, we list the $\mathrm{C}_6$ coefficients of the Rydberg vdW interaction for $H_{\mathrm{vdW}}$ [Eq.~(\ref{HVdw})] in Table \ref{tab:C6}. 

\begin{table}[!htp]
	\centering
	\begin{tabular}{|c|c|c|c|}
		\hline  ${\mathrm{C}}_{6,P1P1}$  & ${\mathrm{C}}_{6,P2P2}$ & ${\mathrm{C}}_{6,P1P2}$ & ${\mathrm{C}}_{6,S1S1}$  \\
		\hline $228.5$ $\mathrm{GHz/\mu m^3}$ & $188.94$ $\mathrm{GHz/\mu m^3}$ & $-148.53$ $\mathrm{GHz/\mu m^3}$ & $-36.46$ $\mathrm{GHz/\mu m^3}$ \\
		\hline
	\end{tabular} 
	\caption{The $\mathrm{C}_6$ coefficients for $H_{\mathrm{vdW}}$ in Eq.~(\ref{HVdw}).} \label{tab:C6}
\end{table}

\section{Perturbation estimations of $J_{\mathrm{ex}}$ and $J_z$}
\label{app:perturbation}

In this section, we give an estimation of $J_{\mathrm{ex}}$ and $J_z$ by perturbation theory. The total Hamiltonian $H_J^{\mathrm{tot}}$ is decomposed into two parts: $H_{\mathrm{DDI}}+H_{\Delta}$ serves as the unperturbed Hamiltonian, while $H_{\Omega}$ is treated as the perturbation. Only even-order perturbation processes contribute to the low-energy effective Hamiltonian $H_J^{\mathrm{eff}}$ within the low-energy subspace composed of spin states $\lvert \uparrow\rangle$ and $\lvert \downarrow\rangle$. Noting that the second-order perturbation processes constitute pure single-particle energy corrections, their contribution adds up to a constant due to the conservation of particle number for both $\lvert\uparrow\rangle$ and $\lvert\downarrow\rangle$. Therefore, the low-energy effective Hamiltonian is formulated as
\begin{equation}
	H_J^{\mathrm{eff}}=\sum_{n=4,6,8,\cdots}H_{\Omega}(GH_{\Omega})^{n-1},
\end{equation}
where $G=(1-P_0)(E_0-H_{\Delta}-H_{\mathrm{DDI}})^{-1}(1-P_0)$. Here $P_0$ is the projection operator in the low-energy subspace, and $E_0$ is the unperturbed energy. We begin by considering only the leading-order term in the effective Hamiltonian, namely the fourth-order process of the form $H_{\Omega}G H_{\Omega}G H_{\Omega}GH_{\Omega}$ in $H_J^{\mathrm{eff}}$.

To obtain the unperturbed eigenstates, we need to diagonalize  $H_{\Delta}+H_{\mathrm{DDI}}$. However, if all the relevant twelve Rydberg pairs are considered, the full matrix can not be diagonalized analytically. As a first estimation of $J_{\mathrm{ex}}$ and $J_z$, we reserve only four main bases with the ordering as $\lvert P1S1\rangle$, $ \lvert S1P1\rangle$, $\lvert P3S2\rangle$, $\lvert	S2P3\rangle$. The matrix representation of $H_{\Delta}+H_{\mathrm{DDI}}$ with these four main bases is
\begin{equation}
	\begin{aligned}
		H_{\Delta}+H_{\mathrm{DDI}}&=
		\begin{pmatrix}
			\Delta_{P1}+\Delta_{S1} & W_1(x) & 0 & W_2(x) \\
			W_{1}(x) & \Delta_{P1}+\Delta_{S1} & W_{2}(x) & 0\\
			0 & W_{2}(x) & \Delta_{P3}+\Delta_{S2} & W_{1}(x)\\
			W_{2}(x) & 0 & W_{1}(x) & \Delta_{P3}+\Delta_{S2}
		\end{pmatrix}\\
		& = \Delta_{+}I\otimes I+\Delta_{-}\sigma_{z}\otimes I+W_{1}(x)I\otimes\sigma_{x}+W_{2}(x)\sigma_{x}\otimes\sigma_{x}.
	\end{aligned}\label{ddi_matrix}
\end{equation}
Here $\Delta_{\pm}=[(\Delta_{P1}+\Delta_{S1})\pm(\Delta_{P3}+\Delta_{S2})]/2$, $W_1(x) ={\mathrm{C}}_{3,(1,1,1,1)}/x^3$, $W_2(x)=\mathrm{C}_{3,(1,1,3,2)}/x^3$, and $x$ is the two-atom distance. The four eigenstates and corresponding energies of matrix (\ref{ddi_matrix}) are
\begin{equation}
	\begin{array}{ll}
		\lvert\psi_{+,1}\rangle=\cos(\frac{\theta}{2})\lvert P1S1_{+}\rangle+\sin(\frac{\theta}{2})\lvert P3S2_{+}\rangle,& E_{+,1}=\Delta_{+}+W_{1}+\sqrt{\Delta_{-}^{2}+W_{2}^{2}}, \\
		\lvert\psi_{+,2}\rangle=-\sin(\frac{\theta}{2})\lvert P1S1_{+}\rangle+\cos(\frac{\theta}{2})\lvert P3S2_{+}\rangle, & E_{+,2}=\Delta_{+}+W_{1}-\sqrt{\Delta_{-}^{2}+W_{2}^{2}},\\
		\lvert\psi_{-,1}\rangle=\cos(\frac{\theta}{2})\lvert P1S1_{-}\rangle+\sin(\frac{\theta}{2})\lvert P3S2_{-}\rangle ,& E_{-,1}=\Delta_{+}-W_{1}+\sqrt{\Delta_{-}^{2}+W_{2}^{2}}, \\
		\lvert\psi_{-,2}\rangle=-\sin(\frac{\theta}{2})\lvert P1S1_{-}\rangle+\cos(\frac{\theta}{2})\lvert P3S2_{-}\rangle, &
		E_{-,2}=\Delta_{+}-W_{1}-\sqrt{\Delta_{-}^{2}+W_{2}^{2}}.
	\end{array}
\end{equation}
Here $\tan(\theta)=W_{2}/\Delta_{-}$, and $\lvert {P1S1}_{\pm}\rangle=(\lvert P1S1\rangle\pm\lvert S1P1\rangle)/\sqrt{2}$, $\lvert {P3S2}_{\pm}\rangle=(\pm\lvert P3S2\rangle+\lvert S2P3\rangle)/\sqrt{2}$.

Then $J_{\mathrm{ex}}$ and $J_z$ can be estimated. Since $J_{\mathrm{ex},ij}(S_{i,x}S_{j,x}+S_{i,y}S_{j,y})$ leads to the exchange process between $\lvert\uparrow\downarrow\rangle$ and $\lvert\downarrow\uparrow\rangle$, the specific value of $J_{\mathrm{ex}}$ can be determined as
\begin{equation}
	J_{\mathrm{ex}}=2\langle \downarrow \uparrow\rvert H^{\mathrm{eff}}_J\lvert \uparrow \downarrow\rangle.
\end{equation}
And the fourth-order off-diagonal channels of $H_J^{\mathrm{eff}}$ generating $J_{\mathrm{ex}}$ are 
\begin{equation}
	\lvert\uparrow \downarrow\rangle \stackrel{H_{\Omega}}{\longleftrightarrow} \lvert\uparrow \!S1\rangle, \lvert P1(2)\! \downarrow\rangle \stackrel{H_{\Omega}} {\longleftrightarrow} \lvert\psi_{+,1(2)} \rangle, \lvert\psi_{-,1(2)} \rangle\stackrel{H_{\Omega}}{\longleftrightarrow}\lvert S1\!\uparrow\rangle, \lvert \downarrow\! P1(2)\rangle\stackrel{H_{\Omega}}{\longleftrightarrow}\lvert\downarrow \uparrow\rangle.\label{channel_jex}
\end{equation}
The detailed calculations provide an analytical estimation of $J_{\mathrm{ex}}$ as
\begin{equation}
	J_{\mathrm{ex}}= 2(\frac{\Omega_{P1}}{2\Delta_{P1}})^{2}(\frac{\Omega_{S1}}{2\Delta_{S1}})^{2}	\frac{(\Delta_{P1}+\Delta_{S1})^{2}(\Delta_{+}^{2}W_{1}-W_{1}^{3}+W_{1}\Delta_{-}^{2}+W_{1}W_{2}^{2}-2\Delta_{-}\Delta_{+}W_{1})}{[(\Delta_{+}-W_{1})^{2}-(\Delta_{-}^{2}+W_{2}^{2})][(\Delta_{+}+W_{1})^{2}-(\Delta_{-}^{2}+W_{2}^{2})]}.
	\label{eq:jex}
\end{equation}
The Ising interaction $J_{z,ij}(S_{i,z}S_{j,z}-1/4n_in_j)$ can be reformulate as the density-density interaction $-J_{z,ij}/2(n_{i,\uparrow}n_{j,\downarrow}+n_{i,\downarrow}n_{j,\uparrow})$ with no double occupancy. However, the matrix element $\langle \uparrow\downarrow\rvert H^{\mathrm{eff}}_J\lvert\uparrow \downarrow\rangle$ contains the contribution of single-particle energy corrections which add up to a constant. The specific value of $J_z$ should be determined as
\begin{equation}
	J_z=-2(\langle \uparrow\downarrow\rvert H^{\mathrm{eff}}_J\lvert \uparrow\downarrow \rangle-\langle \uparrow\downarrow\rvert H^{\mathrm{eff}}_{J}(W=0)\lvert \uparrow\downarrow\rangle).
\end{equation}
Here $H^{\mathrm{eff}}_{J}(W=0)$ represents the low-energy effective Hamiltonian when $H_{\mathrm{DDI}}=0$, which contains all the contribution of single-particle corrections. And the fourth-order diagonal channels of $H_J^{\mathrm{eff}}$ generating $J_{z}$ are
\begin{equation}
	\begin{aligned}
		&\lvert\uparrow \downarrow\rangle \stackrel{H_{\Omega}}{\longrightarrow} \lvert\uparrow \!S1\rangle,\lvert P1(2)\! \downarrow\rangle \stackrel{H_{\Omega}}{\longrightarrow} \lvert\psi_{+,1(2)} \rangle, \lvert\psi_{-,1(2)}\rangle\stackrel{H_{\Omega}}{\longrightarrow}\lvert\uparrow \!S1\rangle,\lvert P1(2)\!\downarrow\rangle \stackrel{H_{\Omega}}{\longrightarrow} \lvert\uparrow \downarrow\rangle, \\
		&\lvert\downarrow \uparrow\rangle \stackrel{H_{\Omega}}{\longrightarrow} \lvert\downarrow \!P1(2)\rangle,\lvert S1\! \uparrow\rangle \stackrel{H_{\Omega}}{\longrightarrow} \lvert\psi_{+,1(2)} \rangle, \lvert\psi_{-,1(2)}\rangle\stackrel{H_{\Omega}}{\longrightarrow}\lvert S1\! \uparrow\rangle,\lvert\downarrow \!P1(2)\rangle \stackrel{H_{\Omega}}{\longrightarrow} \lvert\downarrow \uparrow\rangle.
	\end{aligned}\label{Jz_channel}	
\end{equation}
Detailed calculations also provide an analytical estimation of $J_z$ as
\begin{equation}
	J_z= - (\frac{\Omega_{P1}}{2\Delta_{P1}})^{2}(\frac{\Omega_{S1}}{2\Delta_{S1}})^{2}(\Delta_{P1}+\Delta_{S1})^{2}
	[\frac{\Delta_{-}-\Delta_{+}+W_{1}}{(\Delta_{+}+W_{1})^{2}-(\Delta_{-}^{2}+W_{2}^{2})}+\frac{\Delta_{-}-\Delta_{+}-W_{1}}{(\Delta_{+}-W_{1})^{2}-(\Delta_{-}^{2}+W_{2}^{2})}+\frac{2}{\Delta_{+}+\Delta_{-}}].
	\label{eq:ud}
\end{equation}

\begin{figure}[H]
	\begin{centering}
		\includegraphics[scale=0.95]{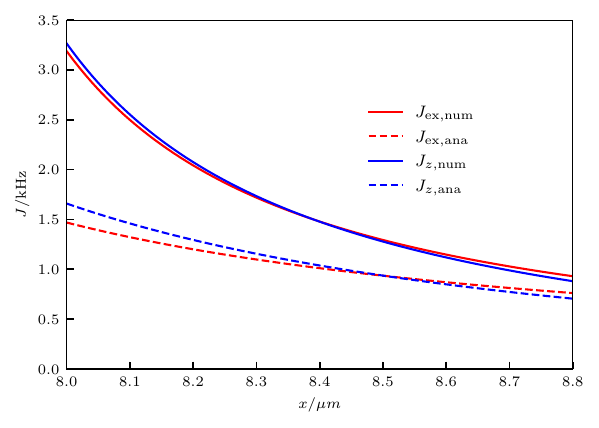}		\caption{Comparison between the numerical simulations and the analytical estimations for $J_{\mathrm{ex}}$ and $J_z$. The solid lines represent the $J_{\mathrm{ex}}$ and $J_z$ determined by numerical method explained in Sec.~\ref{app:ED}, while the dashed lines represent the $J_{\mathrm{ex}}$ and $J_z$ determined by analytical estimation in this section.}
		\label{fig: jperturb}
		\par\end{centering}
\end{figure}

The estimations derived in Eq.~(\ref{eq:jex}) and (\ref{eq:ud}) are compared with the numerical simulations shown in Sec.~\ref{app:ED} (See Fig. \ref{fig: jperturb}), using the experimental parameters in Sec.~\ref{app:parameters}. As illustrated in Fig.~\ref{fig: jperturb}, the perturbation analysis actually provide a preliminary estimation for $J_{\mathrm{ex}}$ and $J_z$ at the two-atom distance $x\approx8.2$ $\mu m$, corresponding to the selected lattice spacing. The discrepancies between analytical estimations and  numerical simulations arise from the contribution of higher-order perturbations and those omitted Rydberg pairs.

The physical origin of \( J_{\mathrm{ex}} \) and \( J_z \) can be understood in terms of the dynamical evolution of Rydberg pair states. Since the interaction strength is comparable to the detuning under the chosen parameters, the DDI cannot be treated perturbatively. Instead, the DDI induces coherent oscillations between the Rydberg pair states \( \lvert PS\rangle \) and \( \lvert SP\rangle \). Starting from the ground state \( \lvert \uparrow\downarrow\rangle \), the coupling light drives second-order transitions to \( \lvert PS\rangle \). Then DDI subsequently generates a coherent oscillation between \( \lvert PS\rangle \) and \( \lvert SP\rangle \), which are then coupled back to the ground-state manifold via second-order processes: the return to \( \lvert \uparrow\downarrow\rangle \) yields \( J_z \), while the transition to \( \lvert \downarrow\uparrow\rangle \) yields \( J_{\mathrm{ex}} \). The near-equality of \( J_{\mathrm{ex}} \) and \( J_z \) observed in Fig.~\ref{fig: jperturb} reflects underlying coherent ocillation dynamics so wherein two types of transitions channels play equally essential roles.

\section{Numerical simulations of $J_{\mathrm{ex}}$, $J_z$, $V_{\uparrow\uparrow}$ and $V_{\downarrow\downarrow}$ }
\label{app:ED}
\begin{figure}[H]
	\begin{centering}
		\includegraphics[scale=1]{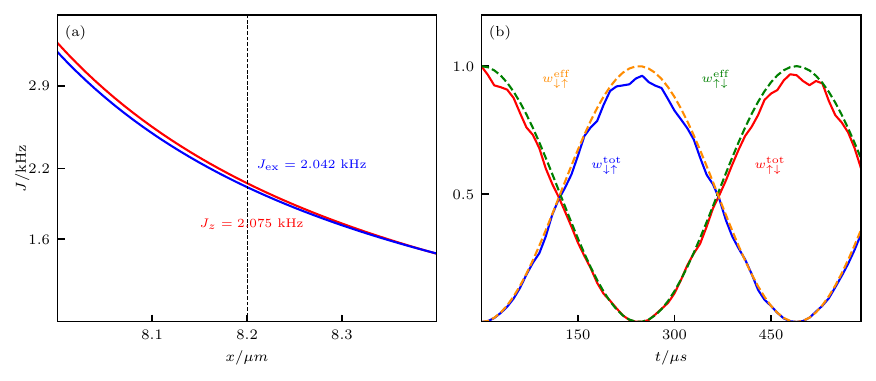}
		\caption{Estimation of $J_{\mathrm{ex}}$, $J_z$, and validation of the effective Hamiltonian $H_J^\mathrm{eff}$. (a) The effective interaction strengths $J_{\mathrm{ex}}$ and $J_z$ are plotted as functions of the two-atom distance $x$. An appropriate distance of $8.2~\mu\mathrm{m}$ is selected, yielding $J_{\mathrm{ex}} \approx J_z \approx 2~\mathrm{kHz}$.  (b) The time evolution of the system is compared using the total Hamiltonian $H_J^{\mathrm{tot}}$ (solid line) and the effective Hamiltonian $H_J^\mathrm{eff}$ (dashed line). The initial state is chosen as the two-particle state $\lvert\uparrow\downarrow\rangle$. The evolved wave function is expressed as  $\lvert\psi(t)\rangle = c_{\uparrow\downarrow} (t) \lvert \uparrow \downarrow \rangle + c_{\downarrow\uparrow} (t) \lvert \downarrow \uparrow \rangle + \lvert\text{others}\rangle$, then the probabilities $w_{\uparrow\downarrow} = |c_{\uparrow \downarrow}|^2$ and $w_{\downarrow \uparrow} = |c_{\downarrow \uparrow}|^2$ are shown as functions of time.}
		\label{fig: jexjz}
		\par\end{centering}
\end{figure}
In this section, we explain the details of numerical simulation for $J_{\mathrm{ex}}$, $J_z$, and the strength of  density-density interaction $V_{\uparrow\uparrow}$, $V_{\downarrow\downarrow}$ induced by the Rydberg vdW interaction. For $J_{\mathrm{ex}}$ and $J_z$, the total Hamiltonian $H_J^{\mathrm{tot}}$ is diagonalized in a two-site system within the subspace composed of $\lvert\uparrow\downarrow\rangle$, $\lvert\downarrow\uparrow\rangle$ and all the relevant two-site states. Two special eigenstates are identified as $\lvert s_{\pm}\rangle \approx(\lvert\uparrow\downarrow\rangle\pm\lvert\downarrow\uparrow\rangle)/\sqrt{2}$ with the eigenvalues $\gamma_{\pm}$. And the matrix representation of the effective Hamiltonian $H_J^{\mathrm{eff}}$, describing the quantum dynamics within subspace composed of $\lvert\uparrow\downarrow\rangle$ and $\lvert\downarrow\uparrow\rangle$, is
\begin{equation}
	H_{J}^{\mathrm{eff}} =
	\begin{pmatrix} 
		-\frac{J_z}{2}+\delta_{\mathrm{single}}	& \frac{J_\mathrm{ex}}{2} \\
		\frac{J_\mathrm{ex}}{2}	& -\frac{J_z}{2}+\delta_{\mathrm{single}}
	\end{pmatrix}.\label{Hg1g2}
\end{equation}
Here $\delta_{\mathrm{single}}$ is the single-particle energy shift with $H_{\mathrm{DDI}}=0$ for $\lvert\uparrow\downarrow\rangle$ and $\lvert\downarrow\uparrow\rangle$. Through matching the eigenvalues of $H_J^{\mathrm{eff}}$ and $H_{J}^{\mathrm{tot}}$, $J_\mathrm{ex}$ and $J_z$ are determined as
\begin{equation}
	\begin{aligned}
		J_{\mathrm{ex}}&=\gamma_+-\gamma_- ,\\J_z&=-\gamma_+-\gamma_-+2\delta_{\mathrm{single}}.
	\end{aligned}
\end{equation}

Fig. \ref{fig: jexjz}(a) shows the numerically determined $J_{\mathrm{ex}}$ and $J_z$ as functions of two-atom distance $x$, using the experimental parameters in Sec. \ref{app:parameters}. To choose a suitable lattice spacing, an important consideration is that the energies of the relevant Rydberg pairs may be resonant with the energy of $\lvert\uparrow\downarrow\rangle$ at some special two-atom distances, resulting in the breakdown of perturbation theory and the divergence of $J_{\mathrm{ex}}$ and $J_z$. To avoid the undesirable resonance and generate reasonable $J_{\mathrm{ex}}$ and $J_z$, the lattice spacing $a$ is selected as 8.2 $\mu$m. And the numerically determined $J_{\mathrm{ex}}$ and $J_z$ under different two-atom distances are displayed in Table \ref{tab:jexjz}. As mentioned in Sec.~\ref{app:oH}, the two-photon coupling between $\lvert\downarrow\rangle$ and $\lvert S3\rangle$ may lead to undesirable outcomes. The numerical simulations including atom states relevant to $\lvert S3\rangle$ reveal that, using the experimental parameters in Sec.~\ref{app:parameters}, the impact of the new coupling channel is just a tiny correction of $J_{\mathrm{ex}}$ and $J_z$, as shown in Table \ref{tab:jexjz2}. The validity of the effective Hamiltonian is further confirmed by comparing the time evolution of a two-site system using the total Hamiltonian $H^{\mathrm{tot}}_J$ (atom states relevant to $\lvert S3\rangle$ are included here) and the effective Hamiltonian $H^{\mathrm{eff}}_J$ in Eq.~(\ref{Hg1g2}), as illustrated in Fig. \ref{fig: jexjz}(b). 
\begin{table}[!htp]
	\centering
	\begin{tabular}{|c|c|c|c|c|c|}
		\hline $x$ & $a$ $(8.2$ $\mu\mathrm{m})$ & $\sqrt{2}a$ & $2a$ & $3a$ \\
		\hline $J_{\mathrm{ex}}$ & $2.0548$ $\text{kHz}$ & $0.2199$ $\text{kHz}$ & $0.0698$ $\text{kHz}$ & $0.0204$ $\text{kHz}$ \\
		\hline $J_{z}$ & $2.0879$ $\text{kHz}$ & $0.1125$ $\text{kHz}$ & $0.0132$ $\text{kHz}$ & $0.00115$ $\text{kHz}$
		\\
		\hline
	\end{tabular} 
	\caption{ $J_{\mathrm{ex}}$ and $J_z$ as functions of two-atom distance $x$, obtained by diagonalization of $H_J^{\mathrm{tot}}$ in a two-site system.} \label{tab:jexjz}
\end{table}

\begin{table}[!htp]
	\centering
	\begin{tabular}{|c|c|c|c|c|c|}
		\hline $x$ & $a$ $(8.2$ $\mu\mathrm{m})$ & $\sqrt{2}a$ & $2a$ & $3a$\\
		\hline $J_{\mathrm{ex}}$ & $2.0421$ $\text{kHz}$ & $0.2188$ $\text{kHz}$ & $0.0695$ $\text{kHz}$ & $0.0203$ $\text{kHz}$ \\
		\hline $J_{z}$ & $2.0752$ $\text{kHz}$ & $0.1119$ $\text{kHz}$ & $0.0131$ $\text{kHz}$ & $0.00114$ $\text{kHz}$
		\\
		\hline
	\end{tabular} 
	\caption{$J_{\mathrm{ex}}$ and $J_z$ as functions of two-atom distance $x$ with states relevant to $\lvert S3\rangle$.} \label{tab:jexjz2}
\end{table}

For the numerical simulations of $V_{\uparrow\uparrow}$ and $V_{\downarrow\downarrow}$, the established method involves comparing the energies of $\lvert\uparrow\uparrow\rangle$ and $\lvert\downarrow\downarrow\rangle$ with and without the Rydberg Vdw interactions \cite{PhysRevA.82.033412}. The summarized outcomes are presented in Table \ref{tab:vuudd}, using the model parameters in Sec.~\ref{app:parameters}. And these results indicate that $V_{\uparrow\uparrow},V_{\downarrow\downarrow}\ll J_{\mathrm{ex}},J_z$ in general.
\begin{table}[!htp]
	\centering
	\begin{tabular}{|c|c|c|c|}
		\hline $x$ & $a$ $(8.2$ $\text{\ensuremath{\mu }m})$ & $\sqrt{2}a$ & $2a$\\
		\hline $V_{\uparrow\uparrow}$ & $0.174$ $\text{kHz}$ & $0.0232$ $\text{kHz}$ & $2.93\times10^{-3}$ $\text{kHz}$\\
		\hline $V_{\downarrow\downarrow}$ & $-2.78\times10^{-3}$ $\text{kHz}$ & $-3.46\times10^{-4}$ $\text{kHz}$ & $-4.32\times10^{-5}$ $\text{kHz}$ \\
		\hline 
	\end{tabular} 
	\caption{$V_{\uparrow\uparrow}$ and $V_{\downarrow\downarrow}$ as functions of two-atom distance $x$.} \label{tab:vuudd}
\end{table}

\section{Lifetime estimation}
\label{app:lifetime}
\subsection{Preservation of coherent oscillation: our scheme v.s. optical lattice }

Firstly, we analyze the coupling time restricted by the lifetime of Rydberg-dressed state (denoted as $\tau_{\mathrm{dress}}$), through comparison with the Fermi-Hubbard model in optical lattices in the harmonic trap. $J\tau_{\mathrm{L}}$, where $J$ is the strength of antiferromagnetic interaction and $\tau_\mathrm{L}$ is the characteristic lifetime, can serve as a measure to quantify how well the coherent oscillations are preserved. In our scheme, the characteristic lifetime $\tau_{\mathrm{L}}$ is the lifetime of the Rydberg-dressed states $\tau_{\mathrm{dress}}$, which is approximately 60 ms according to Sec.~\ref{app:parameters}. And the strength of NN spin-spin interaction, $J_{\mathrm{ex}}$ and $J_z$, is about 2 kHz from the calculation in Sec.~\ref{app:ED}, which gives 
\begin{equation}
	J\tau_{\mathrm{L}}=J_{\mathrm{ex}}(J_z)\tau_{\mathrm{dress}}\approx2\pi\times120\hbar.
\end{equation}

In contrast, the characteristic lifetime $\tau_\mathrm{L}$ in optical lattices is the lifetime of optical lattice denoted as $\tau_{\mathrm{OL}}$, reported to be on the order of $1$ s \cite{annurev-conmatphys-070909-104059}. To realize the two-dimensional $t$-$J$ model which excludes double occupancy, the ratio of on-site repulsion strength $U$ to the NN hopping amplitude $t$ must be tuned to drive the Fermi-Hubbard model into Mott regime, generating the effective spin-spin interaction $J=4t^2/U$. For typical optical lattices in the harmonic trap, the Mott transition occurs at $U/t\approx20$ on the edge of system, while the entire system enters into Mott phase at $U/t\approx80$~\cite{PhysRevLett.116.235301,doi:10.1126/science.aad9041}. The typical value for $t$ can be estimated as 1 kHz, which gives

\begin{equation}
	J\tau_{\mathrm{L}}=\frac{4t^2}{U}\tau_{\mathrm{OL}}\approx2\pi\times50\hbar\sim2\pi\times200\hbar.
\end{equation}

The comparison demonstrates that our scheme performs comparably to the Fermi-Hubbard model in optical lattice in the harmonic trap, sufficient for the simulation of $t$-$J$ model.

\subsection{Total operation time v.s. lifetime of tweezers}
Secondly, we analyze the restriction from the lifetime of tweezers (denoted as $\tau_{\mathrm{tweezer}}$). The total operation time can be divided into two parts: time for gate operations (denoted as $\tau_{\mathrm{gate}}$) and time for tweezer movements (denoted as $\tau_{\mathrm{move}}$). We estimate the time for gate operation, $\tau_{\mathrm{gate}}$, as
\begin{equation}  \tau_{\mathrm{gate}}\approx(1+\frac{2\tau_{t,\mathrm{N}}+2\tau_{t^{\prime},\mathrm{NN}}}{\tau_J})\tau_{\mathrm{dress}},
\end{equation}
where $2\tau_{t,\mathrm{N}}$, $2\tau_{t^{\prime},\mathrm{NN}}$ and $\tau_J$ are the operation times for $H_{t,\mathrm{N}}$, $H_{t^{\prime},\mathrm{NN}}$ and $H_J$ per cycle, and $\tau_{\mathrm{dress}}\approx 60$ ms is the lifetime of the Rydberg-dressed states. We analysis the ratio $2\tau_{t,\mathrm{N}}/\tau_J$, $2\tau_{t^{\prime},\mathrm{NN}}/\tau_J$ in two main applications of our scheme: probing the d-wave superconductivity and high $T_{\mathrm{c}}$ physics, where $t>J$, and simulating the non-trivial dynamics related to the Hilbert space fragmentation, where $t\ll J$. Based on the previous experiment \cite{doi:10.1126/science.1250057}, the inter-tweezer hopping amplitude can be tuned to exceed 0.3 kHz, while the strength of $J$ in our realization is about 2 kHz, as shown in Sec.~\ref{app:ED}. To realize typical model parameters for high-$T_\mathrm{c}$ superconductivity, it requires $2\tau_{t,\mathrm{N}}/\tau_J\approx10$ and  $2\tau_{t^{\prime},\mathrm{NN}}/\tau_J\approx1$, leading to a gate operation time $\tau_{\mathrm{gate}}\approx$ 0.7 s. On the other hand, for the $J/t\gg 1$ regime to realize Hilbert space fragmentation, only $2\tau_{t,\mathrm{N}}/\tau_J\approx1$ and $\tau_{t^{\prime},\mathrm{NN}}=0$ are required, resulting in a significantly shorter gate operation time $\tau_{\mathrm{gate}}\approx$ 0.12 s.

Next, we estimate the time for tweezer movement, $\tau_{\mathrm{move}}$, as 
\begin{equation}
	\tau_{\mathrm{move}}\approx N_{\mathrm{move}}\tau_{1\mathrm{move}},
\end{equation}
where $N_{\mathrm{move}}$ is the total number of tweezer movements, and $\tau_{1\mathrm{move}}$ is the time required for a single movement. To ensure that the higher-order contributions in the  Suzuki-Trotter
decomposition are negligible, we divide the total simulation procedure into about 3000 cycles with $\tau_J\approx$ 20 $\mu \mathrm{s}$, yielding $N_{\mathrm{move}}\approx10^4$. For the estimation of $\tau_{1\mathrm{move}}$, we consider a typical lattice configuration, a $10\times10$ square lattice. The length scale for a single tweezer movement within this lattice is approximately 100 $\mu\mathrm{m}$ with the selected lattice spacing $a=8.2$ $\mu\mathrm{m}$ in Sec.~\ref{app:ED}, which gives $\tau_{1\mathrm{move}}\approx$200 $\mu\mathrm{s}$ to maintain quantum coherence~\cite{Bluvstein2022,Bluvstein2024}. Thus, we estimate $\tau_{\mathrm{move}}\approx$ 2 $\mathrm{s}$ for a $10\times10$ lattice. 

Consequently, the total time required for our scheme, estimated as $\tau_{\mathrm{gate}}+\tau_{\mathrm{move}}$, is on the order of several seconds for moderate-sized lattices, which is significantly shorter than the lifetime of the tweezers, $\tau_{\mathrm{tweezer}} \sim 10^3$ s at 4 K~\cite{PhysRevApplied.16.034013,PhysRevApplied.22.024073}. 

\section{Hilbert space fragmentation and parameter conditions}
\label{app:require_HSF}
In the main text, we demonstrate that when $J/t\gg1$, Hilbert space fragmentation (HSF) arises. In this section, we derive more detailed criteria for $t$ and $J$ to ensure robust HSF by comparing the effective transition amplitude and energy splitting. Specifically, we investigate the Krylov subspace of the one-dimensional $t$-$J$ model in the $J/t\gg1$ regime, composed of entangled spin bound states $\lvert\mathrm{A}\rangle$, $\lvert\mathrm{B}\rangle$ and $\lvert\mathrm{C}\rangle$. The kinetic part of the Hamiltonian is given by 
\begin{equation}
	H_{t} = \sum_{i,\mathrm{\sigma}} (-t P_{i}c_{i,\mathrm{\sigma}}^{\dagger}c_{i+1,\mathrm{\sigma}}P_{i+1}+\mathrm{H.c.})
\end{equation}
where $P_i=1-n_{i,\uparrow}n_{i,\downarrow}$ is the projection operator excluding double occupancy on site $i$, and $n_{i,\uparrow}$ and $n_{i,\downarrow}$ are the density operators of spin-up and spin-down states. This kinetic term can induce transitions from the initial states $\lvert \mathrm{I}\rangle$ within the Krylov subspace to final states $\lvert\mathrm{F}\rangle$ outside the Krylov subspace. These undesirable transitions may disrupt the fragmented Krylov subspace and therefore should be suppressed. Our analysis focuses on the potential first-order and second-order undesirable transitions, as the higher-order transitions are weaker effects and will be typically blocked by the effective energy splitting, as elucidated by the self-pinning effect in Sec.~\ref{app: theory for self pin}.

\subsection{\texorpdfstring{First-order transition channels}%
	{First-order transition channels}}
\begin{table}[!h]
	\centering
	\begin{tabular}{|c|c|c|c|}
		\hline $\lvert\mathrm{I}\rangle$ & $\lvert\mathrm{F}\rangle$ & $\Omega_{\mathrm{I}\rightarrow \mathrm{F}}^{(1)}$ & $\delta E^{(0)}$\\
		\hline $\lvert\cdots0\mathrm{B}0\cdots\rangle$ & $\lvert\cdots0\mathrm{A}0\!\downarrow\!0\cdots\rangle$ & $\frac{\sqrt{3}}{2}t$ & $\frac{1}{2}J$ \\
		\hline $\lvert\cdots0\mathrm{A}0\mathrm{A}0\cdots\rangle$ & $\lvert\cdots0\mathrm{B}0\!\uparrow\!0\cdots\rangle$ & $-\frac{\sqrt{6}}{4}t$ &$\frac{1}{2}J$ \\
		\hline $\lvert\cdots0\mathrm{B}0\mathrm{C}0\cdots\rangle$ & $\lvert\cdots0\mathrm{D}_{1}0\mathrm{A}0\cdots\rangle$ & $-0.5915t$ &$\-0.3660J$ \\
		\hline $\lvert\cdots0\mathrm{B}0\mathrm{C}0\cdots\rangle$ & $\lvert\cdots0\mathrm{D}_{1}0\mathrm{T}0\cdots\rangle$ & $0.1972t$ &$0.6340J$ \\
		\hline $\lvert\cdots0\mathrm{B}0\mathrm{C}0\cdots\rangle$ & $\lvert\cdots0\mathrm{D}_{2}0\mathrm{A}0\cdots\rangle$ & $0.5576t$ &$0.2929J$ \\
		\hline $\lvert\cdots0\mathrm{B}0\mathrm{B}0\cdots\rangle$ & $\lvert\cdots0\mathrm{D}_{1}0\!\downarrow\downarrow\!0\cdots\rangle$ & $-0.2788t$ &$0.6340J$ \\
		\hline $\lvert\cdots0\mathrm{B}0\mathrm{B}0\cdots\rangle$ & $\lvert\cdots0\mathrm{D}_{3}0\mathrm{A}0\cdots\rangle$ & $-0.7886t$ &$0.2929J$ \\
		\hline 
	\end{tabular} 
	\caption{Relevant first-order transition channels between the initial states $\lvert\mathrm{I}\rangle$ and the final states $\lvert\mathrm{F}\rangle$ with the corresponding transition amplitude $\Omega_{\mathrm{I}\rightarrow\mathrm{F}}^{(1)}$ and the difference of bound energies $\delta E^{(0)}$. For channels related by spin inversion symmetry and spatial inversion symmetry, only one representative is listed. Here $\lvert\mathrm{T}\rangle=(\lvert\uparrow\downarrow\rangle-\lvert\downarrow\uparrow\rangle)/\sqrt{2}$ is the spin triplet state, $\mathrm{D}_{1}$ and $\mathrm{D}_{2}$ are two kinds of four-body spin bound states consisting of two $\lvert\uparrow\rangle$ and two $\lvert\downarrow\rangle$ states, and $\mathrm{D}_{3}$ is a four-body spin bound state consisting of one $\lvert\uparrow\rangle$ and three $\lvert\downarrow\rangle$ states.  } \label{tab:refined_parameters}
\end{table}

Firstly, we investigate the first-order transition channels between initial states $\lvert\mathrm{I}\rangle$ and $\lvert\mathrm{F}\rangle$ with the bound energies $E_{\mathrm{I}}$ and $E_{\mathrm{F}}$, respectively.
The first-order effective transition amplitude is given by
\begin{equation}
	\Omega_{\mathrm{I}\rightarrow\mathrm{F}}^{(1)}=\langle \mathrm{F}\rvert H_t\lvert\mathrm{I}\rangle.
\end{equation}
The effective energy splitting $\Delta E$ between $\lvert\mathrm{I}\rangle$ and $\lvert\mathrm{F}\rangle$ mainly results from the difference of bound energies $\delta E^{(0)}$:
\begin{equation}
	\Delta E=\delta E^{(0)}=E_{\mathrm{F}}-E_{\mathrm{I}}.
\end{equation}
The higher-order energy corrections of $\lvert\mathrm{I}\rangle$ and $\lvert\mathrm{F}\rangle$ induced by $H_t$ are small compared to $\delta E^{(0)}$ for the relevant first-order transition channels, so $\delta E^{(0)}$ is the primary contributor to the effective energy splitting. The suppression of undesirable first-order transition channels requires 
\begin{equation}
	|\Omega_{\mathrm{I}\rightarrow \mathrm{F}}^{(1)}| \ll |\Delta E|=|\delta E^{(0)}|.
\end{equation}
We list all the relevant first-order transition channels satisfying $|\delta E^{(0)}|<J$ with the corresponding effective transition amplitude $\Omega_{\mathrm{I}\rightarrow\mathrm{F}}^{(1)}$ and the difference between bound energies $\delta E^{(0)}$ in Table \ref{tab:refined_parameters}, which gives the  criteria 
\begin{equation}
	\frac{t}{J}\ll 0.3714.\label{re1}
\end{equation}

\subsection{\texorpdfstring{Second-order transition channels}%
	{Second-order transition channels}}
\begin{table}[!h]
	\centering
	\begin{tabular}{|c|c|c|c|c|c|}
		\hline $\lvert\mathrm{I}\rangle$ & $\lvert\mathrm{F}\rangle$ & $\Omega_{\mathrm{I}\rightarrow \mathrm{F}}^{(2)}$ & $\delta E^{(0)}$ & $\delta E_{\mathrm{I}}^{(2)}$ & $\delta E_{\mathrm{F}}^{(2)}$\\
		\hline $\lvert\cdots00\mathrm{B}0\mathrm{C}00\cdots\rangle$ & $\lvert\cdots00\mathrm{E}0\!\downarrow\!00\cdots\rangle$ & $ -0.252 t^2/J$ & $-0.0721J$ & $-3.8938t^2/J$ & $-1.7557t^2/J$\\
		\hline $\lvert\cdots00\mathrm{C}0\mathrm{B}00\cdots\rangle$ & $\lvert\cdots00\mathrm{E}0\!\downarrow\!00\cdots\rangle$ & $ -0.252 t^2/J$ & $-0.0721J$ & $-3.8938 t^2/J$ & $-1.7557 t^2/J$ \\
		\hline $\lvert\cdots00\mathrm{C}0\mathrm{C}00\cdots\rangle$ &
		$\lvert\cdots00\mathrm{E}0\!\uparrow\!00\cdots\rangle$ & $ -0.183 t^2/J$ & $-0.0721J$ & $-8.1399 t^2/J$ & $-3.2479 t^2/J$ \\
		\hline 
	\end{tabular} 
	\caption{Relevant second-order transition channels between the inital  states $\lvert\mathrm{I}\rangle$ and the final states $\lvert\mathrm{F}\rangle$ with the corresponding transition amplitude $\Omega_{\mathrm{I}\rightarrow\mathrm{F}}^{(2)}$, difference of bound energies $\delta E^{(0)}$, and the second-order energy corrections $\delta E_{\mathrm{I}}^{(2)}$ and $\delta E_{\mathrm{F}}^{(2)}$. For channels related by spin inversion symmetry and spatial inversion symmetry, only one representative is listed. Here $\mathrm{E}$ is a five-body spin bound state consisting of three $\lvert\uparrow\rangle$ and two $\lvert\downarrow\rangle$ states. } \label{tab:refined_parameters2}
\end{table}
Secondly, we investigate the second-order transition channels between initial states $\lvert\mathrm{I}\rangle$ and $\lvert\mathrm{F}\rangle$ with the bound energies $E_{\mathrm{I}}$ and $E_{\mathrm{F}}$, respectively. Since the second-order transitions are weaker compared to the first-order ones, only the nearly resonant second-order transition channels are considered here. The second-order effective transition amplitude is given by
\begin{equation}  \Omega_{\mathrm{I}\rightarrow\mathrm{F}}^{(2)}=\sum_{E_{\mathrm{IS}_1}\neq E_{\mathrm{I}}} \frac{\langle \mathrm{F}\rvert H_{t}\lvert\mathrm{IS}_1\rangle \langle \mathrm{IS}_1\rvert H_{t}\lvert\mathrm{I}\rangle}{E_{\mathrm{I}}-E_{\mathrm{IS_1}}},
\end{equation}
where $\lvert\mathrm{IS}_1\rangle$ represents the intermediate states involved in the perturbation calculation of $\Omega_{\mathrm{I}\rightarrow\mathrm{F}}^{(2)}$ with bound energy $E_{\mathrm{IS}_1}$. For the nearly resonant transitions, the energy correction induced by $H_t$ can not be neglected in comparison to the difference of bound energies $\delta E^{(0)}=E_{\mathrm{F}}-E_{\mathrm{I}}$, and the second-order energy corrections for $\lvert\mathrm{I}\rangle$ and $\lvert\mathrm{F}\rangle$ are given by
\begin{equation}
	\begin{aligned}
		\delta E_{\mathrm{I}}^{(2)}=&\sum_{E_{\mathrm{IS}_2}\neq E_{\mathrm{I}}} \frac{\langle \mathrm{I}\rvert H_{t}\lvert\mathrm{IS}_2\rangle \langle \mathrm{IS}_2\rvert H_{t}\lvert\mathrm{I}\rangle}{E_{\mathrm{I}}-E_{\mathrm{IS_2}}},\\
		\delta E_{\mathrm{F}}^{(2)}=&\sum_{E_{\mathrm{IS}_3}\neq E_{\mathrm{I}}} \frac{\langle \mathrm{F}\rvert H_{t}\lvert\mathrm{IS}_3\rangle \langle \mathrm{IS}_3\rvert H_{t}\lvert\mathrm{F}\rangle}{E_{\mathrm{I}}-E_{\mathrm{IS_3}}},
	\end{aligned}
\end{equation}
where $\lvert\mathrm{IS}_2\rangle$ and $\lvert\mathrm{IS}_3\rangle$ are the intermediate states with energies $E_{\mathrm{IS}_2}$ and $E_{\mathrm{IS}_3}$, relevant in the perturbative calculation of $\delta E_{\mathrm{I}}^{(2)}$ and $\delta E_{\mathrm{F}}^{(2)}$ respectively. We include the contribution of $\delta E_{\mathrm{I}}^{(2)}$ and $\delta E_{\mathrm{F}}^{(2)}$ in the effective energy splitting. The comparison between the effective transition amplitude and the effective energy splitting requires
\begin{equation}
	|\Omega_{\mathrm{I}\rightarrow \mathrm{F}}^{(2)}| \ll |\Delta E|=|\delta E^{(0)}+\delta E_{\mathrm{F}}^{(2)}-\delta E_{\mathrm{I}}^{(2)}|.
\end{equation}
We list all the relevant second-order transition channels satisfying $|\delta E^{(0)}|<0.1J$ with the corresponding effective transition amplitude $\Omega_{\mathrm{I}\rightarrow\mathrm{F}}^{(2)}$, difference of bound energies $\delta E^{(0)}$ and the second-order energy corrections $\delta E_{\mathrm{I}}^{(2)}$ and $\delta E_{\mathrm{F}}^{(2)}$ in Table \ref{tab:refined_parameters2}, which gives the  criteria 
\begin{equation}
	\frac{t^2}{J} \ll \min(|0.2861J-8.4845\frac{t^2}{J}|,|0.394J-26.7322\frac{t^2}{J}|).\label{re2}
\end{equation}
Assuming $0.2861J-8.4845t^2/J,~0.394J-26.7322t^2/J>0$, the Ineq. (\ref{re2}) can be further simplified as
\begin{equation}
	\frac{t^2}{J^2}\ll 0.01474. \label{re3}
\end{equation}
For most cases, if Ineq.~(\ref{re3}) is satisfied,  Ineq.~(\ref{re1}) is also satisfied. But for accuaracy, we choose to keep both Ineq.~(\ref{re1}) and Ineq.~(\ref{re3}) as the criteria to ensure robust HSF and associated many-body self-pinning effect.

\section{General theory of the many-body self-pinning effect}
\label{app: theory for self pin}
In this section, we present a general theory for the many-body self-pinning effect. A general interacting Hamiltonian with kinetic part $H_{\mathrm{kin}}$ and interaction part $H_{\mathrm{int}}$ is given by 
\begin{equation}
	H=\alpha_{\mathrm{kin}} H_\mathrm{kin}+\alpha_{\mathrm{int}} H_\mathrm{int},   
\end{equation}
where $\alpha_{\mathrm{kin}}$ and $\alpha_{\mathrm{int}}$ are the tuning parameters controlling the relative contributions of the kinetic and interaction terms.  We analyze this Hamiltonian in the strong coupling regime $|\alpha_\mathrm{int}/\alpha_{\mathrm{kin}}\gg 1|$. In this regime, the interaction term dominates the Hamiltonian, and $\alpha_{\mathrm{int}}$ emerges as the dominant energy scale. The bound states $\lvert Q \rangle$, defined as the eigenstate of $\alpha_\mathrm{int}H_{\mathrm{int}}$ without empty sites, thus serve as the fundamental units in quantum dynamics. The general form of the eigenstates $\lvert\phi_{\mathrm{I}}\rangle$ of $\alpha_{\mathrm{int}} H_\mathrm{int}$ can be represented as a series of bound states $Q_i$ separated by empty states, which is given by
\begin{equation}
	\lvert\phi_\mathrm{I}\rangle=\lvert\cdots0Q_{1}0\cdots0Q_{2}0\cdots\cdots0Q_{i}0\cdots\rangle.
\end{equation}
The term $\alpha_{\mathrm{kin}}H_\mathrm{kin}$ generates transitions that preserve the energies of the eigenstates. The dynamics involving $\lvert\phi_{\mathrm{I}}\rangle$ induced by $\alpha_{\mathrm{kin}}H_{\mathrm{kin}}$ can be described by the effective Hamiltonian 
\begin{equation}
	H^{\mathrm{eff}}_{\mathrm{\phi}}=\sum_{n=1,2,3,\cdots}\alpha_\mathrm{kin}H_{\mathrm{kin}}(G_{\mathrm{int}}\alpha_\mathrm{kin}H_{\mathrm{kin}})^{n-1}=\sum_{n=1,2,3,\cdots} (\alpha_{\mathrm{kin}})^{n}H_{\mathrm{kin}}(G_{\mathrm{int}}H_{\mathrm{kin}})^{n-1},\label{h_eff_selfpin}
\end{equation}
where $G_{\mathrm{int}}=(1-P_{\phi})(E_\phi-\alpha_{\mathrm{int}}H_{\mathrm{int}})^{-1}(1-P_{\phi})$. Here, $P_{\phi}$ is the projection operator in the subspace spanned by $\lvert\phi_{\mathrm{I}}\rangle$, and $E_{\phi}$ is the unperturbed energy. Since $G_{\mathrm{int}} \propto \alpha_{\mathrm{int}}^{-1}$, the nth-order term of $H_{\phi}^{\mathrm{eff}}$ scales as 
\begin{equation}
	H_{\phi}^{\mathrm{eff}}\propto \frac{\alpha_{\mathrm{kin}}^n}{\alpha_{\mathrm{int}}^{n-1}}.
\end{equation}
For two resonant eigenstates $\lvert\phi_{\mathrm{I}1}\rangle$ and $\lvert\phi_{\mathrm{I}2}\rangle$, the effective transition amplitude $\lambda_{\mathrm{eff}}$ and effective energy splitting $\delta_{\mathrm{eff}}$ are derived from the off-diagonal and diagonal matrix elments of $H_{\phi}^{\mathrm{eff}}$ as
\begin{equation}
	\begin{aligned}
		& \lambda_{\mathrm{eff}}=\langle\phi_{\mathrm{I}1}\rvert H^{\mathrm{eff}}_{\mathrm{\phi}}\lvert\phi_{\mathrm{I}2}\rangle=\sum_{n=n_0^{\lambda},n_0^{\lambda}+1,\cdots} g_n^{\lambda} \frac{\alpha_{\mathrm{kin}}^n}{\alpha_{\mathrm{int}}^{n-1}}, \\
		& \delta_{\mathrm{eff}}=\langle\phi_{\mathrm{I}1}|H^{\mathrm{eff}}_{\phi}|\phi_{\mathrm{I}1}\rangle-\langle\phi_{\mathrm{I2}}|H^{\mathrm{eff}}_{\phi}|\phi_{\mathrm{I}2}\rangle=\sum_{n=n_0^{\delta},n_0^{\delta}+1,\cdots} g_n^{\delta} \frac{\alpha_{\mathrm{kin}}^n}{\alpha_{\mathrm{int}}^{n-1}},
	\end{aligned}\label{lambda_delta}
\end{equation}
where $n_0^{\lambda}$ and $n_0^{\delta}$ are the leading perturbative orders of $\lambda_{\mathrm{eff}}$ and $\delta_{\mathrm{eff}}$, respectively. The quantum dynamics is determined by the comparison between $\delta_{\mathrm{eff}}$ and $\lambda_{\mathrm{eff}}$:
\begin{equation}      \frac{\delta_{\mathrm{eff}}}{\lambda_{\mathrm{eff}}}=f(g_n^\lambda,g_n^\delta)\times
	\frac{g_{n_0^{\delta}}^{\delta}}{g_{n_0^{\lambda}}^{\lambda}}(\frac{\alpha_{\mathrm{kin}}}{\alpha_{\mathrm{int}}})^{n_0^{\delta}-n_0^{\lambda}},\quad 
	f(g_n^\lambda,g_n^\delta) = \frac{1+\sum_{i=1,2,\cdots}(g_{n_0^{\delta}+i}^{\delta}/g_{n_0^{\delta}}^{\delta})(\alpha_{\mathrm{kin}}/\alpha_{int})^i}{1+\sum_{i=1,2,\cdots}(g_{n_0^{\lambda}+i}^{\lambda}/g_{n_0^{\lambda}}^{\lambda})(\alpha_{\mathrm{kin}}/\alpha_{int})^i}.\label{ratio}
\end{equation}
Here $f(g_{n}^{\lambda},g_{n}^{\delta})$ summarizes the contribution of higher-order perturbations. In the strong coupling regime $|\alpha_{\mathrm{int}}/\alpha_{\mathrm{kin}}|\gg 1$, we have $f(g_{n}^{\lambda},g_{n}^{\delta})\approx 1$, and the ratio $\delta_{\mathrm{eff}}/\lambda_{\mathrm{eff}}$ in Eq.~(\ref{ratio}) simplifies to
\begin{equation}
	\lim_{|\alpha_{\mathrm{I}}/\alpha_{K}|\rightarrow\infty}\frac{\delta_{\mathrm{eff}}}{\lambda_{\mathrm{eff}}}= \left\{
	\begin{array}{c c}
		\infty, \quad & n_0^{\lambda} > n_0^{\delta},\\
		g_{n_0^{\delta}}^{\delta}/g_{n_0^{\lambda}}^{\lambda}, \quad & n_0^{\lambda} = n_0^{\delta},\\
		0, \quad & n_0^{\lambda} < n_0^{\delta}.
	\end{array}
	\right. \label{ratio2}
\end{equation}
When $ n_0^{\lambda} > n_0^{\delta}$, the effective transition amplitude $\lambda_{\mathrm{eff}}$ is negligible compared to the effective energy splitting $\delta_{\mathrm{eff}}$, and the transition between $\lvert\phi_{\mathrm{I}1}\rangle$ and $\lvert\phi_{\mathrm{I}2}\rangle$ is completely blocked. When $ n_0^{\lambda} = n_0^{\delta}$, the effective transition amplitude $\lambda_{\mathrm{eff}}$ is comparable to the effective energy splitting $\delta_{\mathrm{eff}}$, and the transition between $\lvert\phi_{\mathrm{I}1}\rangle$ and $\lvert\phi_{\mathrm{I}2}\rangle$ can occur but is partially blocked. Only when $ n_0^{\lambda} < n_0^{\delta}$ does the transition between $|\phi_{\mathrm{I}1}\rangle$ and $|\phi_{\mathrm{I}2}\rangle$ occur freely. Thus, when $ n_0^{\lambda} \geq n_0^{\delta}$, quantum states will be completely or partially pinned to its initial configurations, despite the presence of other resonant states. We term this novel phenomenon the many-body self-pinning effect.

This fact partly explains why, in Sec.~\ref{app:require_HSF}, only first-order and second-order undesirable transition channels are considered. The effective energy splitting is generally a second-order process with $n_0^{\delta}=2$. Therefore, higher-order transitions with $n_0^{\lambda}>2$ will be completely blocked since $n_0^{\lambda}>n_0^{\delta}$. Consequently, these channels do not participate in the quantum dynamics relevant to spin bound states $\lvert\mathrm{A}\rangle$, $\lvert\mathrm{B}\rangle$ and $\lvert\mathrm{C}\rangle$ in the $J/t\gg1$ regime.

\section{\texorpdfstring{Many-body elf-pinning effect in the $t$-$J$ and $t$-$J_z$ model}
	{Self-pinning effect in the t-J and t-Jz models}}
\label{app:self-pinning}
In this section, we apply the general theory outlined in Sec.~\ref{app: theory for self pin} to a typical Krylov subspace of the one-dimensional $t$-$J$ model and $t$-$J_z$ model in the strong coupling regime. The different behaviours exhibited in these two models indicate that local entanglement enforces the occurrence of the many-body self-pinning effect.

\subsection{\texorpdfstring{Self-Pinning in the $t$-$J$ Model}%
	{Self-Pinning in the t-J Model}}
Firstly, we investigate the self-pinning effect in the Krylov subspaces formed by entangled spin bound states $\lvert\mathrm{A}\rangle$, $\lvert\mathrm{B}\rangle$ and $\lvert\mathrm{C}\rangle$. The transitions permitted by the emergent conserved quantities occur between $3n$ $\lvert\mathrm{A}\rangle$ states and $n$ $\lvert\mathrm{B}\rangle$, $n$ $\lvert\mathrm{C}\rangle$ states.  The effective energy splitting, typically a second-order process with $n_0^{\delta}=2$, limits the unblocked transition channels to those with $n_0^{\lambda}\leq2$. 

A typical transition with $n_0^{\lambda}=2$ between $\lvert\cdots 00\mathrm{A}0\mathrm{A}0\mathrm{A}00 \cdots\rangle$ and $\lvert\cdots 00\mathrm{B}00\mathrm{C}00 \cdots\rangle$ is analyzed according to Eq.~(\ref{lambda_delta}) and (\ref{ratio2}). Further 
calculations yield
\begin{equation}
	g_2^{\lambda}=-3\sqrt{2}/2, \quad g_2^{\delta}=4/3, \quad \delta_{\mathrm{eff}}/\lambda_{\mathrm{eff}}\approx0.63.
\end{equation}
This indicates that the effective energy splitting $\delta_{\mathrm{eff}}$ and the effective transition amplitude $\lambda_{\mathrm{eff}}$ are comparable. Therefore, many-body self-pinning effects is anticipated in this Krylov subspace of the $t$-$J$ model, partially obstructing transitions between resonant states $\lvert\cdots 00\mathrm{A}0\mathrm{A}0\mathrm{A}00 \cdots\rangle$ and $\lvert\cdots 00\mathrm{B}00\mathrm{C}00 \cdots\rangle$.

\subsection{\texorpdfstring{Absence of Self-Pinning in the $t$-$J_z$ Model}%
	{Absence of Self-Pinning in the t-Jz Model}}
Before delving into the investigation of thee many-body self-pinning effect within $t$-$J_z$ model, we fist analyze the Hamiltonian and its associated Hilbert space fragmentation (HSF). The $t$-$J_z$ Hamiltonian is given by 
\begin{equation}
	H_{t\mathrm{-}J_z} = \sum_{i,\mathrm{\sigma}} (-t P_{i}c_{i,\mathrm{\sigma}}^{\dagger}c_{i+1,\mathrm{\sigma}}P_{i+1}+\mathrm{H.c.})+\sum_{i} J_z S_{z,i}S_{z,i+1},
\end{equation}
where $P_i=1-n_{i,\uparrow}n_{i,\downarrow}$ is the projection operator excluding double occupancy on site $i$, and $n_{i,\uparrow}$ and $n_{i,\downarrow}$ are the density operators of spin-up and spin-down states, respectively. Due to the kinetic constraint of the $t\mathrm{-}J_z$ model, the spin pattern of a product state becomes a conserved quantity \cite{PhysRevB.101.125126}. We consider the Krylov subspace characterized by the spin pattern ``$\uparrow\downarrow\uparrow\downarrow\cdots$'', comprising an equal number of $\lvert\uparrow\rangle$ and $\lvert\downarrow\rangle$ states. In the $J_z/t\gg 1$ regime, the bound states of $t$-$J_z$ model are product states, and the Krylov subspace with spin pattern ``$\uparrow\downarrow\uparrow\downarrow\cdots$'' further splits into smaller Krylov subspaces with the emergent conserved quantity $N_\mathrm{BS}$, representing the number of product spin bound states. For instance, if there are five lattice sites and the conserved spin pattern is ``$\uparrow\downarrow\uparrow\downarrow$'', the three bases of the Krylov subspace labeled by emerged conserved quantity $N_{\mathrm{BS}}=2$ are $\lvert\uparrow\!0\!\downarrow\uparrow\downarrow\rangle$, $\lvert\uparrow\downarrow\!0\!\uparrow\downarrow\rangle$ and $\lvert\uparrow\downarrow\uparrow\!0\!\downarrow\rangle$.

Subsequently, we demonstrate the absence of self-pinning in the typical Krylov subspace characterized by spin pattern ``$\uparrow\downarrow\uparrow\downarrow\cdots$'' and emergent conserved quantity $N_{\mathrm{BS}}$. The leading perturbative orders satisfy $n_0^{\lambda}<n_0^{\delta}$, with 
\begin{equation}
	n_0^{\lambda}=1, \quad n_0^{\delta}\geq2.
\end{equation}
Here, the typical transitions occur via the alteration of a single spin at the boundary of the product spin bound states, exemplified by the resonant transition between 
\begin{equation}
	\lvert\cdots\uparrow\downarrow\!0\!\uparrow\downarrow\cdots\rangle \longleftrightarrow \lvert\cdots\uparrow\downarrow\uparrow\!0\!\downarrow\cdots\rangle.
\end{equation}
Consequently, the typical transition corresponds to $n_0^{\lambda}=1$. But the leading perturbative order of the effective energy splitting $n_0^{\delta}$ remains equal to two, e.g., the transition between 
\begin{equation}
	\lvert\cdots 00\!\uparrow\downarrow\!0\!\uparrow\downarrow\!00\cdots\rangle \longleftrightarrow \lvert\cdots 00\!\uparrow\!0\!\downarrow\uparrow\downarrow\!00\cdots\rangle,
\end{equation}
or even larger than two, e.g., the transition between 
\begin{equation}
	\lvert\cdots 00\!\uparrow\downarrow\uparrow\!0\!\downarrow\uparrow\downarrow\!00\cdots\rangle \longleftrightarrow \lvert\cdots 00\!\uparrow\downarrow\!0\!\uparrow\downarrow\uparrow\downarrow\!00\cdots\rangle,
\end{equation}
resulting $n_0^{\lambda}<n_0^{\delta}$. Therefore, the many-body self-pinning effect is predicted to vanish in this Krylov subspace of the $t$-$J_z$ model, allowing free transitions between different product spin bound states.

\subsection{\texorpdfstring{Comparison Between $t$-$J$ and $t$-$J_z$ Models}%
	{Comparison Between t-J and t-Jz Models}}

Upon comparing the $t$-$J$ and $t$-$J_z$ models, we observe that local entanglement of the bound states in the $t$-$J$ model effectively suppresses low-order transition channels, resulting in a higher $n_0^{\lambda}$ and facilitating the self-pinning. This is because entangled states involve multi-particle correlations, typically requiring transitions beyond single-particle transfers to change configurations. For instance, the energy mismatch in the $t$-$J$ model between entangled bound states typically inhibits first-order resonant transitions with $n_0^{\lambda}=1$. In contrast, for product bound states in the $t$-$J_z$ model, resonant transitions remain feasible as long as the local bond structure remains unchanged between the initial and final configurations. This demonstrates the local entanglement enforces the many-body self-pinning effect.

\section{Supplementary simulations on thermalization related dynamics}
\label{app:numerical_thermal}
In this section, we provide additional numerical simulations for the $t$-$J$ and $t$-$J_z$ model in the $J/t,J_z/t\gg1$ regime on the quantum dynamics related to the Krylov-restricted thermalization, as a supplementary for the numerical simulation presented in the main text.
All simulations in this section are based on the time-evolving block decimation (TEBD) method in the Tenpy library \cite{Tenpy}.

\begin{figure}[H]
	\begin{centering}
		\includegraphics[scale=1.21]{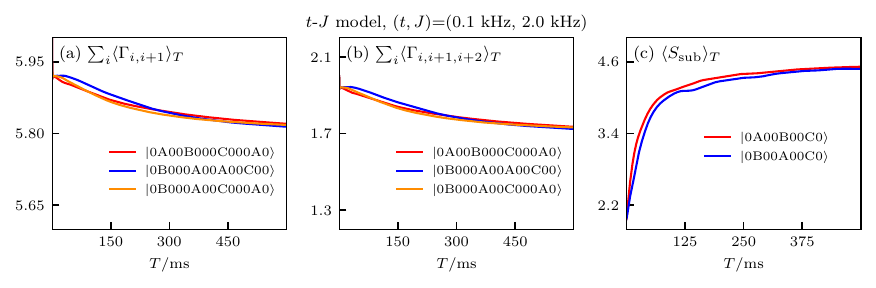}
		\caption{Comparison for $t$-$J$ model across initial states composed of the same spin bound states but arranged in different orders. Long-time averages of the density correlations ($\Gamma_{i,i+1}=n_i n_{i+1}$, $\Gamma_{i,i+1,i+2}=n_i n_{i+1}n_{i+2}$) and sublattice entanglement entropy (EE) are calculated with parameters
			$t$=0.1 kHz and $J$=2.0 kHz. (a-b) shows the long-time averages of density correlations for a 20-site $t$-$J$ chain. (c) shows the long-time averages of sublattice EE for a 14-site $t$-$J$ chain. The long-time average values converge to similar values. The bond dimension for TEBD simulation keeps up to 800 for (a-b), 400 for (c).}
		\label{fig: S3}
		\par\end{centering}
\end{figure}

Firstly, we compare the long-time averages of the multi-site density correlations and sublattice entanglement entropy (EE) in $t$-$J$ model. The different resonant initial states is composed of the same spin bound states, while arranged in different orders. As shown Fig.~\ref{fig: S3}, the long-time averages converge to similar values for these initial states, indicating the outcome of long-time evolution of these states may be described by a single statistical ensemble. This is reasonable since for states composed of the same bound states, the effective energy corrections are the same, leading to the vanishing of effective energy splitting.

\begin{figure}[H]
	\begin{centering}
		\includegraphics[scale=1.1]{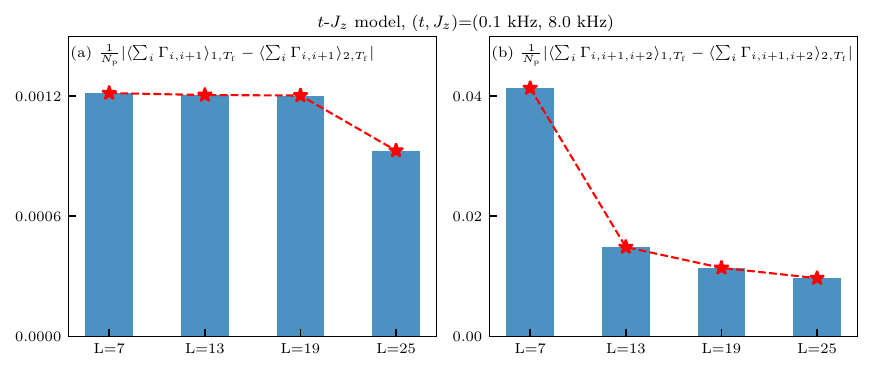}
		\caption{Normalized discrepancies of the long-time averages of the density correlations in the $t$-$J_z$ model with different number of sites. Two kinds of initial states are $|\textbf{a}\textbf{a}\cdots\textbf{a}0\rangle$ and $|\textbf{b}\textbf{b}\cdots\textbf{b}0\rangle$, where $\textbf{a}$ and $\textbf{b}$ represent the spin pattern ``0$\uparrow\downarrow$0$\uparrow\downarrow$'' and ``0$\uparrow$0$\downarrow\uparrow\downarrow$''.
			And the site number $L$ is chosen as $7,13,19,25$ with parameters $t=0.1$ kHz and $J=8.0$ kHz. The total time of evolution $T_{\mathrm{F}}$ is 600 ms. (a) The normalized discrepancies of long-time averages of $\Gamma_{i,i+1}=n_i n_{i+1}$ (b) The normalized discrepancies of long-time average of $\Gamma_{i,i+1,i+2}=n_i n_{i+1}n_{i+2}$. The bond dimension for TEBD simulation keeps up to  8,50,306,500 for $L=7,13,19,25$.}
		\label{fig: density_tjz}
		\par\end{centering}
\end{figure}

Secondly, we compare the normalized discrepancies of the long-time averages of the density correlations in $t$-$J_z$ model with different number of sites and find some signatures of Krylov-restricted thermalization. The normalized discrepancy of operator $O$ is defined as
\begin{equation}
	\frac{1}{N_{\mathrm{p}}}|\langle O\rangle_{1,T_f}-\langle O\rangle_{2,T_f}|= \frac{1}{N_{\mathrm{p}}}|\frac{1}{T_{\mathrm{F}}}\int_0^{T_{\mathrm{F}}}dt(\langle\psi_1(t)\lvert O\lvert\psi_1(t)\rangle-\langle\psi_2(t)\lvert O\lvert\psi_2(t)\rangle)|,
\end{equation}
where $\lvert\psi_1(0)\rangle$ and $\lvert\psi_2(0)\rangle$ are two resonant initial states with particle number $N_{\mathrm{P}}$, and $T_{\mathrm{F}}$ is the total time of evolution. For $t$-$J_z$ model, we choose two kinds of resonant initial states denoted as $\lvert\textbf{a}\textbf{a}\cdots\textbf{a}0\rangle$ and $\lvert\textbf{b}\textbf{b}\cdots\textbf{b}0\rangle$, where $\textbf{a}$ and $\textbf{b}$ represent the spin pattern ``0$\uparrow\downarrow$0$\uparrow\downarrow$'' and ``0$\uparrow$0$\downarrow\uparrow\downarrow$''. As shown in Fig.~\ref{fig: density_tjz}, the normalized discrepancies diminish as the number of sites increase. This result confirms that non-thermalization is mitigated in the $t$-$J_z$ model compared to the $t$-$J$ model, and suggests that Krylov-restricted thermalization may occur for $t$-$J_z$ model.

\section{Supplementary simulations on entanglement related dynamics}
\label{app: other_numerics}
\begin{figure}[H]
	\begin{centering}
		\includegraphics[scale=1]{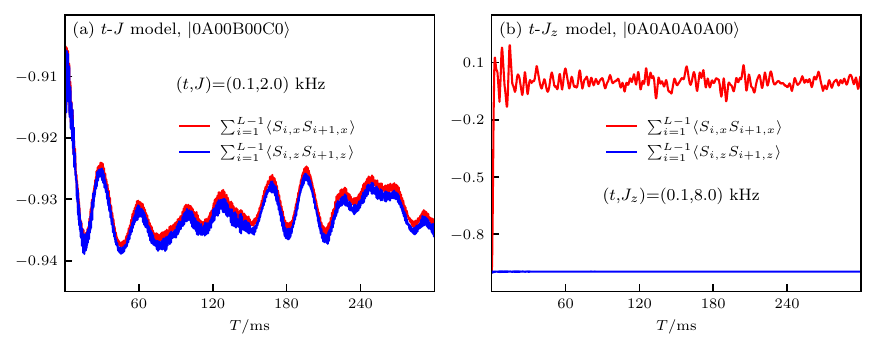}
		\caption{ Comparison between $t$-$J$ and $t$-$J_z$ model: $S_x$ and $S_z$ correlations. For $t$-$J$ model, $S_x$ and $S_z$ correlations remain isotropic. In contrast, the $S_x$ correlations decay to zero while the $S_z$ correlations remain finite in the $t$-$J_z$ model. (a) corresponds to a 14-site $t$-$J$ chain with parameters $t=0.1$ kHz and $J=2.0$ kHz. The initial state is $\lvert 0\mathrm{A}00\mathrm{B}00\mathrm{C}0\rangle$. (b) corresponds to a 14-site $t$-$J_z$ chain with parameters $t=0.1$ kHz and $J_z=8.0$ kHz. The initial state is $\lvert 0\mathrm{A}0\mathrm{A}0\mathrm{A}0\mathrm{A}00\rangle$. For both (a) and (b), The bond dimension for TEBD simulation keeps up to 500.}
		\label{fig:ss}
		\par\end{centering}
\end{figure}

In this section, we provide additional numerical simulations for the $t$-$J$ and $t$-$J_z$ model in the $J/t,J_z/t\gg1$ regime on some non-trivial quantum dynamics, highlighting the significance of the local spin entanglement. All simulations
in this section are also based on the time-evolving block decimation (TEBD) method in the Tenpy library \cite{Tenpy}.

Firstly, we compare the expectation values of $S_x$ and $S_z$ correlations (defined as $\langle S_{i,x}S_{i+1,x}\rangle$ and $\langle S_{i,z}S_{i+1,z}\rangle$) in the time evolution of $t$-$J$ and $t$-$J_z$ model, demonstrating the key influence of local spin entanglement. Special initial states are selected to satisfy $\sum_i\langle S_{i,x}S_{i+1,x}\rangle = \sum_i\langle S_{i,z}S_{i+1,z}\rangle$) for both models. As depicted in Fig. \ref{fig:ss}(a), the $S_x$ and $S_z$ correlations remain isotropic in the $t$-$J$ model:
\begin{equation}
	\sum_i\langle S_{i,x}S_{i+1,x}\rangle = \sum_i\langle S_{i,z}S_{i+1,z}\rangle.
\end{equation}
In contrast, for the $t$-$J_z$ model in Fig. \ref{fig:ss}(b), the $S_x$ correlations decay to zero while the $S_z$ correlations remain finite: 
\begin{equation}
	\sum_i\langle S_{i,x}S_{i+1,x}\rangle \rightarrow 0, \quad \sum_i\langle S_{i,z}S_{i+1,z}\rangle \rightarrow \mathrm{Const}.
\end{equation}

The discrepancy between $t$-$J$ and $t$-$J_z$ model can be attributed to the local spin entanglement. For $t$-$J$ model, all entangled spin bound states $\lvert\mathrm{A}\rangle$, $\lvert\mathrm{B}\rangle$ and $\lvert\mathrm{C}\rangle$ satisfy $\langle\sum_{i}S_{i,x}S_{i+1,x}\rangle=\langle\sum_{i}S_{i+1,z}S_{i,z}\rangle$, thereby preserving information of short-range spin entanglement through these bound states. Conversely, for the $t$-$J_z$ model, while the singlet state $\lvert\mathrm{A}\rangle$ is an eigenstate of the interaction part of the Hamiltonian, the product state $\lvert\uparrow\downarrow\rangle$ and $\lvert\downarrow\uparrow\rangle$ are also degenerate eigenstates. When two singlet states $\lvert\mathrm{A}\rangle$ approach each other, the spin-spin interaction introduces two components $\lvert\uparrow\downarrow\rangle$ and $\lvert\downarrow\uparrow\rangle$ with different phase factors. The accumulated chaotic phase factors leads to the decoherence of the system and the breakdown of the singlet state $\lvert\mathrm{A}
\rangle$. Consequently, the system can be described by a statistical mixture of product states $\lvert\uparrow\downarrow\rangle$ and $\lvert\downarrow\uparrow\rangle$, resulting in the loss of information on short-range spin entanglement and a reduction in $S_x$ correlation. In contrast, for the $t$-$J$ model, destruction of the singlet state $\lvert\mathrm{A}\rangle$ into $\lvert\uparrow\downarrow\rangle$ and $\lvert\downarrow\uparrow\rangle$ suffers an energy cost due to local spin entanglement, thus suppressing the decoherence process.

\begin{figure}[H]
	\begin{centering}
		\includegraphics[scale=1.1]{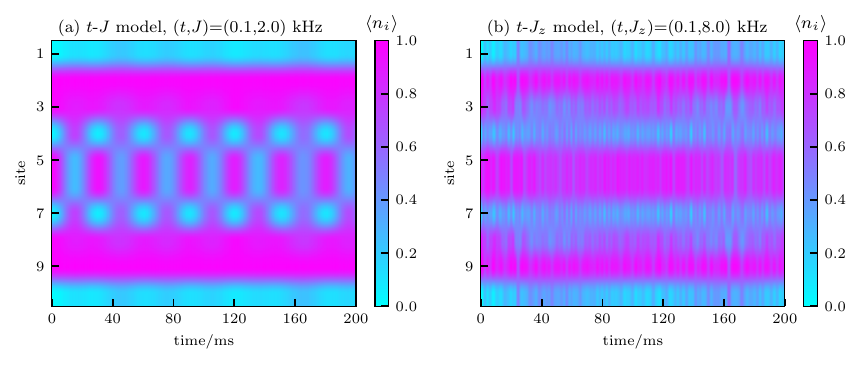}
		\caption{ 
			Comparison between $t$-$J$ model and $t$-$J_z$ model: density distributions. The density distribution behaves more coherently than that of $t$-$J_z$ model. (a) corresponds to a 10-site $t$-$J$ chain with parameters $t=0.1$ kHz and $J=2.0$ kHz. (b) corresponds to a 10-site $t$-$J_z$ chain with parameters $t=0.1$ kHz and $J_z=8.0$ kHz. For both (a) and (b), the initial state is $\lvert0\mathrm{A}0\mathrm{A}0\mathrm{A}0\rangle$. And the bond dimension for TEBD simulation keeps up to 300 for (a), 96 for (b).}
		\label{fig:d}
		\par\end{centering}
\end{figure}

This mechanism also significantly influences the density distribution during the time evolution of $t$-$J$ and $t$-$J_z$ model. As illustrated in Fig. \ref{fig:d}(a), the time evolution of the density distribution with the $t$-$J$ model exhibits a more coherent feature than that with the $t$-$J_z$ model in Fig. \ref{fig:d}(b). This observation further confirms that the outcome of long-time evolution of the initial spin bound states in the $t$-$J_z$ model resembles a statistical mixture of different product states, while the long-time evolution in the $t$-$J$ model retains more information due to local spin entanglement.

\section{Numerical simulation of the experimentally relevant model }
\label{app:non_ideal_model}
\begin{figure}[H]
	\begin{centering}
		\includegraphics[scale=1.1]{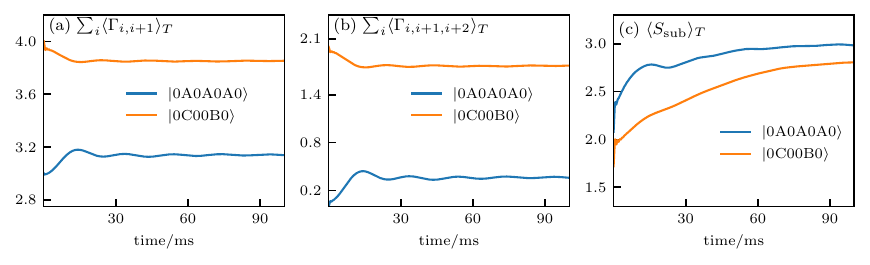}
		\caption{ Breakdown of Krylov-restricted thermalization in the experimentally relevant 10-site extended $t$-$J$ chain with parameters $t=0.1$ kHz. Other parameters are shown in Table \ref{tab:jexjz2} and Table \ref{tab:vuudd}. Similar phenomena discussed in the main text is also observed here.  (a) and (b) correspond to the long-time averages of the density correlations for different quasi-resonant initial states. For (a), $\Gamma_{i,i+1}=n_i n_{i+1}$, and for (b), $\Gamma_{i,i+1,i+2}=n_i n_{i+1}n_{i+2}$. (c) corresponds to the long-time averages of the sublattice EE for different quasi-resonant initial states.}
		\label{fig:real_model}
		\par\end{centering}
\end{figure}
In the main text, we demonstrate the HSF and nonthermal dynamics in ideal $t$-$J$ model. In this section, we demonstrate that the HSF and the nonthermal dynamics also exists in the experimentally relevant extended $t$-$J$ chain. The main differences between the experimentally relevant model and the ideal $t$-$J$ model is the density interaction $V_{\uparrow\uparrow}$, $V_{\downarrow\downarrow}$ induced by Rydberg vdW interaction, long-range interaction beyond the nearest-neighbour site, and a small anisotropy between $J_{\mathrm{ex}}$ and $J_z$. In Sec.~\ref{app:ED}, we list the strength of spin-spin interaction $J_{\mathrm{ex}}$, $J_z$ and density-density interaction $V_{\uparrow\uparrow}$, $V_{\downarrow\downarrow}$ between different sites, with the experimentally reasonable lattice spacing $a=8.2$ $\mu\mathrm{m}$. For the lattice geometry of one-dimensional chain, it is reasonable to keep $J_{\mathrm{ex}}$ and $J_z$ up to the next-nearest-neighbour interaction, and keep $V_{\uparrow\uparrow}$ and $V_{\downarrow\downarrow}$ up to the nearest-neighbour interaction, since longer-range terms are negligible.

Then we investigate the spin bound states $\lvert\mathrm{A}\rangle$, $\lvert\mathrm{B}\rangle$ and $\lvert\mathrm{C}\rangle$ in the experimentally relevant model, as the eigenstates of all the relevant interactions $J_{\mathrm{ex},\mathrm{N}}$, $J_{\mathrm{ex},\mathrm{NN}}$, $J_{z,\mathrm{N}}$, $J_{z,\mathrm{NN}}$, $V_{\uparrow\uparrow,\mathrm{N}}$, $V_{\downarrow\downarrow,\mathrm{N}}$. Here the subscripts $\mathrm{N}$ represent the nearest-neighbour interaction with the two-atom distance $x=8.2$ $\mu\mathrm{m}$, and the subscripts $\mathrm{NN}$ represent the next-nearest-neighbour interaction with the two-atom distance $x=16.4$ $\mu\mathrm{m}$. And the specific values of these interactions can be found in Table~\ref{tab:jexjz2} and Table~\ref{tab:vuudd}. In the experimentally relevant model, the corresponding wave functions and bound energies of spin bound states $\lvert\mathrm{A}\rangle$, $\lvert\mathrm{B}\rangle$ and $\lvert\mathrm{C}\rangle$ are
\begin{equation}
	\begin{aligned}
		&\lvert\mathrm{A}\rangle=(\lvert\uparrow\downarrow\rangle-\lvert\downarrow\uparrow\rangle)/
		\sqrt{2},&\ E_{\mathrm{A}}=-2.05865~\mathrm{kHz},\\
		&\lvert\mathrm{B}\rangle=0.4107\lvert\uparrow\downarrow\downarrow\rangle-0.8140\lvert\downarrow\uparrow\downarrow\rangle+0.4107\lvert\downarrow\downarrow\uparrow\rangle,&\ E_{\mathrm{B}}=-3.10548~\mathrm{kHz},\\
		&\lvert\mathrm{C}\rangle=0.3951\lvert\downarrow\downarrow\uparrow\rangle-0.8293\lvert\uparrow\downarrow\uparrow\rangle+0.3951\lvert\uparrow\uparrow\downarrow\rangle,&\ E_{\mathrm{c}}=-3.04809~\mathrm{kHz}.  
	\end{aligned}
\end{equation}
The spin bound state $\lvert\mathrm{A}\rangle$ remains the spin singlet state with minor corrections to its energy. In contrast, both the energies and wave functions of the spin bound state $\lvert\mathrm{B}\rangle$ and $\lvert\mathrm{C}\rangle$ will be slightly modified. The degeneracy relation $3E_{\mathrm{A}}=E_{\mathrm{B}}+E_{\mathrm{C}}$ is not precisely established with a small discrepancy as $|3E_{\mathrm{A}}-E_{\mathrm{B}}-E_{\mathrm{C}}|=$0.0224 kHz $\approx0.01097J_{\mathrm{ex},\mathrm{N}}$. If the amplitude of nearest-neighbour hopping $t$ is not too small compared to the 0.0224 kHz discrepancy, three $\lvert\mathrm{A}\rangle$ and one $\lvert\mathrm{B}\rangle$, one $\lvert\mathrm{C}\rangle$ states can still be treated as quasi-degenerate states. Consequently, the nonthermal dynamics demonstrated in the main text can still be observed in the experimentally relevant model.

In Fig. \ref{fig:real_model}, we repeat the numerical simulation of long-time average of density correlations and sublattice EE in the experimentally relevant extended $t$-$J$ chain by exact diagonalization (ED) method in the QuSpin library \cite{10.21468/SciPostPhys.2.1.003,10.21468/SciPostPhys.7.2.020}. In real experiment, the lifetime of the dressed-Rydberg states is about $60$ ms as listed in Sec.~\ref{app:parameters}. Therefore, we select a ten-site chain to ensure that our system can evolve into a regime near a steady state before the lifetime of the dressed-Rydberg state is exhausted. Despite the limited system size, we observe non-trivial nonthermal dynamics similar to those discussed in the main text, which may be detectable in future experiments.


\begin{thebibliography}{99}%
\makeatletter
\providecommand \@ifxundefined [1]{%
 \@ifx{#1\undefined}
}%
\providecommand \@ifnum [1]{%
 \ifnum #1\expandafter \@firstoftwo
 \else \expandafter \@secondoftwo
 \fi
}%
\providecommand \@ifx [1]{%
 \ifx #1\expandafter \@firstoftwo
 \else \expandafter \@secondoftwo
 \fi
}%
\providecommand \natexlab [1]{#1}%
\providecommand \enquote  [1]{``#1''}%
\providecommand \bibnamefont  [1]{#1}%
\providecommand \bibfnamefont [1]{#1}%
\providecommand \citenamefont [1]{#1}%
\providecommand \href@noop [0]{\@secondoftwo}%
\providecommand \href [0]{\begingroup \@sanitize@url \@href}%
\providecommand \@href[1]{\@@startlink{#1}\@@href}%
\providecommand \@@href[1]{\endgroup#1\@@endlink}%
\providecommand \@sanitize@url [0]{\catcode `\\12\catcode `\$12\catcode
  `\&12\catcode `\#12\catcode `\^12\catcode `\_12\catcode `\%12\relax}%
\providecommand \@@startlink[1]{}%
\providecommand \@@endlink[0]{}%
\providecommand \url  [0]{\begingroup\@sanitize@url \@url }%
\providecommand \@url [1]{\endgroup\@href {#1}{\urlprefix }}%
\providecommand \urlprefix  [0]{URL }%
\providecommand \Eprint [0]{\href }%
\providecommand \doibase [0]{https://doi.org/}%
\providecommand \selectlanguage [0]{\@gobble}%
\providecommand \bibinfo  [0]{\@secondoftwo}%
\providecommand \bibfield  [0]{\@secondoftwo}%
\providecommand \translation [1]{[#1]}%
\providecommand \BibitemOpen [0]{}%
\providecommand \bibitemStop [0]{}%
\providecommand \bibitemNoStop [0]{.\EOS\space}%
\providecommand \EOS [0]{\spacefactor3000\relax}%
\providecommand \BibitemShut  [1]{\csname bibitem#1\endcsname}%
\let\auto@bib@innerbib\@empty
\bibitem [{\citenamefont {Bednorz}\ and\ \citenamefont
  {M{\"u}ller}(1986)}]{Bednorz1986}%
  \BibitemOpen
  \bibfield  {author} {\bibinfo {author} {\bibfnamefont {J.~G.}\ \bibnamefont
  {Bednorz}}\ and\ \bibinfo {author} {\bibfnamefont {K.~A.}\ \bibnamefont
  {M{\"u}ller}},\ }\bibfield  {title} {\emph {\bibinfo {title} {Possible high
  ${T}_{c}$ superconductivity in the {B}a-{L}a-{C}u-{O} system}},\ }\href
  {https://doi.org/10.1007/BF01303701} {\bibfield  {journal} {\bibinfo
  {journal} {Zeitschrift f{\"u}r Physik B Condensed Matter}\ }\textbf {\bibinfo
  {volume} {64}},\ \bibinfo {pages} {189} (\bibinfo {year} {1986})}\BibitemShut
  {NoStop}%
\bibitem [{\citenamefont {Keimer}\ \emph {et~al.}(2015)\citenamefont {Keimer},
  \citenamefont {Kivelson}, \citenamefont {Norman}, \citenamefont {Uchida},\
  and\ \citenamefont {Zaanen}}]{Keimer2015}%
  \BibitemOpen
  \bibfield  {author} {\bibinfo {author} {\bibfnamefont {B.}~\bibnamefont
  {Keimer}}, \bibinfo {author} {\bibfnamefont {S.~A.}\ \bibnamefont
  {Kivelson}}, \bibinfo {author} {\bibfnamefont {M.~R.}\ \bibnamefont
  {Norman}}, \bibinfo {author} {\bibfnamefont {S.}~\bibnamefont {Uchida}},\
  and\ \bibinfo {author} {\bibfnamefont {J.}~\bibnamefont {Zaanen}},\
  }\bibfield  {title} {\emph {\bibinfo {title} {From quantum matter to
  high-temperature superconductivity in copper oxides}},\ }\href
  {https://doi.org/10.1038/nature14165} {\bibfield  {journal} {\bibinfo
  {journal} {Nature}\ }\textbf {\bibinfo {volume} {518}},\ \bibinfo {pages}
  {179} (\bibinfo {year} {2015})}\BibitemShut {NoStop}%
\bibitem [{\citenamefont {Zhou}\ \emph {et~al.}(2021)\citenamefont {Zhou},
  \citenamefont {Lee}, \citenamefont {Imada}, \citenamefont {Trivedi},
  \citenamefont {Phillips}, \citenamefont {Kee}, \citenamefont
  {T{\"o}rm{\"a}},\ and\ \citenamefont {Eremets}}]{Zhou2021}%
  \BibitemOpen
  \bibfield  {author} {\bibinfo {author} {\bibfnamefont {X.}~\bibnamefont
  {Zhou}}, \bibinfo {author} {\bibfnamefont {W.-S.}\ \bibnamefont {Lee}},
  \bibinfo {author} {\bibfnamefont {M.}~\bibnamefont {Imada}}, \bibinfo
  {author} {\bibfnamefont {N.}~\bibnamefont {Trivedi}}, \bibinfo {author}
  {\bibfnamefont {P.}~\bibnamefont {Phillips}}, \bibinfo {author}
  {\bibfnamefont {H.-Y.}\ \bibnamefont {Kee}}, \bibinfo {author} {\bibfnamefont
  {P.}~\bibnamefont {T{\"o}rm{\"a}}},\ and\ \bibinfo {author} {\bibfnamefont
  {M.}~\bibnamefont {Eremets}},\ }\bibfield  {title} {\emph {\bibinfo {title}
  {High-temperature superconductivity}},\ }\href
  {https://doi.org/10.1038/s42254-021-00324-3} {\bibfield  {journal} {\bibinfo
  {journal} {Nature Reviews Physics}\ }\textbf {\bibinfo {volume} {3}},\
  \bibinfo {pages} {462} (\bibinfo {year} {2021})}\BibitemShut {NoStop}%
\bibitem [{\citenamefont {Cirac}\ and\ \citenamefont
  {Zoller}(2012)}]{Cirac2012}%
  \BibitemOpen
  \bibfield  {author} {\bibinfo {author} {\bibfnamefont {J.~I.}\ \bibnamefont
  {Cirac}}\ and\ \bibinfo {author} {\bibfnamefont {P.}~\bibnamefont {Zoller}},\
  }\bibfield  {title} {\emph {\bibinfo {title} {Goals and opportunities in
  quantum simulation}},\ }\href {https://doi.org/10.1038/nphys2275} {\bibfield
  {journal} {\bibinfo  {journal} {Nature Physics}\ }\textbf {\bibinfo {volume}
  {8}},\ \bibinfo {pages} {264} (\bibinfo {year} {2012})}\BibitemShut {NoStop}%
\bibitem [{\citenamefont {Georgescu}\ \emph {et~al.}(2014)\citenamefont
  {Georgescu}, \citenamefont {Ashhab},\ and\ \citenamefont
  {Nori}}]{RevModPhys.86.153}%
  \BibitemOpen
  \bibfield  {author} {\bibinfo {author} {\bibfnamefont {I.~M.}\ \bibnamefont
  {Georgescu}}, \bibinfo {author} {\bibfnamefont {S.}~\bibnamefont {Ashhab}},\
  and\ \bibinfo {author} {\bibfnamefont {F.}~\bibnamefont {Nori}},\ }\bibfield
  {title} {\emph {\bibinfo {title} {Quantum simulation}},\ }\href
  {https://doi.org/10.1103/RevModPhys.86.153} {\bibfield  {journal} {\bibinfo
  {journal} {Rev. Mod. Phys.}\ }\textbf {\bibinfo {volume} {86}},\ \bibinfo
  {pages} {153} (\bibinfo {year} {2014})}\BibitemShut {NoStop}%
\bibitem [{\citenamefont {Daley}\ \emph {et~al.}(2022)\citenamefont {Daley},
  \citenamefont {Bloch}, \citenamefont {Kokail}, \citenamefont {Flannigan},
  \citenamefont {Pearson}, \citenamefont {Troyer},\ and\ \citenamefont
  {Zoller}}]{Daley2022}%
  \BibitemOpen
  \bibfield  {author} {\bibinfo {author} {\bibfnamefont {A.~J.}\ \bibnamefont
  {Daley}}, \bibinfo {author} {\bibfnamefont {I.}~\bibnamefont {Bloch}},
  \bibinfo {author} {\bibfnamefont {C.}~\bibnamefont {Kokail}}, \bibinfo
  {author} {\bibfnamefont {S.}~\bibnamefont {Flannigan}}, \bibinfo {author}
  {\bibfnamefont {N.}~\bibnamefont {Pearson}}, \bibinfo {author} {\bibfnamefont
  {M.}~\bibnamefont {Troyer}},\ and\ \bibinfo {author} {\bibfnamefont
  {P.}~\bibnamefont {Zoller}},\ }\bibfield  {title} {\emph {\bibinfo {title}
  {Practical quantum advantage in quantum simulation}},\ }\href
  {https://doi.org/10.1038/s41586-022-04940-6} {\bibfield  {journal} {\bibinfo
  {journal} {Nature}\ }\textbf {\bibinfo {volume} {607}},\ \bibinfo {pages}
  {667} (\bibinfo {year} {2022})}\BibitemShut {NoStop}%
\bibitem [{\citenamefont {Chao}\ \emph {et~al.}(1978)\citenamefont {Chao},
  \citenamefont {Spa\l{}ek},\ and\ \citenamefont {Ole\ifmmode~\acute{s}\else
  \'{s}\fi{}}}]{PhysRevB.18.3453}%
  \BibitemOpen
  \bibfield  {author} {\bibinfo {author} {\bibfnamefont {K.~A.}\ \bibnamefont
  {Chao}}, \bibinfo {author} {\bibfnamefont {J.}~\bibnamefont {Spa\l{}ek}},\
  and\ \bibinfo {author} {\bibfnamefont {A.~M.}\ \bibnamefont
  {Ole\ifmmode~\acute{s}\else \'{s}\fi{}}},\ }\bibfield  {title} {\emph
  {\bibinfo {title} {Canonical perturbation expansion of the {H}ubbard
  model}},\ }\href {https://doi.org/10.1103/PhysRevB.18.3453} {\bibfield
  {journal} {\bibinfo  {journal} {Phys. Rev. B}\ }\textbf {\bibinfo {volume}
  {18}},\ \bibinfo {pages} {3453} (\bibinfo {year} {1978})}\BibitemShut
  {NoStop}%
\bibitem [{\citenamefont {Anderson}(1987)}]{doi:10.1126/science.235.4793.1196}%
  \BibitemOpen
  \bibfield  {author} {\bibinfo {author} {\bibfnamefont {P.~W.}\ \bibnamefont
  {Anderson}},\ }\bibfield  {title} {\emph {\bibinfo {title} {The {R}esonating
  {V}alence {B}ond {S}tate in ${La}_2${C}u${O}_{4}$ and {S}uperconductivity}},\
  }\href {https://doi.org/10.1126/science.235.4793.1196} {\bibfield  {journal}
  {\bibinfo  {journal} {Science}\ }\textbf {\bibinfo {volume} {235}},\ \bibinfo
  {pages} {1196} (\bibinfo {year} {1987})}\BibitemShut {NoStop}%
\bibitem [{\citenamefont {Zhang}\ and\ \citenamefont
  {Rice}(1988)}]{PhysRevB.37.3759}%
  \BibitemOpen
  \bibfield  {author} {\bibinfo {author} {\bibfnamefont {F.~C.}\ \bibnamefont
  {Zhang}}\ and\ \bibinfo {author} {\bibfnamefont {T.~M.}\ \bibnamefont
  {Rice}},\ }\bibfield  {title} {\emph {\bibinfo {title} {Effective hamiltonian
  for the superconducting {C}u oxides}},\ }\href
  {https://doi.org/10.1103/PhysRevB.37.3759} {\bibfield  {journal} {\bibinfo
  {journal} {Phys. Rev. B}\ }\textbf {\bibinfo {volume} {37}},\ \bibinfo
  {pages} {3759} (\bibinfo {year} {1988})}\BibitemShut {NoStop}%
\bibitem [{\citenamefont {Izyumov}(1997)}]{YuriiAIzyumov_1997}%
  \BibitemOpen
  \bibfield  {author} {\bibinfo {author} {\bibfnamefont {Y.~A.}\ \bibnamefont
  {Izyumov}},\ }\bibfield  {title} {\emph {\bibinfo {title} {Strongly
  correlated electrons: the t-{J} model}},\ }\href
  {https://doi.org/10.1070/PU1997v040n05ABEH000234} {\bibfield  {journal}
  {\bibinfo  {journal} {Physics-Uspekhi}\ }\textbf {\bibinfo {volume} {40}},\
  \bibinfo {pages} {445} (\bibinfo {year} {1997})}\BibitemShut {NoStop}%
\bibitem [{\citenamefont {Lee}\ \emph {et~al.}(2006)\citenamefont {Lee},
  \citenamefont {Nagaosa},\ and\ \citenamefont {Wen}}]{RevModPhys.78.17}%
  \BibitemOpen
  \bibfield  {author} {\bibinfo {author} {\bibfnamefont {P.~A.}\ \bibnamefont
  {Lee}}, \bibinfo {author} {\bibfnamefont {N.}~\bibnamefont {Nagaosa}},\ and\
  \bibinfo {author} {\bibfnamefont {X.-G.}\ \bibnamefont {Wen}},\ }\bibfield
  {title} {\emph {\bibinfo {title} {{Doping a Mott insulator: Physics of
  high-temperature superconductivity}}},\ }\href
  {https://doi.org/10.1103/RevModPhys.78.17} {\bibfield  {journal} {\bibinfo
  {journal} {Rev. Mod. Phys.}\ }\textbf {\bibinfo {volume} {78}},\ \bibinfo
  {pages} {17} (\bibinfo {year} {2006})}\BibitemShut {NoStop}%
\bibitem [{\citenamefont {Spałek}(2007)}]{10.12693/aphyspola.111.409}%
  \BibitemOpen
  \bibfield  {author} {\bibinfo {author} {\bibfnamefont {J.}~\bibnamefont
  {Spałek}},\ }\bibfield  {title} {\emph {\bibinfo {title} {t-{J} model then
  and now: a personal perspective from the pioneering times}},\ }\href
  {https://doi.org/10.12693/aphyspola.111.409} {\bibfield  {journal} {\bibinfo
  {journal} {Acta Physica Polonica A}\ }\textbf {\bibinfo {volume} {111}},\
  \bibinfo {pages} {409} (\bibinfo {year} {2007})}\BibitemShut {NoStop}%
\bibitem [{\citenamefont {Ogata}\ and\ \citenamefont
  {Fukuyama}(2008)}]{Ogata_2008}%
  \BibitemOpen
  \bibfield  {author} {\bibinfo {author} {\bibfnamefont {M.}~\bibnamefont
  {Ogata}}\ and\ \bibinfo {author} {\bibfnamefont {H.}~\bibnamefont
  {Fukuyama}},\ }\bibfield  {title} {\emph {\bibinfo {title} {The t-{J} model
  for the oxide high-${T}_{c}$ superconductors}},\ }\href
  {https://doi.org/10.1088/0034-4885/71/3/036501} {\bibfield  {journal}
  {\bibinfo  {journal} {Reports on Progress in Physics}\ }\textbf {\bibinfo
  {volume} {71}},\ \bibinfo {pages} {036501} (\bibinfo {year}
  {2008})}\BibitemShut {NoStop}%
\bibitem [{\citenamefont {Esslinger}(2010)}]{annurev-conmatphys-070909-104059}%
  \BibitemOpen
  \bibfield  {author} {\bibinfo {author} {\bibfnamefont {T.}~\bibnamefont
  {Esslinger}},\ }\bibfield  {title} {\emph {\bibinfo {title} {Fermi-{H}ubbard
  {P}hysics with {A}toms in an {O}ptical {L}attice}},\ }\href
  {https://doi.org/https://doi.org/10.1146/annurev-conmatphys-070909-104059}
  {\bibfield  {journal} {\bibinfo  {journal} {Annual Review of Condensed Matter
  Physics}\ }\textbf {\bibinfo {volume} {1}},\ \bibinfo {pages} {129} (\bibinfo
  {year} {2010})}\BibitemShut {NoStop}%
\bibitem [{\citenamefont {Strohmaier}\ \emph {et~al.}(2010)\citenamefont
	{Strohmaier}, \citenamefont {Greif}, \citenamefont {J\"ordens},
	\citenamefont {Tarruell}, \citenamefont {Moritz}, \citenamefont
	{Esslinger}, \citenamefont {Sensarma}, \citenamefont {Pekker}, \citenamefont
	{Altman},\ and\ \citenamefont {Demler}}]{PhysRevLett.104.080401}%
\BibitemOpen
\bibfield  {author} {\bibinfo {author} {\bibfnamefont {Niels}\ \bibnamefont
		{Strohmaier}}, \bibinfo {author} {\bibfnamefont {Daniel}\ \bibnamefont
		{Greif}}, \bibinfo {author} {\bibfnamefont {Robert}\ \bibnamefont
		{J\"ordens}}, \bibinfo {author} {\bibfnamefont {Leticia}\ \bibnamefont
		{Tarruell}}, \bibinfo {author} {\bibfnamefont {Henning}\ \bibnamefont
		{Moritz}}, \bibinfo {author} {\bibfnamefont {Tilman}\ \bibnamefont
		{Esslinger}}, \bibinfo {author} {\bibfnamefont {Rajdeep}\ \bibnamefont
		{Sensarma}}, \bibinfo {author} {\bibfnamefont {David}\ \bibnamefont
		{Pekker}}, \bibinfo {author} {\bibfnamefont {Ehud}\ \bibnamefont {Altman}},\
	and\ \bibinfo {author} {\bibfnamefont {Eugene}\ \bibnamefont {Demler}},\
}\bibfield  {title} {\emph {\bibinfo {title} {Observation of Elastic Doublon Decay in the Fermi-Hubbard Model}},\ }\href
{https://doi.org/10.1103/PhysRevLett.104.080401} {\bibfield  {journal}
	{\bibinfo  {journal} {Phys. Rev. Lett.}\ }\textbf {\bibinfo {volume} {104}},\
	\bibinfo {pages} {080401} (\bibinfo {year} {2010})}\BibitemShut {NoStop}%
\bibitem [{\citenamefont {Hart}\ \emph {et~al.}(2015)\citenamefont {Hart},
  \citenamefont {Duarte}, \citenamefont {Yang}, \citenamefont {Liu},
  \citenamefont {Paiva}, \citenamefont {Khatami}, \citenamefont {Scalettar},
  \citenamefont {Trivedi}, \citenamefont {Huse},\ and\ \citenamefont
  {Hulet}}]{Hart2015}%
  \BibitemOpen
  \bibfield  {author} {\bibinfo {author} {\bibfnamefont {R.~A.}\ \bibnamefont
  {Hart}}, \bibinfo {author} {\bibfnamefont {P.~M.}\ \bibnamefont {Duarte}},
  \bibinfo {author} {\bibfnamefont {T.-L.}\ \bibnamefont {Yang}}, \bibinfo
  {author} {\bibfnamefont {X.}~\bibnamefont {Liu}}, \bibinfo {author}
  {\bibfnamefont {T.}~\bibnamefont {Paiva}}, \bibinfo {author} {\bibfnamefont
  {E.}~\bibnamefont {Khatami}}, \bibinfo {author} {\bibfnamefont {R.~T.}\
  \bibnamefont {Scalettar}}, \bibinfo {author} {\bibfnamefont {N.}~\bibnamefont
  {Trivedi}}, \bibinfo {author} {\bibfnamefont {D.~A.}\ \bibnamefont {Huse}},\
  and\ \bibinfo {author} {\bibfnamefont {R.~G.}\ \bibnamefont {Hulet}},\
  }\bibfield  {title} {\emph {\bibinfo {title} {{Observation of
  antiferromagnetic correlations in the Hubbard model with ultracold atoms}}},\
  }\href {https://doi.org/10.1038/nature14223} {\bibfield  {journal} {\bibinfo
  {journal} {Nature}\ }\textbf {\bibinfo {volume} {519}},\ \bibinfo {pages}
  {211} (\bibinfo {year} {2015})}\BibitemShut {NoStop}%
\bibitem [{\citenamefont {Cheuk}\ \emph
  {et~al.}(2016{\natexlab{a}})\citenamefont {Cheuk}, \citenamefont {Nichols},
  \citenamefont {Lawrence}, \citenamefont {Okan}, \citenamefont {Zhang},
  \citenamefont {Khatami}, \citenamefont {Trivedi}, \citenamefont {Paiva},
  \citenamefont {Rigol},\ and\ \citenamefont
  {Zwierlein}}]{doi:10.1126/science.aag3349}%
  \BibitemOpen
  \bibfield  {author} {\bibinfo {author} {\bibfnamefont {L.~W.}\ \bibnamefont
  {Cheuk}}, \bibinfo {author} {\bibfnamefont {M.~A.}\ \bibnamefont {Nichols}},
  \bibinfo {author} {\bibfnamefont {K.~R.}\ \bibnamefont {Lawrence}}, \bibinfo
  {author} {\bibfnamefont {M.}~\bibnamefont {Okan}}, \bibinfo {author}
  {\bibfnamefont {H.}~\bibnamefont {Zhang}}, \bibinfo {author} {\bibfnamefont
  {E.}~\bibnamefont {Khatami}}, \bibinfo {author} {\bibfnamefont
  {N.}~\bibnamefont {Trivedi}}, \bibinfo {author} {\bibfnamefont
  {T.}~\bibnamefont {Paiva}}, \bibinfo {author} {\bibfnamefont
  {M.}~\bibnamefont {Rigol}},\ and\ \bibinfo {author} {\bibfnamefont {M.~W.}\
  \bibnamefont {Zwierlein}},\ }\bibfield  {title} {\emph {\bibinfo {title}
  {{O}bservation of spatial charge and spin correlations in the 2{D}
  {F}ermi-{H}ubbard model}},\ }\href {https://doi.org/10.1126/science.aag3349}
  {\bibfield  {journal} {\bibinfo  {journal} {Science}\ }\textbf {\bibinfo
  {volume} {353}},\ \bibinfo {pages} {1260} (\bibinfo {year}
  {2016}{\natexlab{a}})}\BibitemShut {NoStop}%
\bibitem [{\citenamefont {Mazurenko}\ \emph {et~al.}(2017)\citenamefont
  {Mazurenko}, \citenamefont {Chiu}, \citenamefont {Ji}, \citenamefont
  {Parsons}, \citenamefont {Kan{\'a}sz-Nagy}, \citenamefont {Schmidt},
  \citenamefont {Grusdt}, \citenamefont {Demler}, \citenamefont {Greif},\ and\
  \citenamefont {Greiner}}]{Mazurenko2017}%
  \BibitemOpen
  \bibfield  {author} {\bibinfo {author} {\bibfnamefont {A.}~\bibnamefont
  {Mazurenko}}, \bibinfo {author} {\bibfnamefont {C.~S.}\ \bibnamefont {Chiu}},
  \bibinfo {author} {\bibfnamefont {G.}~\bibnamefont {Ji}}, \bibinfo {author}
  {\bibfnamefont {M.~F.}\ \bibnamefont {Parsons}}, \bibinfo {author}
  {\bibfnamefont {M.}~\bibnamefont {Kan{\'a}sz-Nagy}}, \bibinfo {author}
  {\bibfnamefont {R.}~\bibnamefont {Schmidt}}, \bibinfo {author} {\bibfnamefont
  {F.}~\bibnamefont {Grusdt}}, \bibinfo {author} {\bibfnamefont
  {E.}~\bibnamefont {Demler}}, \bibinfo {author} {\bibfnamefont
  {D.}~\bibnamefont {Greif}},\ and\ \bibinfo {author} {\bibfnamefont
  {M.}~\bibnamefont {Greiner}},\ }\bibfield  {title} {\emph {\bibinfo {title}
  {{A} cold-atom {F}ermi--{H}ubbard antiferromagnet}},\ }\href
  {https://doi.org/10.1038/nature22362} {\bibfield  {journal} {\bibinfo
  {journal} {Nature}\ }\textbf {\bibinfo {volume} {545}},\ \bibinfo {pages}
  {462} (\bibinfo {year} {2017})}\BibitemShut {NoStop}%
\bibitem [{\citenamefont {Tarruell}\ and\ \citenamefont
  {Sanchez-Palencia}(2018)}]{TARRUELL2018365}%
  \BibitemOpen
  \bibfield  {author} {\bibinfo {author} {\bibfnamefont {L.}~\bibnamefont
  {Tarruell}}\ and\ \bibinfo {author} {\bibfnamefont {L.}~\bibnamefont
  {Sanchez-Palencia}},\ }\bibfield  {title} {\emph {\bibinfo {title} {Quantum
  simulation of the {H}ubbard model with ultracold fermions in optical
  lattices}},\ }\href
  {https://doi.org/https://doi.org/10.1016/j.crhy.2018.10.013} {\bibfield
  {journal} {\bibinfo  {journal} {Comptes Rendus Physique}\ }\textbf {\bibinfo
  {volume} {19}},\ \bibinfo {pages} {365} (\bibinfo {year} {2018})}\BibitemShut
  {NoStop}%
\bibitem [{\citenamefont {Koepsell}\ \emph {et~al.}(2019)\citenamefont
  {Koepsell}, \citenamefont {Vijayan}, \citenamefont {Sompet}, \citenamefont
  {Grusdt}, \citenamefont {Hilker}, \citenamefont {Demler}, \citenamefont
  {Salomon}, \citenamefont {Bloch},\ and\ \citenamefont
  {Gross}}]{Koepsell2019}%
  \BibitemOpen
  \bibfield  {author} {\bibinfo {author} {\bibfnamefont {J.}~\bibnamefont
  {Koepsell}}, \bibinfo {author} {\bibfnamefont {J.}~\bibnamefont {Vijayan}},
  \bibinfo {author} {\bibfnamefont {P.}~\bibnamefont {Sompet}}, \bibinfo
  {author} {\bibfnamefont {F.}~\bibnamefont {Grusdt}}, \bibinfo {author}
  {\bibfnamefont {T.~A.}\ \bibnamefont {Hilker}}, \bibinfo {author}
  {\bibfnamefont {E.}~\bibnamefont {Demler}}, \bibinfo {author} {\bibfnamefont
  {G.}~\bibnamefont {Salomon}}, \bibinfo {author} {\bibfnamefont
  {I.}~\bibnamefont {Bloch}},\ and\ \bibinfo {author} {\bibfnamefont
  {C.}~\bibnamefont {Gross}},\ }\bibfield  {title} {\emph {\bibinfo {title}
  {{I}maging magnetic polarons in the doped {F}ermi--{H}ubbard model}},\ }\href
  {https://doi.org/10.1038/s41586-019-1463-1} {\bibfield  {journal} {\bibinfo
  {journal} {Nature}\ }\textbf {\bibinfo {volume} {572}},\ \bibinfo {pages}
  {358} (\bibinfo {year} {2019})}\BibitemShut {NoStop}%
\bibitem [{\citenamefont {Bohrdt}\ \emph {et~al.}(2021)\citenamefont {Bohrdt},
  \citenamefont {Homeier}, \citenamefont {Reinmoser}, \citenamefont {Demler},\
  and\ \citenamefont {Grusdt}}]{BOHRDT2021168651}%
  \BibitemOpen
  \bibfield  {author} {\bibinfo {author} {\bibfnamefont {A.}~\bibnamefont
  {Bohrdt}}, \bibinfo {author} {\bibfnamefont {L.}~\bibnamefont {Homeier}},
  \bibinfo {author} {\bibfnamefont {C.}~\bibnamefont {Reinmoser}}, \bibinfo
  {author} {\bibfnamefont {E.}~\bibnamefont {Demler}},\ and\ \bibinfo {author}
  {\bibfnamefont {F.}~\bibnamefont {Grusdt}},\ }\bibfield  {title} {\emph
  {\bibinfo {title} {Exploration of doped quantum magnets with ultracold
  atoms}},\ }\href {https://doi.org/https://doi.org/10.1016/j.aop.2021.168651}
  {\bibfield  {journal} {\bibinfo  {journal} {Annals of Physics}\ }\textbf
  {\bibinfo {volume} {435}},\ \bibinfo {pages} {168651} (\bibinfo {year}
  {2021})}\BibitemShut {NoStop}%
\bibitem [{\citenamefont {Xu}\ \emph {et~al.}(2023)\citenamefont {Xu},
  \citenamefont {Kendrick}, \citenamefont {Kale}, \citenamefont {Gang},
  \citenamefont {Ji}, \citenamefont {Scalettar}, \citenamefont {Lebrat},\ and\
  \citenamefont {Greiner}}]{Xu2023}%
  \BibitemOpen
  \bibfield  {author} {\bibinfo {author} {\bibfnamefont {M.}~\bibnamefont
  {Xu}}, \bibinfo {author} {\bibfnamefont {L.~H.}\ \bibnamefont {Kendrick}},
  \bibinfo {author} {\bibfnamefont {A.}~\bibnamefont {Kale}}, \bibinfo {author}
  {\bibfnamefont {Y.}~\bibnamefont {Gang}}, \bibinfo {author} {\bibfnamefont
  {G.}~\bibnamefont {Ji}}, \bibinfo {author} {\bibfnamefont {R.~T.}\
  \bibnamefont {Scalettar}}, \bibinfo {author} {\bibfnamefont {M.}~\bibnamefont
  {Lebrat}},\ and\ \bibinfo {author} {\bibfnamefont {M.}~\bibnamefont
  {Greiner}},\ }\bibfield  {title} {\emph {\bibinfo {title} {{F}rustration- and
  doping-induced magnetism in a {F}ermi--{H}ubbard simulator}},\ }\href
  {https://doi.org/10.1038/s41586-023-06280-5} {\bibfield  {journal} {\bibinfo
  {journal} {Nature}\ }\textbf {\bibinfo {volume} {620}},\ \bibinfo {pages}
  {971} (\bibinfo {year} {2023})}\BibitemShut {NoStop}%
\bibitem [{\citenamefont {Shao}\ \emph {et~al.}(2024)\citenamefont {Shao},
  \citenamefont {Wang}, \citenamefont {Zhu}, \citenamefont {Zhu}, \citenamefont
  {Sun}, \citenamefont {Chen}, \citenamefont {Zhang}, \citenamefont {Fan},
  \citenamefont {Deng}, \citenamefont {Yao}, \citenamefont {Chen},\ and\
  \citenamefont {Pan}}]{Shao2024}%
  \BibitemOpen
  \bibfield  {author} {\bibinfo {author} {\bibfnamefont {H.-J.}\ \bibnamefont
  {Shao}}, \bibinfo {author} {\bibfnamefont {Y.-X.}\ \bibnamefont {Wang}},
  \bibinfo {author} {\bibfnamefont {D.-Z.}\ \bibnamefont {Zhu}}, \bibinfo
  {author} {\bibfnamefont {Y.-S.}\ \bibnamefont {Zhu}}, \bibinfo {author}
  {\bibfnamefont {H.-N.}\ \bibnamefont {Sun}}, \bibinfo {author} {\bibfnamefont
  {S.-Y.}\ \bibnamefont {Chen}}, \bibinfo {author} {\bibfnamefont
  {C.}~\bibnamefont {Zhang}}, \bibinfo {author} {\bibfnamefont {Z.-J.}\
  \bibnamefont {Fan}}, \bibinfo {author} {\bibfnamefont {Y.}~\bibnamefont
  {Deng}}, \bibinfo {author} {\bibfnamefont {X.-C.}\ \bibnamefont {Yao}},
  \bibinfo {author} {\bibfnamefont {Y.-A.}\ \bibnamefont {Chen}},\ and\
  \bibinfo {author} {\bibfnamefont {J.-W.}\ \bibnamefont {Pan}},\ }\bibfield
  {title} {\emph {\bibinfo {title} {{Antiferromagnetic phase transition in a 3D
  fermionic Hubbard model}}},\ }\href
  {https://doi.org/10.1038/s41586-024-07689-2} {\bibfield  {journal} {\bibinfo
  {journal} {Nature}\ }\textbf {\bibinfo {volume} {632}},\ \bibinfo {pages}
  {267} (\bibinfo {year} {2024})}\BibitemShut {NoStop}%
\bibitem [{\citenamefont {Jiang}\ and\ \citenamefont
  {Devereaux}(2019)}]{doi:10.1126/science.aal5304}%
  \BibitemOpen
  \bibfield  {author} {\bibinfo {author} {\bibfnamefont {H.-C.}\ \bibnamefont
  {Jiang}}\ and\ \bibinfo {author} {\bibfnamefont {T.~P.}\ \bibnamefont
  {Devereaux}},\ }\bibfield  {title} {\emph {\bibinfo {title}
  {Superconductivity in the doped {H}ubbard model and its interplay with
  next-nearest hopping $t^{\prime}$}},\ }\href
  {https://doi.org/10.1126/science.aal5304} {\bibfield  {journal} {\bibinfo
  {journal} {Science}\ }\textbf {\bibinfo {volume} {365}},\ \bibinfo {pages}
  {1424} (\bibinfo {year} {2019})}\BibitemShut {NoStop}%
\bibitem [{\citenamefont {Gong}\ \emph {et~al.}(2021)\citenamefont {Gong},
  \citenamefont {Zhu},\ and\ \citenamefont {Sheng}}]{PhysRevLett.127.097003}%
  \BibitemOpen
  \bibfield  {author} {\bibinfo {author} {\bibfnamefont {S.}~\bibnamefont
  {Gong}}, \bibinfo {author} {\bibfnamefont {W.}~\bibnamefont {Zhu}},\ and\
  \bibinfo {author} {\bibfnamefont {D.~N.}\ \bibnamefont {Sheng}},\ }\bibfield
  {title} {\emph {\bibinfo {title} {Robust $d$-{W}ave {S}uperconductivity in
  the {S}quare-{L}attice $t\text{\ensuremath{-}}{J}$ model}},\ }\href
  {https://doi.org/10.1103/PhysRevLett.127.097003} {\bibfield  {journal}
  {\bibinfo  {journal} {Phys. Rev. Lett.}\ }\textbf {\bibinfo {volume} {127}},\
  \bibinfo {pages} {097003} (\bibinfo {year} {2021})}\BibitemShut {NoStop}%
\bibitem [{\citenamefont {Jiang}\ \emph {et~al.}(2021)\citenamefont {Jiang},
  \citenamefont {Scalapino},\ and\ \citenamefont
  {White}}]{doi:10.1073/pnas.2109978118}%
  \BibitemOpen
  \bibfield  {author} {\bibinfo {author} {\bibfnamefont {S.}~\bibnamefont
  {Jiang}}, \bibinfo {author} {\bibfnamefont {D.~J.}\ \bibnamefont
  {Scalapino}},\ and\ \bibinfo {author} {\bibfnamefont {S.~R.}\ \bibnamefont
  {White}},\ }\bibfield  {title} {\emph {\bibinfo {title} {Ground-state phase
  diagram of the $t-t^{\prime}-{J}$ model}},\ }\href
  {https://doi.org/10.1073/pnas.2109978118} {\bibfield  {journal} {\bibinfo
  {journal} {Proceedings of the National Academy of Sciences}\ }\textbf
  {\bibinfo {volume} {118}},\ \bibinfo {pages} {e2109978118} (\bibinfo {year}
  {2021})}\BibitemShut {NoStop}%
\bibitem [{\citenamefont {Jiang}\ and\ \citenamefont
  {Kivelson}(2021)}]{PhysRevLett.127.097002}%
  \BibitemOpen
  \bibfield  {author} {\bibinfo {author} {\bibfnamefont {H.-C.}\ \bibnamefont
  {Jiang}}\ and\ \bibinfo {author} {\bibfnamefont {S.~A.}\ \bibnamefont
  {Kivelson}},\ }\bibfield  {title} {\emph {\bibinfo {title} {High
  {T}emperature {S}uperconductivity in a {L}ightly {D}oped {Q}uantum {S}pin
  {L}iquid}},\ }\href {https://doi.org/10.1103/PhysRevLett.127.097002}
  {\bibfield  {journal} {\bibinfo  {journal} {Phys. Rev. Lett.}\ }\textbf
  {\bibinfo {volume} {127}},\ \bibinfo {pages} {097002} (\bibinfo {year}
  {2021})}\BibitemShut {NoStop}%
\bibitem [{\citenamefont {Lu}\ \emph {et~al.}(2024)\citenamefont {Lu},
  \citenamefont {Chen}, \citenamefont {Zhu}, \citenamefont {Sheng},\ and\
  \citenamefont {Gong}}]{PhysRevLett.132.066002}%
  \BibitemOpen
  \bibfield  {author} {\bibinfo {author} {\bibfnamefont {X.}~\bibnamefont
  {Lu}}, \bibinfo {author} {\bibfnamefont {F.}~\bibnamefont {Chen}}, \bibinfo
  {author} {\bibfnamefont {W.}~\bibnamefont {Zhu}}, \bibinfo {author}
  {\bibfnamefont {D.~N.}\ \bibnamefont {Sheng}},\ and\ \bibinfo {author}
  {\bibfnamefont {S.-S.}\ \bibnamefont {Gong}},\ }\bibfield  {title} {\emph
  {\bibinfo {title} {{E}mergent {S}uperconductivity and {C}ompeting {C}harge
  {O}rders in {H}ole-{D}oped {S}quare-{L}attice $t\text{\ensuremath{-}}{J}$
  model}},\ }\href {https://doi.org/10.1103/PhysRevLett.132.066002} {\bibfield
  {journal} {\bibinfo  {journal} {Phys. Rev. Lett.}\ }\textbf {\bibinfo
  {volume} {132}},\ \bibinfo {pages} {066002} (\bibinfo {year}
  {2024})}\BibitemShut {NoStop}%
\bibitem [{\citenamefont {Xu}\ \emph {et~al.}(2024)\citenamefont {Xu},
  \citenamefont {Chung}, \citenamefont {Qin}, \citenamefont {Schollwöck},
  \citenamefont {White},\ and\ \citenamefont
  {Zhang}}]{doi:10.1126/science.adh7691}%
  \BibitemOpen
  \bibfield  {author} {\bibinfo {author} {\bibfnamefont {H.}~\bibnamefont
  {Xu}}, \bibinfo {author} {\bibfnamefont {C.-M.}\ \bibnamefont {Chung}},
  \bibinfo {author} {\bibfnamefont {M.}~\bibnamefont {Qin}}, \bibinfo {author}
  {\bibfnamefont {U.}~\bibnamefont {Schollwöck}}, \bibinfo {author}
  {\bibfnamefont {S.~R.}\ \bibnamefont {White}},\ and\ \bibinfo {author}
  {\bibfnamefont {S.}~\bibnamefont {Zhang}},\ }\bibfield  {title} {\emph
  {\bibinfo {title} {Coexistence of superconductivity with partially filled
  stripes in the {H}ubbard model}},\ }\href
  {https://doi.org/10.1126/science.adh7691} {\bibfield  {journal} {\bibinfo
  {journal} {Science}\ }\textbf {\bibinfo {volume} {384}},\ \bibinfo {pages}
  {eadh7691} (\bibinfo {year} {2024})}\BibitemShut {NoStop}%
\bibitem [{\citenamefont {Qu}\ \emph {et~al.}(2024)\citenamefont {Qu},
  \citenamefont {Li}, \citenamefont {Gong}, \citenamefont {Qi}, \citenamefont
  {Li},\ and\ \citenamefont {Su}}]{PhysRevLett.133.256003}%
  \BibitemOpen
  \bibfield  {author} {\bibinfo {author} {\bibfnamefont {D.-W.}\ \bibnamefont
  {Qu}}, \bibinfo {author} {\bibfnamefont {Q.}~\bibnamefont {Li}}, \bibinfo
  {author} {\bibfnamefont {S.-S.}\ \bibnamefont {Gong}}, \bibinfo {author}
  {\bibfnamefont {Y.}~\bibnamefont {Qi}}, \bibinfo {author} {\bibfnamefont
  {W.}~\bibnamefont {Li}},\ and\ \bibinfo {author} {\bibfnamefont
  {G.}~\bibnamefont {Su}},\ }\bibfield  {title} {\emph {\bibinfo {title} {Phase
  {D}iagram, $d$-wave {S}uperconductivity, and {P}seudogap of the
  $t\ensuremath{-}{t}^{\ensuremath{'}}\ensuremath{-}{J}$ {M}odel at {F}inite
  {T}emperature}},\ }\href {https://doi.org/10.1103/PhysRevLett.133.256003}
  {\bibfield  {journal} {\bibinfo  {journal} {Phys. Rev. Lett.}\ }\textbf
  {\bibinfo {volume} {133}},\ \bibinfo {pages} {256003} (\bibinfo {year}
  {2024})}\BibitemShut {NoStop}%
\bibitem [{\citenamefont {Jiang}\ \emph {et~al.}(2020)\citenamefont {Jiang},
  \citenamefont {Zaanen}, \citenamefont {Devereaux},\ and\ \citenamefont
  {Jiang}}]{PhysRevResearch.2.033073}%
  \BibitemOpen
  \bibfield  {author} {\bibinfo {author} {\bibfnamefont {Y.-F.}\ \bibnamefont
  {Jiang}}, \bibinfo {author} {\bibfnamefont {J.}~\bibnamefont {Zaanen}},
  \bibinfo {author} {\bibfnamefont {T.~P.}\ \bibnamefont {Devereaux}},\ and\
  \bibinfo {author} {\bibfnamefont {H.-C.}\ \bibnamefont {Jiang}},\ }\bibfield
  {title} {\emph {\bibinfo {title} {{Ground state phase diagram of the doped
  Hubbard model on the four-leg cylinder}}},\ }\href
  {https://doi.org/10.1103/PhysRevResearch.2.033073} {\bibfield  {journal}
  {\bibinfo  {journal} {Phys. Rev. Res.}\ }\textbf {\bibinfo {volume} {2}},\
  \bibinfo {pages} {033073} (\bibinfo {year} {2020})}\BibitemShut {NoStop}%
\bibitem [{\citenamefont {Chen}\ \emph {et~al.}(2025)\citenamefont {Chen},
	\citenamefont {Haldane},\ and\ \citenamefont {Sheng}}]{Chen2025}%
\BibitemOpen
\bibfield  {author} {\bibinfo {author} {\bibfnamefont {Feng}\ \bibnamefont
		{Chen}}, \bibinfo {author} {\bibfnamefont {F.~D.~M.}\ \bibnamefont {Haldane}},\
	and\ \bibinfo {author} {\bibfnamefont {D.~N.}\ \bibnamefont {Sheng}},\
}\bibfield  {title} {\emph {\bibinfo {title} {Global phase diagram of D-wave
			superconductivity in the square-lattice \textit{t-J} model}},\ }\href
{https://doi.org/10.1073/pnas.2420963122} {\bibfield  {journal} {\bibinfo
		{journal} {Proc. Natl. Acad. Sci. U.S.A.}\ }\textbf {\bibinfo {volume} {122}},\
	\bibinfo {pages} {e2420963122} (\bibinfo {year} {2025})}\BibitemShut {NoStop}%
\bibitem [{\citenamefont {Miller}\ \emph {et~al.}(2024)\citenamefont {Miller},
  \citenamefont {Carroll}, \citenamefont {Lin}, \citenamefont {Hirzler},
  \citenamefont {Gao}, \citenamefont {Zhou}, \citenamefont {Lukin},\ and\
  \citenamefont {Ye}}]{Miller2024}%
  \BibitemOpen
  \bibfield  {author} {\bibinfo {author} {\bibfnamefont {C.}~\bibnamefont
  {Miller}}, \bibinfo {author} {\bibfnamefont {A.~N.}\ \bibnamefont {Carroll}},
  \bibinfo {author} {\bibfnamefont {J.}~\bibnamefont {Lin}}, \bibinfo {author}
  {\bibfnamefont {H.}~\bibnamefont {Hirzler}}, \bibinfo {author} {\bibfnamefont
  {H.}~\bibnamefont {Gao}}, \bibinfo {author} {\bibfnamefont {H.}~\bibnamefont
  {Zhou}}, \bibinfo {author} {\bibfnamefont {M.~D.}\ \bibnamefont {Lukin}},\
  and\ \bibinfo {author} {\bibfnamefont {J.}~\bibnamefont {Ye}},\ }\bibfield
  {title} {\emph {\bibinfo {title} {Two-axis twisting using floquet-engineered
  xyz spin models with polar molecules}},\ }\href
  {https://doi.org/10.1038/s41586-024-07883-2} {\bibfield  {journal} {\bibinfo
  {journal} {Nature}\ }\textbf {\bibinfo {volume} {633}},\ \bibinfo {pages}
  {332} (\bibinfo {year} {2024})}\BibitemShut {NoStop}%
\bibitem [{\citenamefont {Carroll}\ \emph {et~al.}(2025)\citenamefont {Carroll},
	\citenamefont {Hirzler}, \citenamefont {Miller}, \citenamefont {Wellnitz},
	\citenamefont {Muleady}, \citenamefont {Lin}, \citenamefont {Zamarski},
	\citenamefont {Wang}, \citenamefont {Bohn}, \citenamefont {Rey},\ and\
	\citenamefont {Ye}}]{Carroll2025}%
\BibitemOpen
\bibfield  {author} {\bibinfo {author} {\bibfnamefont {A.~N.}\
		\bibnamefont {Carroll}}, \bibinfo {author} {\bibfnamefont {H.}\
		\bibnamefont {Hirzler}}, \bibinfo {author} {\bibfnamefont {C.}\
		\bibnamefont {Miller}}, \bibinfo {author} {\bibfnamefont {D.}\
		\bibnamefont {Wellnitz}}, \bibinfo {author} {\bibfnamefont {S.~R.}\
		\bibnamefont {Muleady}}, \bibinfo {author} {\bibfnamefont {J.}\
		\bibnamefont {Lin}}, \bibinfo {author} {\bibfnamefont {K.~P.}\
		\bibnamefont {Zamarski}}, \bibinfo {author} {\bibfnamefont {R.~R.~W.}\
		\bibnamefont {Wang}}, \bibinfo {author} {\bibfnamefont {J.~L.}\
		\bibnamefont {Bohn}}, \bibinfo {author} {\bibfnamefont {A.~M.}\
		\bibnamefont {Rey}}, \ and\ \bibinfo {author} {\bibfnamefont {J.}\
		\bibnamefont {Ye}},\ }\bibfield  {title} {\emph {\bibinfo {title} {Observation
			of generalized \textit{t-J} spin dynamics with tunable dipolar
			interactions}},\ }\href {https://doi.org/10.1126/science.adq0911} {\bibfield
	{journal} {\bibinfo  {journal} {Science}\ }\textbf {\bibinfo {volume} {388}},\
	\bibinfo {pages} {381--386} (\bibinfo {year} {2025})}\BibitemShut {NoStop}%
\bibitem [{\citenamefont {Endres}\ \emph {et~al.}(2016)\citenamefont {Endres},
  \citenamefont {Bernien}, \citenamefont {Keesling}, \citenamefont {Levine},
  \citenamefont {Anschuetz}, \citenamefont {Krajenbrink}, \citenamefont
  {Senko}, \citenamefont {Vuletic}, \citenamefont {Greiner},\ and\
  \citenamefont {Lukin}}]{doi:10.1126/science.aah3752}%
  \BibitemOpen
  \bibfield  {author} {\bibinfo {author} {\bibfnamefont {M.}~\bibnamefont
  {Endres}}, \bibinfo {author} {\bibfnamefont {H.}~\bibnamefont {Bernien}},
  \bibinfo {author} {\bibfnamefont {A.}~\bibnamefont {Keesling}}, \bibinfo
  {author} {\bibfnamefont {H.}~\bibnamefont {Levine}}, \bibinfo {author}
  {\bibfnamefont {E.~R.}\ \bibnamefont {Anschuetz}}, \bibinfo {author}
  {\bibfnamefont {A.}~\bibnamefont {Krajenbrink}}, \bibinfo {author}
  {\bibfnamefont {C.}~\bibnamefont {Senko}}, \bibinfo {author} {\bibfnamefont
  {V.}~\bibnamefont {Vuletic}}, \bibinfo {author} {\bibfnamefont
  {M.}~\bibnamefont {Greiner}},\ and\ \bibinfo {author} {\bibfnamefont {M.~D.}\
  \bibnamefont {Lukin}},\ }\bibfield  {title} {\emph {\bibinfo {title}
  {Atom-by-atom assembly of defect-free one-dimensional cold atom arrays}},\
  }\href {https://doi.org/10.1126/science.aah3752} {\bibfield  {journal}
  {\bibinfo  {journal} {Science}\ }\textbf {\bibinfo {volume} {354}},\ \bibinfo
  {pages} {1024} (\bibinfo {year} {2016})}\BibitemShut {NoStop}%
\bibitem [{\citenamefont {Barredo}\ \emph {et~al.}(2016)\citenamefont
  {Barredo}, \citenamefont {de~Léséleuc}, \citenamefont {Lienhard},
  \citenamefont {Lahaye},\ and\ \citenamefont
  {Browaeys}}]{doi:10.1126/science.aah3778}%
  \BibitemOpen
  \bibfield  {author} {\bibinfo {author} {\bibfnamefont {D.}~\bibnamefont
  {Barredo}}, \bibinfo {author} {\bibfnamefont {S.}~\bibnamefont
  {de~Léséleuc}}, \bibinfo {author} {\bibfnamefont {V.}~\bibnamefont
  {Lienhard}}, \bibinfo {author} {\bibfnamefont {T.}~\bibnamefont {Lahaye}},\
  and\ \bibinfo {author} {\bibfnamefont {A.}~\bibnamefont {Browaeys}},\
  }\bibfield  {title} {\emph {\bibinfo {title} {An atom-by-atom assembler of
  defect-free arbitrary two-dimensional atomic arrays}},\ }\href
  {https://doi.org/10.1126/science.aah3778} {\bibfield  {journal} {\bibinfo
  {journal} {Science}\ }\textbf {\bibinfo {volume} {354}},\ \bibinfo {pages}
  {1021} (\bibinfo {year} {2016})}\BibitemShut {NoStop}%
\bibitem [{\citenamefont {Barredo}\ \emph {et~al.}(2018)\citenamefont
  {Barredo}, \citenamefont {Lienhard}, \citenamefont {de~L{\'e}s{\'e}leuc},
  \citenamefont {Lahaye},\ and\ \citenamefont {Browaeys}}]{Barredo2018}%
  \BibitemOpen
  \bibfield  {author} {\bibinfo {author} {\bibfnamefont {D.}~\bibnamefont
  {Barredo}}, \bibinfo {author} {\bibfnamefont {V.}~\bibnamefont {Lienhard}},
  \bibinfo {author} {\bibfnamefont {S.}~\bibnamefont {de~L{\'e}s{\'e}leuc}},
  \bibinfo {author} {\bibfnamefont {T.}~\bibnamefont {Lahaye}},\ and\ \bibinfo
  {author} {\bibfnamefont {A.}~\bibnamefont {Browaeys}},\ }\bibfield  {title}
  {\emph {\bibinfo {title} {Synthetic three-dimensional atomic structures
  assembled atom by atom}},\ }\href {https://doi.org/10.1038/s41586-018-0450-2}
  {\bibfield  {journal} {\bibinfo  {journal} {Nature}\ }\textbf {\bibinfo
  {volume} {561}},\ \bibinfo {pages} {79} (\bibinfo {year} {2018})}\BibitemShut
  {NoStop}%
\bibitem [{\citenamefont {Saffman}\ \emph {et~al.}(2010)\citenamefont
  {Saffman}, \citenamefont {Walker},\ and\ \citenamefont
  {M\o{}lmer}}]{RevModPhys.82.2313}%
  \BibitemOpen
  \bibfield  {author} {\bibinfo {author} {\bibfnamefont {M.}~\bibnamefont
  {Saffman}}, \bibinfo {author} {\bibfnamefont {T.~G.}\ \bibnamefont
  {Walker}},\ and\ \bibinfo {author} {\bibfnamefont {K.}~\bibnamefont
  {M\o{}lmer}},\ }\bibfield  {title} {\emph {\bibinfo {title} {Quantum
  information with {R}ydberg atoms}},\ }\href
  {https://doi.org/10.1103/RevModPhys.82.2313} {\bibfield  {journal} {\bibinfo
  {journal} {Rev. Mod. Phys.}\ }\textbf {\bibinfo {volume} {82}},\ \bibinfo
  {pages} {2313} (\bibinfo {year} {2010})}\BibitemShut {NoStop}%
\bibitem [{\citenamefont {Adams}\ \emph {et~al.}(2019)\citenamefont {Adams},
  \citenamefont {Pritchard},\ and\ \citenamefont {Shaffer}}]{Adams_2020}%
  \BibitemOpen
  \bibfield  {author} {\bibinfo {author} {\bibfnamefont {C.~S.}\ \bibnamefont
  {Adams}}, \bibinfo {author} {\bibfnamefont {J.~D.}\ \bibnamefont
  {Pritchard}},\ and\ \bibinfo {author} {\bibfnamefont {J.~P.}\ \bibnamefont
  {Shaffer}},\ }\bibfield  {title} {\emph {\bibinfo {title} {Rydberg atom
  quantum technologies}},\ }\href {https://doi.org/10.1088/1361-6455/ab52ef}
  {\bibfield  {journal} {\bibinfo  {journal} {Journal of Physics B: Atomic,
  Molecular and Optical Physics}\ }\textbf {\bibinfo {volume} {53}},\ \bibinfo
  {pages} {012002} (\bibinfo {year} {2019})}\BibitemShut {NoStop}%
\bibitem [{\citenamefont {Wu}\ \emph {et~al.}(2021)\citenamefont {Wu},
  \citenamefont {Liang}, \citenamefont {Tian}, \citenamefont {Yang},
  \citenamefont {Chen}, \citenamefont {Liu}, \citenamefont {Tey},\ and\
  \citenamefont {You}}]{Wu_2021}%
  \BibitemOpen
  \bibfield  {author} {\bibinfo {author} {\bibfnamefont {X.}~\bibnamefont
  {Wu}}, \bibinfo {author} {\bibfnamefont {X.}~\bibnamefont {Liang}}, \bibinfo
  {author} {\bibfnamefont {Y.}~\bibnamefont {Tian}}, \bibinfo {author}
  {\bibfnamefont {F.}~\bibnamefont {Yang}}, \bibinfo {author} {\bibfnamefont
  {C.}~\bibnamefont {Chen}}, \bibinfo {author} {\bibfnamefont {Y.-C.}\
  \bibnamefont {Liu}}, \bibinfo {author} {\bibfnamefont {M.~K.}\ \bibnamefont
  {Tey}},\ and\ \bibinfo {author} {\bibfnamefont {L.}~\bibnamefont {You}},\
  }\bibfield  {title} {\emph {\bibinfo {title} {{A concise review of Rydberg
  atom based quantum computation and quantum simulation}}},\ }\href
  {https://doi.org/10.1088/1674-1056/abd76f} {\bibfield  {journal} {\bibinfo
  {journal} {Chinese Physics B}\ }\textbf {\bibinfo {volume} {30}},\ \bibinfo
  {pages} {020305} (\bibinfo {year} {2021})}\BibitemShut {NoStop}%
\bibitem [{\citenamefont {Browaeys}\ and\ \citenamefont
  {Lahaye}(2020)}]{Browaeys2020}%
  \BibitemOpen
  \bibfield  {author} {\bibinfo {author} {\bibfnamefont {A.}~\bibnamefont
  {Browaeys}}\ and\ \bibinfo {author} {\bibfnamefont {T.}~\bibnamefont
  {Lahaye}},\ }\bibfield  {title} {\emph {\bibinfo {title} {Many-body physics
  with individually controlled {R}ydberg atoms}},\ }\href
  {https://doi.org/10.1038/s41567-019-0733-z} {\bibfield  {journal} {\bibinfo
  {journal} {Nature Physics}\ }\textbf {\bibinfo {volume} {16}},\ \bibinfo
  {pages} {132} (\bibinfo {year} {2020})}\BibitemShut {NoStop}%
\bibitem [{\citenamefont {de~Léséleuc}\ \emph {et~al.}(2019)\citenamefont
  {de~Léséleuc}, \citenamefont {Lienhard}, \citenamefont {Scholl},
  \citenamefont {Barredo}, \citenamefont {Weber}, \citenamefont {Lang},
  \citenamefont {Büchler}, \citenamefont {Lahaye},\ and\ \citenamefont
  {Browaeys}}]{doi:10.1126/science.aav9105}%
  \BibitemOpen
  \bibfield  {author} {\bibinfo {author} {\bibfnamefont {S.}~\bibnamefont
  {de~Léséleuc}}, \bibinfo {author} {\bibfnamefont {V.}~\bibnamefont
  {Lienhard}}, \bibinfo {author} {\bibfnamefont {P.}~\bibnamefont {Scholl}},
  \bibinfo {author} {\bibfnamefont {D.}~\bibnamefont {Barredo}}, \bibinfo
  {author} {\bibfnamefont {S.}~\bibnamefont {Weber}}, \bibinfo {author}
  {\bibfnamefont {N.}~\bibnamefont {Lang}}, \bibinfo {author} {\bibfnamefont
  {H.~P.}\ \bibnamefont {Büchler}}, \bibinfo {author} {\bibfnamefont
  {T.}~\bibnamefont {Lahaye}},\ and\ \bibinfo {author} {\bibfnamefont
  {A.}~\bibnamefont {Browaeys}},\ }\bibfield  {title} {\emph {\bibinfo {title}
  {Observation of a symmetry-protected topological phase of interacting bosons
  with rydberg atoms}},\ }\href {https://doi.org/10.1126/science.aav9105}
  {\bibfield  {journal} {\bibinfo  {journal} {Science}\ }\textbf {\bibinfo
  {volume} {365}},\ \bibinfo {pages} {775} (\bibinfo {year}
  {2019})}\BibitemShut {NoStop}%
\bibitem [{\citenamefont {Yang}\ \emph {et~al.}(2022)\citenamefont {Yang},
  \citenamefont {Wang}, \citenamefont {Zhou},\ and\ \citenamefont
  {Liu}}]{PhysRevA.106.L021101}%
  \BibitemOpen
  \bibfield  {author} {\bibinfo {author} {\bibfnamefont {T.-H.}\ \bibnamefont
  {Yang}}, \bibinfo {author} {\bibfnamefont {B.-Z.}\ \bibnamefont {Wang}},
  \bibinfo {author} {\bibfnamefont {X.-C.}\ \bibnamefont {Zhou}},\ and\
  \bibinfo {author} {\bibfnamefont {X.-J.}\ \bibnamefont {Liu}},\ }\bibfield
  {title} {\emph {\bibinfo {title} {Quantum hall states for {R}ydberg arrays
  with laser-assisted dipole-dipole interactions}},\ }\href
  {https://doi.org/10.1103/PhysRevA.106.L021101} {\bibfield  {journal}
  {\bibinfo  {journal} {Phys. Rev. A}\ }\textbf {\bibinfo {volume} {106}},\
  \bibinfo {pages} {L021101} (\bibinfo {year} {2022})}\BibitemShut {NoStop}%
\bibitem [{\citenamefont {Chen}\ \emph
  {et~al.}(2023{\natexlab{b}})\citenamefont {Chen}, \citenamefont {Bornet},
  \citenamefont {Bintz}, \citenamefont {Emperauger}, \citenamefont {Leclerc},
  \citenamefont {Liu}, \citenamefont {Scholl}, \citenamefont {Barredo},
  \citenamefont {Hauschild}, \citenamefont {Chatterjee}, \citenamefont
  {Schuler}, \citenamefont {L{\"a}uchli}, \citenamefont {Zaletel},
  \citenamefont {Lahaye}, \citenamefont {Yao},\ and\ \citenamefont
  {Browaeys}}]{Chen2023}%
  \BibitemOpen
  \bibfield  {author} {\bibinfo {author} {\bibfnamefont {C.}~\bibnamefont
  {Chen}}, \bibinfo {author} {\bibfnamefont {G.}~\bibnamefont {Bornet}},
  \bibinfo {author} {\bibfnamefont {M.}~\bibnamefont {Bintz}}, \bibinfo
  {author} {\bibfnamefont {G.}~\bibnamefont {Emperauger}}, \bibinfo {author}
  {\bibfnamefont {L.}~\bibnamefont {Leclerc}}, \bibinfo {author} {\bibfnamefont
  {V.~S.}\ \bibnamefont {Liu}}, \bibinfo {author} {\bibfnamefont
  {P.}~\bibnamefont {Scholl}}, \bibinfo {author} {\bibfnamefont
  {D.}~\bibnamefont {Barredo}}, \bibinfo {author} {\bibfnamefont
  {J.}~\bibnamefont {Hauschild}}, \bibinfo {author} {\bibfnamefont
  {S.}~\bibnamefont {Chatterjee}}, \bibinfo {author} {\bibfnamefont
  {M.}~\bibnamefont {Schuler}}, \bibinfo {author} {\bibfnamefont {A.~M.}\
  \bibnamefont {L{\"a}uchli}}, \bibinfo {author} {\bibfnamefont {M.~P.}\
  \bibnamefont {Zaletel}}, \bibinfo {author} {\bibfnamefont {T.}~\bibnamefont
  {Lahaye}}, \bibinfo {author} {\bibfnamefont {N.~Y.}\ \bibnamefont {Yao}},\
  and\ \bibinfo {author} {\bibfnamefont {A.}~\bibnamefont {Browaeys}},\
  }\bibfield  {title} {\emph {\bibinfo {title} {Continuous symmetry breaking in
  a two-dimensional {R}ydberg array}},\ }\href
  {https://doi.org/10.1038/s41586-023-05859-2} {\bibfield  {journal} {\bibinfo
  {journal} {Nature}\ }\textbf {\bibinfo {volume} {616}},\ \bibinfo {pages}
  {691} (\bibinfo {year} {2023}{\natexlab{b}})}\BibitemShut {NoStop}%
\bibitem [{\citenamefont {Chen}\ \emph {et~al.}(2024)\citenamefont {Chen},
  \citenamefont {Wang}, \citenamefont {Poon}, \citenamefont {Zhou},
  \citenamefont {Liu},\ and\ \citenamefont {Liu}}]{PhysRevResearch.6.L042054}%
  \BibitemOpen
  \bibfield  {author} {\bibinfo {author} {\bibfnamefont {Y.-H.}\ \bibnamefont
  {Chen}}, \bibinfo {author} {\bibfnamefont {B.-Z.}\ \bibnamefont {Wang}},
  \bibinfo {author} {\bibfnamefont {T.-F.~J.}\ \bibnamefont {Poon}}, \bibinfo
  {author} {\bibfnamefont {X.-C.}\ \bibnamefont {Zhou}}, \bibinfo {author}
  {\bibfnamefont {Z.-X.}\ \bibnamefont {Liu}},\ and\ \bibinfo {author}
  {\bibfnamefont {X.-J.}\ \bibnamefont {Liu}},\ }\bibfield  {title} {\emph
  {\bibinfo {title} {{P}roposal for realization and detection of {K}itaev
  quantum spin liquid with {R}ydberg atoms}},\ }\href
  {https://doi.org/10.1103/PhysRevResearch.6.L042054} {\bibfield  {journal}
  {\bibinfo  {journal} {Phys. Rev. Res.}\ }\textbf {\bibinfo {volume} {6}},\
  \bibinfo {pages} {L042054} (\bibinfo {year} {2024})}\BibitemShut {NoStop}%
\bibitem [{\citenamefont {Poon}\ \emph {et~al.}(2024)\citenamefont {Poon},
  \citenamefont {Zhou}, \citenamefont {Wang}, \citenamefont {Yang},\ and\
  \citenamefont {Liu}}]{https://doi.org/10.1002/qute.202300356}%
  \BibitemOpen
  \bibfield  {author} {\bibinfo {author} {\bibfnamefont {T.~F.~J.}\
  \bibnamefont {Poon}}, \bibinfo {author} {\bibfnamefont {X.-C.}\ \bibnamefont
  {Zhou}}, \bibinfo {author} {\bibfnamefont {B.-Z.}\ \bibnamefont {Wang}},
  \bibinfo {author} {\bibfnamefont {T.-H.}\ \bibnamefont {Yang}},\ and\
  \bibinfo {author} {\bibfnamefont {X.-J.}\ \bibnamefont {Liu}},\ }\bibfield
  {title} {\emph {\bibinfo {title} {Fractional quantum anomalous hall phase for
  raman superarray of rydberg atoms}},\ }\href
  {https://doi.org/https://doi.org/10.1002/qute.202300356} {\bibfield
  {journal} {\bibinfo  {journal} {Advanced Quantum Technologies}\ }\textbf
  {\bibinfo {volume} {7}},\ \bibinfo {pages} {2300356} (\bibinfo {year}
  {2024})}\BibitemShut {NoStop}%
\bibitem [{\citenamefont {Labuhn}\ \emph {et~al.}(2016)\citenamefont {Labuhn},
  \citenamefont {Barredo}, \citenamefont {Ravets}, \citenamefont
  {de~L{\'e}s{\'e}leuc}, \citenamefont {Macr{\`i}}, \citenamefont {Lahaye},\
  and\ \citenamefont {Browaeys}}]{Labuhn2016}%
  \BibitemOpen
  \bibfield  {author} {\bibinfo {author} {\bibfnamefont {H.}~\bibnamefont
  {Labuhn}}, \bibinfo {author} {\bibfnamefont {D.}~\bibnamefont {Barredo}},
  \bibinfo {author} {\bibfnamefont {S.}~\bibnamefont {Ravets}}, \bibinfo
  {author} {\bibfnamefont {S.}~\bibnamefont {de~L{\'e}s{\'e}leuc}}, \bibinfo
  {author} {\bibfnamefont {T.}~\bibnamefont {Macr{\`i}}}, \bibinfo {author}
  {\bibfnamefont {T.}~\bibnamefont {Lahaye}},\ and\ \bibinfo {author}
  {\bibfnamefont {A.}~\bibnamefont {Browaeys}},\ }\bibfield  {title} {\emph
  {\bibinfo {title} {Tunable two-dimensional arrays of single {R}ydberg atoms
  for realizing quantum {I}sing models}},\ }\href
  {https://doi.org/10.1038/nature18274} {\bibfield  {journal} {\bibinfo
  {journal} {Nature}\ }\textbf {\bibinfo {volume} {534}},\ \bibinfo {pages}
  {667} (\bibinfo {year} {2016})}\BibitemShut {NoStop}%
\bibitem [{\citenamefont {Scholl}\ \emph {et~al.}(2021)\citenamefont {Scholl},
  \citenamefont {Schuler}, \citenamefont {Williams}, \citenamefont
  {Eberharter}, \citenamefont {Barredo}, \citenamefont {Schymik}, \citenamefont
  {Lienhard}, \citenamefont {Henry}, \citenamefont {Lang}, \citenamefont
  {Lahaye}, \citenamefont {L{\"a}uchli},\ and\ \citenamefont
  {Browaeys}}]{Scholl2021}%
  \BibitemOpen
  \bibfield  {author} {\bibinfo {author} {\bibfnamefont {P.}~\bibnamefont
  {Scholl}}, \bibinfo {author} {\bibfnamefont {M.}~\bibnamefont {Schuler}},
  \bibinfo {author} {\bibfnamefont {H.~J.}\ \bibnamefont {Williams}}, \bibinfo
  {author} {\bibfnamefont {A.~A.}\ \bibnamefont {Eberharter}}, \bibinfo
  {author} {\bibfnamefont {D.}~\bibnamefont {Barredo}}, \bibinfo {author}
  {\bibfnamefont {K.-N.}\ \bibnamefont {Schymik}}, \bibinfo {author}
  {\bibfnamefont {V.}~\bibnamefont {Lienhard}}, \bibinfo {author}
  {\bibfnamefont {L.-P.}\ \bibnamefont {Henry}}, \bibinfo {author}
  {\bibfnamefont {T.~C.}\ \bibnamefont {Lang}}, \bibinfo {author}
  {\bibfnamefont {T.}~\bibnamefont {Lahaye}}, \bibinfo {author} {\bibfnamefont
  {A.~M.}\ \bibnamefont {L{\"a}uchli}},\ and\ \bibinfo {author} {\bibfnamefont
  {A.}~\bibnamefont {Browaeys}},\ }\bibfield  {title} {\emph {\bibinfo {title}
  {Quantum simulation of 2{D} antiferromagnets with hundreds of {R}ydberg
  atoms}},\ }\href {https://doi.org/10.1038/s41586-021-03585-1} {\bibfield
  {journal} {\bibinfo  {journal} {Nature}\ }\textbf {\bibinfo {volume} {595}},\
  \bibinfo {pages} {233} (\bibinfo {year} {2021})}\BibitemShut {NoStop}%
\bibitem [{\citenamefont {Bernien}\ \emph {et~al.}(2017)\citenamefont {Bernien},
	\citenamefont {Schwartz}, \citenamefont {Keesling}, \citenamefont {Levine},
	\citenamefont {Omran}, \citenamefont {Pichler}, \citenamefont {Choi},
	\citenamefont {Zibrov}, \citenamefont {Endres}, \citenamefont {Greiner},
	\citenamefont {Vuletić},\ and\ \citenamefont {Lukin}}]{Bernien2017}%
\BibitemOpen
\bibfield  {author} {\bibinfo {author} {\bibfnamefont {H.}~\bibnamefont {Bernien}},
	\bibinfo {author} {\bibfnamefont {S.}~\bibnamefont {Schwartz}},
	\bibinfo {author} {\bibfnamefont {A.}~\bibnamefont {Keesling}},
	\bibinfo {author} {\bibfnamefont {H.}~\bibnamefont {Levine}},
	\bibinfo {author} {\bibfnamefont {A.}~\bibnamefont {Omran}},
	\bibinfo {author} {\bibfnamefont {H.}~\bibnamefont {Pichler}},
	\bibinfo {author} {\bibfnamefont {S.}~\bibnamefont {Choi}},
	\bibinfo {author} {\bibfnamefont {A.~S.}\ \bibnamefont {Zibrov}},
	\bibinfo {author} {\bibfnamefont {M.}~\bibnamefont {Endres}},
	\bibinfo {author} {\bibfnamefont {M.}~\bibnamefont {Greiner}},
	\bibinfo {author} {\bibfnamefont {V.}~\bibnamefont {Vuletić}},\ and\
	\bibinfo {author} {\bibfnamefont {M.~D.}\ \bibnamefont {Lukin}},\ }\bibfield
{title} {\emph {\bibinfo {title} {Probing many-body dynamics on a 51-atom quantum simulator}},\
}\href {https://doi.org/10.1038/nature24622} {\bibfield  {journal} {\bibinfo
		{journal} {Nature}\ }\textbf {\bibinfo {volume} {551}},\ \bibinfo {pages} {579}
	(\bibinfo {year} {2017})}\BibitemShut {NoStop}%
\bibitem [{\citenamefont {Ebadi}\ \emph {et~al.}(2021)\citenamefont {Ebadi},
	\citenamefont {Wang}, \citenamefont {Levine}, \citenamefont {Keesling},
	\citenamefont {Semeghini}, \citenamefont {Omran}, \citenamefont {Bluvstein},
	\citenamefont {Samajdar}, \citenamefont {Pichler}, \citenamefont {Ho},
	\citenamefont {Choi}, \citenamefont {Sachdev}, \citenamefont {Greiner},
	\citenamefont {Vuletić},\ and\ \citenamefont {Lukin}}]{Ebadi2021}%
\BibitemOpen
\bibfield  {author} {\bibinfo {author} {\bibfnamefont {S.}~\bibnamefont {Ebadi}},
	\bibinfo {author} {\bibfnamefont {T.~T.}\ \bibnamefont {Wang}},
	\bibinfo {author} {\bibfnamefont {H.}~\bibnamefont {Levine}},
	\bibinfo {author} {\bibfnamefont {A.}~\bibnamefont {Keesling}},
	\bibinfo {author} {\bibfnamefont {G.}~\bibnamefont {Semeghini}},
	\bibinfo {author} {\bibfnamefont {A.}~\bibnamefont {Omran}},
	\bibinfo {author} {\bibfnamefont {D.}~\bibnamefont {Bluvstein}},
	\bibinfo {author} {\bibfnamefont {R.}~\bibnamefont {Samajdar}},
	\bibinfo {author} {\bibfnamefont {H.}~\bibnamefont {Pichler}},
	\bibinfo {author} {\bibfnamefont {W.~W.}\ \bibnamefont {Ho}},
	\bibinfo {author} {\bibfnamefont {S.}~\bibnamefont {Choi}},
	\bibinfo {author} {\bibfnamefont {S.}~\bibnamefont {Sachdev}},
	\bibinfo {author} {\bibfnamefont {M.}~\bibnamefont {Greiner}},
	\bibinfo {author} {\bibfnamefont {V.}~\bibnamefont {Vuletić}},\ and\
	\bibinfo {author} {\bibfnamefont {M.~D.}\ \bibnamefont {Lukin}},\ }\bibfield
{title} {\emph {\bibinfo {title} {Quantum phases of matter on a 256-atom programmable quantum simulator}},\
}\href {https://doi.org/10.1038/s41586-021-03582-4} {\bibfield  {journal}
	{\bibinfo  {journal} {Nature}\ }\textbf {\bibinfo {volume} {595}},\ \bibinfo
	{pages} {227} (\bibinfo {year} {2021})}\BibitemShut {NoStop}%
\bibitem [{\citenamefont {Johnson}\ and\ \citenamefont
  {Rolston}(2010)}]{PhysRevA.82.033412}%
  \BibitemOpen
  \bibfield  {author} {\bibinfo {author} {\bibfnamefont {J.~E.}\ \bibnamefont
  {Johnson}}\ and\ \bibinfo {author} {\bibfnamefont {S.~L.}\ \bibnamefont
  {Rolston}},\ }\bibfield  {title} {\emph {\bibinfo {title} {Interactions
  between {R}ydberg-dressed atoms}},\ }\href
  {https://doi.org/10.1103/PhysRevA.82.033412} {\bibfield  {journal} {\bibinfo
  {journal} {Phys. Rev. A}\ }\textbf {\bibinfo {volume} {82}},\ \bibinfo
  {pages} {033412} (\bibinfo {year} {2010})}\BibitemShut {NoStop}%
\bibitem [{\citenamefont {Balewski}\ \emph {et~al.}(2014)\citenamefont
  {Balewski}, \citenamefont {Krupp}, \citenamefont {Gaj}, \citenamefont
  {Hofferberth}, \citenamefont {Löw},\ and\ \citenamefont
  {Pfau}}]{Balewski_2014}%
  \BibitemOpen
  \bibfield  {author} {\bibinfo {author} {\bibfnamefont {J.~B.}\ \bibnamefont
  {Balewski}}, \bibinfo {author} {\bibfnamefont {A.~T.}\ \bibnamefont {Krupp}},
  \bibinfo {author} {\bibfnamefont {A.}~\bibnamefont {Gaj}}, \bibinfo {author}
  {\bibfnamefont {S.}~\bibnamefont {Hofferberth}}, \bibinfo {author}
  {\bibfnamefont {R.}~\bibnamefont {Löw}},\ and\ \bibinfo {author}
  {\bibfnamefont {T.}~\bibnamefont {Pfau}},\ }\bibfield  {title} {\emph
  {\bibinfo {title} {Rydberg dressing: understanding of collective many-body
  effects and implications for experiments}},\ }\href
  {https://doi.org/10.1088/1367-2630/16/6/063012} {\bibfield  {journal}
  {\bibinfo  {journal} {New Journal of Physics}\ }\textbf {\bibinfo {volume}
  {16}},\ \bibinfo {pages} {063012} (\bibinfo {year} {2014})}\BibitemShut
  {NoStop}%
\bibitem [{\citenamefont {Henkel}\ \emph {et~al.}(2010)\citenamefont {Henkel},
  \citenamefont {Nath},\ and\ \citenamefont {Pohl}}]{PhysRevLett.104.195302}%
  \BibitemOpen
  \bibfield  {author} {\bibinfo {author} {\bibfnamefont {N.}~\bibnamefont
  {Henkel}}, \bibinfo {author} {\bibfnamefont {R.}~\bibnamefont {Nath}},\ and\
  \bibinfo {author} {\bibfnamefont {T.}~\bibnamefont {Pohl}},\ }\bibfield
  {title} {\emph {\bibinfo {title} {{Three-Dimensional Roton Excitations and
  Supersolid Formation in Rydberg-Excited Bose-Einstein Condensates}}},\ }\href
  {https://doi.org/10.1103/PhysRevLett.104.195302} {\bibfield  {journal}
  {\bibinfo  {journal} {Phys. Rev. Lett.}\ }\textbf {\bibinfo {volume} {104}},\
  \bibinfo {pages} {195302} (\bibinfo {year} {2010})}\BibitemShut {NoStop}%
\bibitem [{\citenamefont {Henkel}\ \emph {et~al.}(2012)\citenamefont {Henkel},
  \citenamefont {Cinti}, \citenamefont {Jain}, \citenamefont {Pupillo},\ and\
  \citenamefont {Pohl}}]{PhysRevLett.108.265301}%
  \BibitemOpen
  \bibfield  {author} {\bibinfo {author} {\bibfnamefont {N.}~\bibnamefont
  {Henkel}}, \bibinfo {author} {\bibfnamefont {F.}~\bibnamefont {Cinti}},
  \bibinfo {author} {\bibfnamefont {P.}~\bibnamefont {Jain}}, \bibinfo {author}
  {\bibfnamefont {G.}~\bibnamefont {Pupillo}},\ and\ \bibinfo {author}
  {\bibfnamefont {T.}~\bibnamefont {Pohl}},\ }\bibfield  {title} {\emph
  {\bibinfo {title} {Supersolid {V}ortex {C}rystals in {R}ydberg-{D}ressed
  {B}ose-{E}instein {C}ondensates}},\ }\href
  {https://doi.org/10.1103/PhysRevLett.108.265301} {\bibfield  {journal}
  {\bibinfo  {journal} {Phys. Rev. Lett.}\ }\textbf {\bibinfo {volume} {108}},\
  \bibinfo {pages} {265301} (\bibinfo {year} {2012})}\BibitemShut {NoStop}%
\bibitem [{\citenamefont {van Bijnen}\ and\ \citenamefont
  {Pohl}(2015)}]{PhysRevLett.114.243002}%
  \BibitemOpen
  \bibfield  {author} {\bibinfo {author} {\bibfnamefont {R.~M.~W.}\
  \bibnamefont {van Bijnen}}\ and\ \bibinfo {author} {\bibfnamefont
  {T.}~\bibnamefont {Pohl}},\ }\bibfield  {title} {\emph {\bibinfo {title}
  {{Quantum Magnetism and Topological Ordering via Rydberg Dressing near
  F\"orster Resonances}}},\ }\href
  {https://doi.org/10.1103/PhysRevLett.114.243002} {\bibfield  {journal}
  {\bibinfo  {journal} {Phys. Rev. Lett.}\ }\textbf {\bibinfo {volume} {114}},\
  \bibinfo {pages} {243002} (\bibinfo {year} {2015})}\BibitemShut {NoStop}%
\bibitem [{\citenamefont {Glaetzle}\ \emph {et~al.}(2015)\citenamefont
  {Glaetzle}, \citenamefont {Dalmonte}, \citenamefont {Nath}, \citenamefont
  {Gross}, \citenamefont {Bloch},\ and\ \citenamefont
  {Zoller}}]{PhysRevLett.114.173002}%
  \BibitemOpen
  \bibfield  {author} {\bibinfo {author} {\bibfnamefont {A.~W.}\ \bibnamefont
  {Glaetzle}}, \bibinfo {author} {\bibfnamefont {M.}~\bibnamefont {Dalmonte}},
  \bibinfo {author} {\bibfnamefont {R.}~\bibnamefont {Nath}}, \bibinfo {author}
  {\bibfnamefont {C.}~\bibnamefont {Gross}}, \bibinfo {author} {\bibfnamefont
  {I.}~\bibnamefont {Bloch}},\ and\ \bibinfo {author} {\bibfnamefont
  {P.}~\bibnamefont {Zoller}},\ }\bibfield  {title} {\emph {\bibinfo {title}
  {{Designing Frustrated Quantum Magnets with Laser-Dressed Rydberg Atoms}}},\
  }\href {https://doi.org/10.1103/PhysRevLett.114.173002} {\bibfield  {journal}
  {\bibinfo  {journal} {Phys. Rev. Lett.}\ }\textbf {\bibinfo {volume} {114}},\
  \bibinfo {pages} {173002} (\bibinfo {year} {2015})}\BibitemShut {NoStop}%
\bibitem [{\citenamefont {Steinert}\ \emph {et~al.}(2023)\citenamefont
  {Steinert}, \citenamefont {Osterholz}, \citenamefont {Eberhard},
  \citenamefont {Festa}, \citenamefont {Lorenz}, \citenamefont {Chen},
  \citenamefont {Trautmann},\ and\ \citenamefont
  {Gross}}]{PhysRevLett.130.243001}%
  \BibitemOpen
  \bibfield  {author} {\bibinfo {author} {\bibfnamefont {L.-M.}\ \bibnamefont
  {Steinert}}, \bibinfo {author} {\bibfnamefont {P.}~\bibnamefont {Osterholz}},
  \bibinfo {author} {\bibfnamefont {R.}~\bibnamefont {Eberhard}}, \bibinfo
  {author} {\bibfnamefont {L.}~\bibnamefont {Festa}}, \bibinfo {author}
  {\bibfnamefont {N.}~\bibnamefont {Lorenz}}, \bibinfo {author} {\bibfnamefont
  {Z.}~\bibnamefont {Chen}}, \bibinfo {author} {\bibfnamefont {A.}~\bibnamefont
  {Trautmann}},\ and\ \bibinfo {author} {\bibfnamefont {C.}~\bibnamefont
  {Gross}},\ }\bibfield  {title} {\emph {\bibinfo {title} {Spatially {T}unable
  {S}pin {I}nteractions in {N}eutral {A}tom {A}rrays}},\ }\href
  {https://doi.org/10.1103/PhysRevLett.130.243001} {\bibfield  {journal}
  {\bibinfo  {journal} {Phys. Rev. Lett.}\ }\textbf {\bibinfo {volume} {130}},\
  \bibinfo {pages} {243001} (\bibinfo {year} {2023})}\BibitemShut {NoStop}%
\bibitem [{\citenamefont {Zeiher}\ \emph {et~al.}(2017)\citenamefont {Zeiher},
  \citenamefont {Choi}, \citenamefont {Rubio-Abadal}, \citenamefont {Pohl},
  \citenamefont {van Bijnen}, \citenamefont {Bloch},\ and\ \citenamefont
  {Gross}}]{PhysRevX.7.041063}%
  \BibitemOpen
  \bibfield  {author} {\bibinfo {author} {\bibfnamefont {J.}~\bibnamefont
  {Zeiher}}, \bibinfo {author} {\bibfnamefont {J.-y.}\ \bibnamefont {Choi}},
  \bibinfo {author} {\bibfnamefont {A.}~\bibnamefont {Rubio-Abadal}}, \bibinfo
  {author} {\bibfnamefont {T.}~\bibnamefont {Pohl}}, \bibinfo {author}
  {\bibfnamefont {R.}~\bibnamefont {van Bijnen}}, \bibinfo {author}
  {\bibfnamefont {I.}~\bibnamefont {Bloch}},\ and\ \bibinfo {author}
  {\bibfnamefont {C.}~\bibnamefont {Gross}},\ }\bibfield  {title} {\emph
  {\bibinfo {title} {Coherent {M}any-{B}ody {S}pin {D}ynamics in a
  {L}ong-{R}ange {I}nteracting {I}sing {C}hain}},\ }\href
  {https://doi.org/10.1103/PhysRevX.7.041063} {\bibfield  {journal} {\bibinfo
  {journal} {Phys. Rev. X}\ }\textbf {\bibinfo {volume} {7}},\ \bibinfo {pages}
  {041063} (\bibinfo {year} {2017})}\BibitemShut {NoStop}%
\bibitem [{\citenamefont {Guardado-Sanchez}\ \emph {et~al.}(2021)\citenamefont
  {Guardado-Sanchez}, \citenamefont {Spar}, \citenamefont {Schauss},
  \citenamefont {Belyansky}, \citenamefont {Young}, \citenamefont {Bienias},
  \citenamefont {Gorshkov}, \citenamefont {Iadecola},\ and\ \citenamefont
  {Bakr}}]{PhysRevX.11.021036}%
  \BibitemOpen
  \bibfield  {author} {\bibinfo {author} {\bibfnamefont {E.}~\bibnamefont
  {Guardado-Sanchez}}, \bibinfo {author} {\bibfnamefont {B.~M.}\ \bibnamefont
  {Spar}}, \bibinfo {author} {\bibfnamefont {P.}~\bibnamefont {Schauss}},
  \bibinfo {author} {\bibfnamefont {R.}~\bibnamefont {Belyansky}}, \bibinfo
  {author} {\bibfnamefont {J.~T.}\ \bibnamefont {Young}}, \bibinfo {author}
  {\bibfnamefont {P.}~\bibnamefont {Bienias}}, \bibinfo {author} {\bibfnamefont
  {A.~V.}\ \bibnamefont {Gorshkov}}, \bibinfo {author} {\bibfnamefont
  {T.}~\bibnamefont {Iadecola}},\ and\ \bibinfo {author} {\bibfnamefont
  {W.~S.}\ \bibnamefont {Bakr}},\ }\bibfield  {title} {\emph {\bibinfo {title}
  {Quench {D}ynamics of a {F}ermi {G}as with {S}trong {N}onlocal
  {I}nteractions}},\ }\href {https://doi.org/10.1103/PhysRevX.11.021036}
  {\bibfield  {journal} {\bibinfo  {journal} {Phys. Rev. X}\ }\textbf {\bibinfo
  {volume} {11}},\ \bibinfo {pages} {021036} (\bibinfo {year}
  {2021})}\BibitemShut {NoStop}%
\bibitem [{\citenamefont {Weckesser}\ \emph {et~al.}(2024)\citenamefont
	{Weckesser}, \citenamefont {Srakaew}, \citenamefont {Blatz}, \citenamefont
	{Wei}, \citenamefont {Adler}, \citenamefont {Agrawal}, \citenamefont
	{Bohrdt}, \citenamefont {Bloch},\ and\ \citenamefont {Zeiher}}]{Weckesser2024}%
\BibitemOpen
\bibfield  {author} {\bibinfo {author} {\bibfnamefont {P.}~\bibnamefont {Weckesser}},
	\bibinfo {author} {\bibfnamefont {K.}~\bibnamefont {Srakaew}},
	\bibinfo {author} {\bibfnamefont {T.}~\bibnamefont {Blatz}},
	\bibinfo {author} {\bibfnamefont {D.}~\bibnamefont {Wei}},
	\bibinfo {author} {\bibfnamefont {D.}~\bibnamefont {Adler}},
	\bibinfo {author} {\bibfnamefont {S.}~\bibnamefont {Agrawal}},
	\bibinfo {author} {\bibfnamefont {A.}~\bibnamefont {Bohrdt}},
	\bibinfo {author} {\bibfnamefont {I.}~\bibnamefont {Bloch}},\ and\
	\bibinfo {author} {\bibfnamefont {J.}~\bibnamefont {Zeiher}},\ }\href
{https://arxiv.org/abs/2405.20128} {\emph {\bibinfo {title} {Realization of a Rydberg-dressed extended Bose Hubbard model}}},\ \bibinfo {howpublished}
{arXiv:2405.20128} (\bibinfo {year} {2024})\BibitemShut {NoStop}%
\bibitem [{\citenamefont {Beugnon}\ \emph {et~al.}(2007)\citenamefont
  {Beugnon}, \citenamefont {Tuchendler}, \citenamefont {Marion}, \citenamefont
  {Ga{\"e}tan}, \citenamefont {Miroshnychenko}, \citenamefont {Sortais},
  \citenamefont {Lance}, \citenamefont {Jones}, \citenamefont {Messin},
  \citenamefont {Browaeys},\ and\ \citenamefont {Grangier}}]{Beugnon2007}%
  \BibitemOpen
  \bibfield  {author} {\bibinfo {author} {\bibfnamefont {J.}~\bibnamefont
  {Beugnon}}, \bibinfo {author} {\bibfnamefont {C.}~\bibnamefont {Tuchendler}},
  \bibinfo {author} {\bibfnamefont {H.}~\bibnamefont {Marion}}, \bibinfo
  {author} {\bibfnamefont {A.}~\bibnamefont {Ga{\"e}tan}}, \bibinfo {author}
  {\bibfnamefont {Y.}~\bibnamefont {Miroshnychenko}}, \bibinfo {author}
  {\bibfnamefont {Y.~R.~P.}\ \bibnamefont {Sortais}}, \bibinfo {author}
  {\bibfnamefont {A.~M.}\ \bibnamefont {Lance}}, \bibinfo {author}
  {\bibfnamefont {M.~P.~A.}\ \bibnamefont {Jones}}, \bibinfo {author}
  {\bibfnamefont {G.}~\bibnamefont {Messin}}, \bibinfo {author} {\bibfnamefont
  {A.}~\bibnamefont {Browaeys}},\ and\ \bibinfo {author} {\bibfnamefont
  {P.}~\bibnamefont {Grangier}},\ }\bibfield  {title} {\emph {\bibinfo {title}
  {Two-dimensional transport and transfer of a single atomic qubit in optical
  tweezers}},\ }\href {https://doi.org/10.1038/nphys698} {\bibfield  {journal}
  {\bibinfo  {journal} {Nature Physics}\ }\textbf {\bibinfo {volume} {3}},\
  \bibinfo {pages} {696} (\bibinfo {year} {2007})}\BibitemShut {NoStop}%
\bibitem [{\citenamefont {Bluvstein}\ \emph {et~al.}(2022)\citenamefont
  {Bluvstein}, \citenamefont {Levine}, \citenamefont {Semeghini}, \citenamefont
  {Wang}, \citenamefont {Ebadi}, \citenamefont {Kalinowski}, \citenamefont
  {Keesling}, \citenamefont {Maskara}, \citenamefont {Pichler}, \citenamefont
  {Greiner}, \citenamefont {Vuleti{\'{c}}},\ and\ \citenamefont
  {Lukin}}]{Bluvstein2022}%
  \BibitemOpen
  \bibfield  {author} {\bibinfo {author} {\bibfnamefont {D.}~\bibnamefont
  {Bluvstein}}, \bibinfo {author} {\bibfnamefont {H.}~\bibnamefont {Levine}},
  \bibinfo {author} {\bibfnamefont {G.}~\bibnamefont {Semeghini}}, \bibinfo
  {author} {\bibfnamefont {T.~T.}\ \bibnamefont {Wang}}, \bibinfo {author}
  {\bibfnamefont {S.}~\bibnamefont {Ebadi}}, \bibinfo {author} {\bibfnamefont
  {M.}~\bibnamefont {Kalinowski}}, \bibinfo {author} {\bibfnamefont
  {A.}~\bibnamefont {Keesling}}, \bibinfo {author} {\bibfnamefont
  {N.}~\bibnamefont {Maskara}}, \bibinfo {author} {\bibfnamefont
  {H.}~\bibnamefont {Pichler}}, \bibinfo {author} {\bibfnamefont
  {M.}~\bibnamefont {Greiner}}, \bibinfo {author} {\bibfnamefont
  {V.}~\bibnamefont {Vuleti{\'{c}}}},\ and\ \bibinfo {author} {\bibfnamefont
  {M.~D.}\ \bibnamefont {Lukin}},\ }\bibfield  {title} {\emph {\bibinfo {title}
  {A quantum processor based on coherent transport of entangled atom arrays}},\
  }\href {https://doi.org/10.1038/s41586-022-04592-6} {\bibfield  {journal}
  {\bibinfo  {journal} {Nature}\ }\textbf {\bibinfo {volume} {604}},\ \bibinfo
  {pages} {451} (\bibinfo {year} {2022})}\BibitemShut {NoStop}%
\bibitem [{\citenamefont {Bluvstein}\ \emph {et~al.}(2024)\citenamefont
  {Bluvstein}, \citenamefont {Evered}, \citenamefont {Geim}, \citenamefont
  {Li}, \citenamefont {Zhou}, \citenamefont {Manovitz}, \citenamefont {Ebadi},
  \citenamefont {Cain}, \citenamefont {Kalinowski}, \citenamefont {Hangleiter},
  \citenamefont {Bonilla~Ataides}, \citenamefont {Maskara}, \citenamefont
  {Cong}, \citenamefont {Gao}, \citenamefont {Sales~Rodriguez}, \citenamefont
  {Karolyshyn}, \citenamefont {Semeghini}, \citenamefont {Gullans},
  \citenamefont {Greiner}, \citenamefont {Vuleti{\'{c}}},\ and\ \citenamefont
  {Lukin}}]{Bluvstein2024}%
  \BibitemOpen
  \bibfield  {author} {\bibinfo {author} {\bibfnamefont {D.}~\bibnamefont
  {Bluvstein}}, \bibinfo {author} {\bibfnamefont {S.~J.}\ \bibnamefont
  {Evered}}, \bibinfo {author} {\bibfnamefont {A.~A.}\ \bibnamefont {Geim}},
  \bibinfo {author} {\bibfnamefont {S.~H.}\ \bibnamefont {Li}}, \bibinfo
  {author} {\bibfnamefont {H.}~\bibnamefont {Zhou}}, \bibinfo {author}
  {\bibfnamefont {T.}~\bibnamefont {Manovitz}}, \bibinfo {author}
  {\bibfnamefont {S.}~\bibnamefont {Ebadi}}, \bibinfo {author} {\bibfnamefont
  {M.}~\bibnamefont {Cain}}, \bibinfo {author} {\bibfnamefont {M.}~\bibnamefont
  {Kalinowski}}, \bibinfo {author} {\bibfnamefont {D.}~\bibnamefont
  {Hangleiter}}, \bibinfo {author} {\bibfnamefont {J.~P.}\ \bibnamefont
  {Bonilla~Ataides}}, \bibinfo {author} {\bibfnamefont {N.}~\bibnamefont
  {Maskara}}, \bibinfo {author} {\bibfnamefont {I.}~\bibnamefont {Cong}},
  \bibinfo {author} {\bibfnamefont {X.}~\bibnamefont {Gao}}, \bibinfo {author}
  {\bibfnamefont {P.}~\bibnamefont {Sales~Rodriguez}}, \bibinfo {author}
  {\bibfnamefont {T.}~\bibnamefont {Karolyshyn}}, \bibinfo {author}
  {\bibfnamefont {G.}~\bibnamefont {Semeghini}}, \bibinfo {author}
  {\bibfnamefont {M.~J.}\ \bibnamefont {Gullans}}, \bibinfo {author}
  {\bibfnamefont {M.}~\bibnamefont {Greiner}}, \bibinfo {author} {\bibfnamefont
  {V.}~\bibnamefont {Vuleti{\'{c}}}},\ and\ \bibinfo {author} {\bibfnamefont
  {M.~D.}\ \bibnamefont {Lukin}},\ }\bibfield  {title} {\emph {\bibinfo {title}
  {Logical quantum processor based on reconfigurable atom arrays}},\ }\href
  {https://doi.org/10.1038/s41586-023-06927-3} {\bibfield  {journal} {\bibinfo
  {journal} {Nature}\ }\textbf {\bibinfo {volume} {626}},\ \bibinfo {pages}
  {58} (\bibinfo {year} {2024})}\BibitemShut {NoStop}%
\bibitem [{\citenamefont {Zhu}\ \emph {et~al.}(2023)\citenamefont {Zhu},
  \citenamefont {Long}, \citenamefont {Gou}, \citenamefont {Pu},\ and\
  \citenamefont {Luo}}]{https://doi.org/10.1002/qute.202300176}%
  \BibitemOpen
  \bibfield  {author} {\bibinfo {author} {\bibfnamefont {S.}~\bibnamefont
  {Zhu}}, \bibinfo {author} {\bibfnamefont {Y.}~\bibnamefont {Long}}, \bibinfo
  {author} {\bibfnamefont {W.}~\bibnamefont {Gou}}, \bibinfo {author}
  {\bibfnamefont {M.}~\bibnamefont {Pu}},\ and\ \bibinfo {author}
  {\bibfnamefont {X.}~\bibnamefont {Luo}},\ }\bibfield  {title} {\emph
  {\bibinfo {title} {{Tunnel-Coupled Optical Microtraps for Ultracold
  Atoms}}},\ }\href {https://doi.org/https://doi.org/10.1002/qute.202300176}
  {\bibfield  {journal} {\bibinfo  {journal} {Advanced Quantum Technologies}\
  }\textbf {\bibinfo {volume} {6}},\ \bibinfo {pages} {2300176} (\bibinfo
  {year} {2023})}\BibitemShut {NoStop}%
\bibitem [{\citenamefont {Kaufman}\ \emph {et~al.}(2014)\citenamefont
  {Kaufman}, \citenamefont {Lester}, \citenamefont {Reynolds}, \citenamefont
  {Wall}, \citenamefont {Foss-Feig}, \citenamefont {Hazzard}, \citenamefont
  {Rey},\ and\ \citenamefont {Regal}}]{doi:10.1126/science.1250057}%
  \BibitemOpen
  \bibfield  {author} {\bibinfo {author} {\bibfnamefont {A.~M.}\ \bibnamefont
  {Kaufman}}, \bibinfo {author} {\bibfnamefont {B.~J.}\ \bibnamefont {Lester}},
  \bibinfo {author} {\bibfnamefont {C.~M.}\ \bibnamefont {Reynolds}}, \bibinfo
  {author} {\bibfnamefont {M.~L.}\ \bibnamefont {Wall}}, \bibinfo {author}
  {\bibfnamefont {M.}~\bibnamefont {Foss-Feig}}, \bibinfo {author}
  {\bibfnamefont {K.~R.~A.}\ \bibnamefont {Hazzard}}, \bibinfo {author}
  {\bibfnamefont {A.~M.}\ \bibnamefont {Rey}},\ and\ \bibinfo {author}
  {\bibfnamefont {C.~A.}\ \bibnamefont {Regal}},\ }\bibfield  {title} {\emph
  {\bibinfo {title} {Two-particle quantum interference in tunnel-coupled
  optical tweezers}},\ }\href {https://doi.org/10.1126/science.1250057}
  {\bibfield  {journal} {\bibinfo  {journal} {Science}\ }\textbf {\bibinfo
  {volume} {345}},\ \bibinfo {pages} {306} (\bibinfo {year}
  {2014})}\BibitemShut {NoStop}%
\bibitem [{\citenamefont {Bergschneider}\ \emph {et~al.}(2019)\citenamefont
  {Bergschneider}, \citenamefont {Klinkhamer}, \citenamefont {Becher},
  \citenamefont {Klemt}, \citenamefont {Palm}, \citenamefont {Z{\"u}rn},
  \citenamefont {Jochim},\ and\ \citenamefont {Preiss}}]{Bergschneider2019}%
  \BibitemOpen
  \bibfield  {author} {\bibinfo {author} {\bibfnamefont {A.}~\bibnamefont
  {Bergschneider}}, \bibinfo {author} {\bibfnamefont {V.~M.}\ \bibnamefont
  {Klinkhamer}}, \bibinfo {author} {\bibfnamefont {J.~H.}\ \bibnamefont
  {Becher}}, \bibinfo {author} {\bibfnamefont {R.}~\bibnamefont {Klemt}},
  \bibinfo {author} {\bibfnamefont {L.}~\bibnamefont {Palm}}, \bibinfo {author}
  {\bibfnamefont {G.}~\bibnamefont {Z{\"u}rn}}, \bibinfo {author}
  {\bibfnamefont {S.}~\bibnamefont {Jochim}},\ and\ \bibinfo {author}
  {\bibfnamefont {P.~M.}\ \bibnamefont {Preiss}},\ }\bibfield  {title} {\emph
  {\bibinfo {title} {Experimental characterization of two-particle entanglement
  through position and momentum correlations}},\ }\href
  {https://doi.org/10.1038/s41567-019-0508-6} {\bibfield  {journal} {\bibinfo
  {journal} {Nature Physics}\ }\textbf {\bibinfo {volume} {15}},\ \bibinfo
  {pages} {640} (\bibinfo {year} {2019})}\BibitemShut {NoStop}%
\bibitem [{\citenamefont {Spar}\ \emph {et~al.}(2022)\citenamefont {Spar},
  \citenamefont {Guardado-Sanchez}, \citenamefont {Chi}, \citenamefont {Yan},\
  and\ \citenamefont {Bakr}}]{PhysRevLett.128.223202}%
  \BibitemOpen
  \bibfield  {author} {\bibinfo {author} {\bibfnamefont {B.~M.}\ \bibnamefont
  {Spar}}, \bibinfo {author} {\bibfnamefont {E.}~\bibnamefont
  {Guardado-Sanchez}}, \bibinfo {author} {\bibfnamefont {S.}~\bibnamefont
  {Chi}}, \bibinfo {author} {\bibfnamefont {Z.~Z.}\ \bibnamefont {Yan}},\ and\
  \bibinfo {author} {\bibfnamefont {W.~S.}\ \bibnamefont {Bakr}},\ }\bibfield
  {title} {\emph {\bibinfo {title} {{Realization of a Fermi-Hubbard Optical
  Tweezer Array}}},\ }\href {https://doi.org/10.1103/PhysRevLett.128.223202}
  {\bibfield  {journal} {\bibinfo  {journal} {Phys. Rev. Lett.}\ }\textbf
  {\bibinfo {volume} {128}},\ \bibinfo {pages} {223202} (\bibinfo {year}
  {2022})}\BibitemShut {NoStop}%
\bibitem [{\citenamefont {Yan}\ \emph {et~al.}(2022)\citenamefont {Yan},
  \citenamefont {Spar}, \citenamefont {Prichard}, \citenamefont {Chi},
  \citenamefont {Wei}, \citenamefont {Ibarra-Garc\'{\i}a-Padilla},
  \citenamefont {Hazzard},\ and\ \citenamefont
  {Bakr}}]{PhysRevLett.129.123201}%
  \BibitemOpen
  \bibfield  {author} {\bibinfo {author} {\bibfnamefont {Z.~Z.}\ \bibnamefont
  {Yan}}, \bibinfo {author} {\bibfnamefont {B.~M.}\ \bibnamefont {Spar}},
  \bibinfo {author} {\bibfnamefont {M.~L.}\ \bibnamefont {Prichard}}, \bibinfo
  {author} {\bibfnamefont {S.}~\bibnamefont {Chi}}, \bibinfo {author}
  {\bibfnamefont {H.-T.}\ \bibnamefont {Wei}}, \bibinfo {author} {\bibfnamefont
  {E.}~\bibnamefont {Ibarra-Garc\'{\i}a-Padilla}}, \bibinfo {author}
  {\bibfnamefont {K.~R.~A.}\ \bibnamefont {Hazzard}},\ and\ \bibinfo {author}
  {\bibfnamefont {W.~S.}\ \bibnamefont {Bakr}},\ }\bibfield  {title} {\emph
  {\bibinfo {title} {{Two-Dimensional Programmable Tweezer Arrays of
  Fermions}}},\ }\href {https://doi.org/10.1103/PhysRevLett.129.123201}
  {\bibfield  {journal} {\bibinfo  {journal} {Phys. Rev. Lett.}\ }\textbf
  {\bibinfo {volume} {129}},\ \bibinfo {pages} {123201} (\bibinfo {year}
  {2022})}\BibitemShut {NoStop}%
\bibitem [{\citenamefont {González-Cuadra}\ \emph {et~al.}(2023)\citenamefont
  {González-Cuadra}, \citenamefont {Bluvstein}, \citenamefont {Kalinowski},
  \citenamefont {Kaubruegger}, \citenamefont {Maskara}, \citenamefont
  {Naldesi}, \citenamefont {Zache}, \citenamefont {Kaufman}, \citenamefont
  {Lukin}, \citenamefont {Pichler}, \citenamefont {Vermersch}, \citenamefont
  {Ye},\ and\ \citenamefont {Zoller}}]{doi:10.1073/pnas.2304294120}%
  \BibitemOpen
  \bibfield  {author} {\bibinfo {author} {\bibfnamefont {D.}~\bibnamefont
  {González-Cuadra}}, \bibinfo {author} {\bibfnamefont {D.}~\bibnamefont
  {Bluvstein}}, \bibinfo {author} {\bibfnamefont {M.}~\bibnamefont
  {Kalinowski}}, \bibinfo {author} {\bibfnamefont {R.}~\bibnamefont
  {Kaubruegger}}, \bibinfo {author} {\bibfnamefont {N.}~\bibnamefont
  {Maskara}}, \bibinfo {author} {\bibfnamefont {P.}~\bibnamefont {Naldesi}},
  \bibinfo {author} {\bibfnamefont {T.~V.}\ \bibnamefont {Zache}}, \bibinfo
  {author} {\bibfnamefont {A.~M.}\ \bibnamefont {Kaufman}}, \bibinfo {author}
  {\bibfnamefont {M.~D.}\ \bibnamefont {Lukin}}, \bibinfo {author}
  {\bibfnamefont {H.}~\bibnamefont {Pichler}}, \bibinfo {author} {\bibfnamefont
  {B.}~\bibnamefont {Vermersch}}, \bibinfo {author} {\bibfnamefont
  {J.}~\bibnamefont {Ye}},\ and\ \bibinfo {author} {\bibfnamefont
  {P.}~\bibnamefont {Zoller}},\ }\bibfield  {title} {\emph {\bibinfo {title}
  {Fermionic quantum processing with programmable neutral atom arrays}},\
  }\href {https://doi.org/10.1073/pnas.2304294120} {\bibfield  {journal}
  {\bibinfo  {journal} {Proceedings of the National Academy of Sciences}\
  }\textbf {\bibinfo {volume} {120}},\ \bibinfo {pages} {e2304294120} (\bibinfo
  {year} {2023})}\BibitemShut {NoStop}%
\bibitem [{\citenamefont {Moreira}\ \emph {et~al.}(2001)\citenamefont
  {Moreira}, \citenamefont {Muñoz}, \citenamefont {Illas}, \citenamefont {{de
  Graaf}},\ and\ \citenamefont {Garcia-Bach}}]{MOREIRA2001183}%
  \BibitemOpen
  \bibfield  {author} {\bibinfo {author} {\bibfnamefont {I.~P.}\ \bibnamefont
  {Moreira}}, \bibinfo {author} {\bibfnamefont {D.}~\bibnamefont {Muñoz}},
  \bibinfo {author} {\bibfnamefont {F.}~\bibnamefont {Illas}}, \bibinfo
  {author} {\bibfnamefont {C.}~\bibnamefont {{de Graaf}}},\ and\ \bibinfo
  {author} {\bibfnamefont {M.}~\bibnamefont {Garcia-Bach}},\ }\bibfield
  {title} {\emph {\bibinfo {title} {{A relationship between electronic
  structure effective parameters and Tc in monolayered cuprate
  superconductors}}},\ }\href
  {https://doi.org/https://doi.org/10.1016/S0009-2614(01)00846-6} {\bibfield
  {journal} {\bibinfo  {journal} {Chemical Physics Letters}\ }\textbf {\bibinfo
  {volume} {345}},\ \bibinfo {pages} {183} (\bibinfo {year}
  {2001})}\BibitemShut {NoStop}%
\bibitem [{\citenamefont {Sala}\ \emph {et~al.}(2020)\citenamefont {Sala},
  \citenamefont {Rakovszky}, \citenamefont {Verresen}, \citenamefont {Knap},\
  and\ \citenamefont {Pollmann}}]{PhysRevX.10.011047}%
  \BibitemOpen
  \bibfield  {author} {\bibinfo {author} {\bibfnamefont {P.}~\bibnamefont
  {Sala}}, \bibinfo {author} {\bibfnamefont {T.}~\bibnamefont {Rakovszky}},
  \bibinfo {author} {\bibfnamefont {R.}~\bibnamefont {Verresen}}, \bibinfo
  {author} {\bibfnamefont {M.}~\bibnamefont {Knap}},\ and\ \bibinfo {author}
  {\bibfnamefont {F.}~\bibnamefont {Pollmann}},\ }\bibfield  {title} {\emph
  {\bibinfo {title} {{Ergodicity Breaking Arising from Hilbert Space
  Fragmentation in Dipole-Conserving Hamiltonians}}},\ }\href
  {https://doi.org/10.1103/PhysRevX.10.011047} {\bibfield  {journal} {\bibinfo
  {journal} {Phys. Rev. X}\ }\textbf {\bibinfo {volume} {10}},\ \bibinfo
  {pages} {011047} (\bibinfo {year} {2020})}\BibitemShut {NoStop}%
\bibitem [{\citenamefont {Khemani}\ \emph {et~al.}(2020)\citenamefont
  {Khemani}, \citenamefont {Hermele},\ and\ \citenamefont
  {Nandkishore}}]{PhysRevB.101.174204}%
  \BibitemOpen
  \bibfield  {author} {\bibinfo {author} {\bibfnamefont {V.}~\bibnamefont
  {Khemani}}, \bibinfo {author} {\bibfnamefont {M.}~\bibnamefont {Hermele}},\
  and\ \bibinfo {author} {\bibfnamefont {R.}~\bibnamefont {Nandkishore}},\
  }\bibfield  {title} {\emph {\bibinfo {title} {Localization from {H}ilbert
  space shattering: {F}rom theory to physical realizations}},\ }\href
  {https://doi.org/10.1103/PhysRevB.101.174204} {\bibfield  {journal} {\bibinfo
   {journal} {Phys. Rev. B}\ }\textbf {\bibinfo {volume} {101}},\ \bibinfo
  {pages} {174204} (\bibinfo {year} {2020})}\BibitemShut {NoStop}%
\bibitem [{\citenamefont {Kohlert}\ \emph {et~al.}(2023)\citenamefont
  {Kohlert}, \citenamefont {Scherg}, \citenamefont {Sala}, \citenamefont
  {Pollmann}, \citenamefont {Hebbe~Madhusudhana}, \citenamefont {Bloch},\ and\
  \citenamefont {Aidelsburger}}]{PhysRevLett.130.010201}%
  \BibitemOpen
  \bibfield  {author} {\bibinfo {author} {\bibfnamefont {T.}~\bibnamefont
  {Kohlert}}, \bibinfo {author} {\bibfnamefont {S.}~\bibnamefont {Scherg}},
  \bibinfo {author} {\bibfnamefont {P.}~\bibnamefont {Sala}}, \bibinfo {author}
  {\bibfnamefont {F.}~\bibnamefont {Pollmann}}, \bibinfo {author}
  {\bibfnamefont {B.}~\bibnamefont {Hebbe~Madhusudhana}}, \bibinfo {author}
  {\bibfnamefont {I.}~\bibnamefont {Bloch}},\ and\ \bibinfo {author}
  {\bibfnamefont {M.}~\bibnamefont {Aidelsburger}},\ }\bibfield  {title} {\emph
  {\bibinfo {title} {{Exploring the Regime of Fragmentation in Strongly Tilted
  Fermi-Hubbard Chains}}},\ }\href
  {https://doi.org/10.1103/PhysRevLett.130.010201} {\bibfield  {journal}
  {\bibinfo  {journal} {Phys. Rev. Lett.}\ }\textbf {\bibinfo {volume} {130}},\
  \bibinfo {pages} {010201} (\bibinfo {year} {2023})}\BibitemShut {NoStop}%
\bibitem [{\citenamefont {Adler}\ \emph {et~al.}(2024)\citenamefont {Adler},
  \citenamefont {Wei}, \citenamefont {Will}, \citenamefont {Srakaew},
  \citenamefont {Agrawal}, \citenamefont {Weckesser}, \citenamefont {Moessner},
  \citenamefont {Pollmann}, \citenamefont {Bloch},\ and\ \citenamefont
  {Zeiher}}]{Adler2024}%
  \BibitemOpen
  \bibfield  {author} {\bibinfo {author} {\bibfnamefont {D.}~\bibnamefont
  {Adler}}, \bibinfo {author} {\bibfnamefont {D.}~\bibnamefont {Wei}}, \bibinfo
  {author} {\bibfnamefont {M.}~\bibnamefont {Will}}, \bibinfo {author}
  {\bibfnamefont {K.}~\bibnamefont {Srakaew}}, \bibinfo {author} {\bibfnamefont
  {S.}~\bibnamefont {Agrawal}}, \bibinfo {author} {\bibfnamefont
  {P.}~\bibnamefont {Weckesser}}, \bibinfo {author} {\bibfnamefont
  {R.}~\bibnamefont {Moessner}}, \bibinfo {author} {\bibfnamefont
  {F.}~\bibnamefont {Pollmann}}, \bibinfo {author} {\bibfnamefont
  {I.}~\bibnamefont {Bloch}},\ and\ \bibinfo {author} {\bibfnamefont
  {J.}~\bibnamefont {Zeiher}},\ }\bibfield  {title} {\emph {\bibinfo {title}
  {{Observation of Hilbert space fragmentation and fractonic excitations in
  2D}}},\ }\href {https://doi.org/10.1038/s41586-024-08188-0} {\bibfield
  {journal} {\bibinfo  {journal} {Nature}\ }\textbf {\bibinfo {volume} {636}},\
  \bibinfo {pages} {80} (\bibinfo {year} {2024})}\BibitemShut {NoStop}%
\bibitem [{\citenamefont {Hahn}\ \emph {et~al.}(2021)\citenamefont {Hahn},
  \citenamefont {McClarty},\ and\ \citenamefont
  {Luitz}}]{Hahn2021InformationDI}%
  \BibitemOpen
  \bibfield  {author} {\bibinfo {author} {\bibfnamefont {D.}~\bibnamefont
  {Hahn}}, \bibinfo {author} {\bibfnamefont {P.~A.}\ \bibnamefont {McClarty}},\
  and\ \bibinfo {author} {\bibfnamefont {D.~J.}\ \bibnamefont {Luitz}},\
  }\bibfield  {title} {\emph {\bibinfo {title} {{Information dynamics in a
  model with Hilbert space fragmentation}}},\ }\href
  {https://api.semanticscholar.org/CorpusID:233004434} {\bibfield  {journal}
  {\bibinfo  {journal} {SciPost Physics}\ } (\bibinfo {year}
  {2021})}\BibitemShut {NoStop}%
\bibitem [{\citenamefont {Moudgalya}\ \emph {et~al.}(2022)\citenamefont
  {Moudgalya}, \citenamefont {Bernevig},\ and\ \citenamefont
  {Regnault}}]{Moudgalya_2022}%
  \BibitemOpen
  \bibfield  {author} {\bibinfo {author} {\bibfnamefont {S.}~\bibnamefont
  {Moudgalya}}, \bibinfo {author} {\bibfnamefont {B.~A.}\ \bibnamefont
  {Bernevig}},\ and\ \bibinfo {author} {\bibfnamefont {N.}~\bibnamefont
  {Regnault}},\ }\bibfield  {title} {\emph {\bibinfo {title} {{Quantum
  many-body scars and Hilbert space fragmentation: a review of exact
  results}}},\ }\href {https://doi.org/10.1088/1361-6633/ac73a0} {\bibfield
  {journal} {\bibinfo  {journal} {Reports on Progress in Physics}\ }\textbf
  {\bibinfo {volume} {85}},\ \bibinfo {pages} {086501} (\bibinfo {year}
  {2022})}\BibitemShut {NoStop}%
\bibitem [{\citenamefont {Zhao}\ \emph {et~al.}(2025)\citenamefont {Zhao},
	\citenamefont {Datla}, \citenamefont {Tian}, \citenamefont {Aliyu},\ and\
	\citenamefont {Loh}}]{PhysRevX.15.011035}%
\BibitemOpen
\bibfield  {author} {\bibinfo {author} {\bibfnamefont {L.}\
		\bibnamefont {Zhao}}, \bibinfo {author} {\bibfnamefont {P.~R.}\
		\bibnamefont {Datla}}, \bibinfo {author} {\bibfnamefont {W.}\
		\bibnamefont {Tian}}, \bibinfo {author} {\bibfnamefont {M.~M.}\
		\bibnamefont {Aliyu}}, \ and\ \bibinfo {author} {\bibfnamefont {H.}\
		\bibnamefont {Loh}},\ }\bibfield  {title} {\emph {\bibinfo {title} {Observation
			of Quantum Thermalization Restricted to Hilbert Space Fragments and
			$\mathbb{Z}_{2k}$ Scars}},\ }\href {https://doi.org/10.1103/PhysRevX.15.011035}
{\bibfield  {journal} {\bibinfo  {journal} {Phys. Rev. X}\ }\textbf {\bibinfo
		{volume} {15}},\ \bibinfo {pages} {011035} (\bibinfo {year} {2025})}\BibitemShut
{NoStop}%
\bibitem [{\citenamefont {Yang}\ \emph {et~al.}(2025)\citenamefont {Yang},
	\citenamefont {Yarloo}, \citenamefont {Zhang}, \citenamefont {M\o{}lmer},\
	and\ \citenamefont {Nielsen}}]{PhysRevB.111.144313}%
\BibitemOpen
\bibfield  {author} {\bibinfo {author} {\bibfnamefont {F.}\ \bibnamefont
		{Yang}}, \bibinfo {author} {\bibfnamefont {H.}\ \bibnamefont {Yarloo}},
	\bibinfo {author} {\bibfnamefont {H.-C.}\ \bibnamefont {Zhang}}, \bibinfo
	{author} {\bibfnamefont {K.}\ \bibnamefont {M\o{}lmer}}, \ and\ \bibinfo
	{author} {\bibfnamefont {A.~E.~B.}\ \bibnamefont {Nielsen}},\ }\bibfield
{title} {\emph {\bibinfo {title} {Probing Hilbert space fragmentation with
			strongly interacting Rydberg atoms}},\ }\href
{https://doi.org/10.1103/PhysRevB.111.144313} {\bibfield  {journal} {\bibinfo
		{journal} {Phys. Rev. B}\ }\textbf {\bibinfo {volume} {111}},\ \bibinfo
	{pages} {144313} (\bibinfo {year} {2025})}\BibitemShut {NoStop}%
\bibitem [{\citenamefont {Moudgalya}\ \emph {et~al.}()\citenamefont
  {Moudgalya}, \citenamefont {Prem}, \citenamefont {Nandkishore}, \citenamefont
  {Regnault},\ and\ \citenamefont {Bernevig}}]{doi:10.1142/9789811231711_0009}%
  \BibitemOpen
  \bibfield  {author} {\bibinfo {author} {\bibfnamefont {S.}~\bibnamefont
  {Moudgalya}}, \bibinfo {author} {\bibfnamefont {A.}~\bibnamefont {Prem}},
  \bibinfo {author} {\bibfnamefont {R.}~\bibnamefont {Nandkishore}}, \bibinfo
  {author} {\bibfnamefont {N.}~\bibnamefont {Regnault}},\ and\ \bibinfo
  {author} {\bibfnamefont {B.~A.}\ \bibnamefont {Bernevig}},\ }\bibinfo {title}
  {{Thermalization and Its Absence within Krylov Subspaces of a Constrained
  Hamiltonian}},\ in\ \href {https://doi.org/10.1142/9789811231711_0009} {\emph
  {\bibinfo {booktitle} {Memorial Volume for Shoucheng Zhang}}},\
  Chap.~\bibinfo {chapter} {7}, pp.\ \bibinfo {pages} {147--209}\BibitemShut
  {NoStop}%
\bibitem [{\citenamefont {Yang}\ \emph {et~al.}(2020)\citenamefont {Yang},
  \citenamefont {Liu}, \citenamefont {Gorshkov},\ and\ \citenamefont
  {Iadecola}}]{PhysRevLett.124.207602}%
  \BibitemOpen
  \bibfield  {author} {\bibinfo {author} {\bibfnamefont {Z.-C.}\ \bibnamefont
  {Yang}}, \bibinfo {author} {\bibfnamefont {F.}~\bibnamefont {Liu}}, \bibinfo
  {author} {\bibfnamefont {A.~V.}\ \bibnamefont {Gorshkov}},\ and\ \bibinfo
  {author} {\bibfnamefont {T.}~\bibnamefont {Iadecola}},\ }\bibfield  {title}
  {\emph {\bibinfo {title} {Hilbert-space fragmentation from strict
  confinement}},\ }\href {https://doi.org/10.1103/PhysRevLett.124.207602}
  {\bibfield  {journal} {\bibinfo  {journal} {Phys. Rev. Lett.}\ }\textbf
  {\bibinfo {volume} {124}},\ \bibinfo {pages} {207602} (\bibinfo {year}
  {2020})}\BibitemShut {NoStop}%
\bibitem [{\citenamefont {De~Tomasi}\ \emph {et~al.}(2019)\citenamefont
  {De~Tomasi}, \citenamefont {Hetterich}, \citenamefont {Sala},\ and\
  \citenamefont {Pollmann}}]{PhysRevB.100.214313}%
  \BibitemOpen
  \bibfield  {author} {\bibinfo {author} {\bibfnamefont {G.}~\bibnamefont
  {De~Tomasi}}, \bibinfo {author} {\bibfnamefont {D.}~\bibnamefont
  {Hetterich}}, \bibinfo {author} {\bibfnamefont {P.}~\bibnamefont {Sala}},\
  and\ \bibinfo {author} {\bibfnamefont {F.}~\bibnamefont {Pollmann}},\
  }\bibfield  {title} {\emph {\bibinfo {title} {{Dynamics of strongly
  interacting systems: From Fock-space fragmentation to many-body
  localization}}},\ }\href {https://doi.org/10.1103/PhysRevB.100.214313}
  {\bibfield  {journal} {\bibinfo  {journal} {Phys. Rev. B}\ }\textbf {\bibinfo
  {volume} {100}},\ \bibinfo {pages} {214313} (\bibinfo {year}
  {2019})}\BibitemShut {NoStop}%
\bibitem [{\citenamefont {Herviou}\ \emph {et~al.}(2021)\citenamefont
  {Herviou}, \citenamefont {Bardarson},\ and\ \citenamefont
  {Regnault}}]{PhysRevB.103.134207}%
  \BibitemOpen
  \bibfield  {author} {\bibinfo {author} {\bibfnamefont {L.}~\bibnamefont
  {Herviou}}, \bibinfo {author} {\bibfnamefont {J.~H.}\ \bibnamefont
  {Bardarson}},\ and\ \bibinfo {author} {\bibfnamefont {N.}~\bibnamefont
  {Regnault}},\ }\bibfield  {title} {\emph {\bibinfo {title} {{Many-body
  localization in a fragmented Hilbert space}}},\ }\href
  {https://doi.org/10.1103/PhysRevB.103.134207} {\bibfield  {journal} {\bibinfo
   {journal} {Phys. Rev. B}\ }\textbf {\bibinfo {volume} {103}},\ \bibinfo
  {pages} {134207} (\bibinfo {year} {2021})}\BibitemShut {NoStop}%
\bibitem [{sup()}]{supplement}%
  \BibitemOpen
  \href@noop {} {}\bibinfo {note} {See more details in Supplementary material.,
  which includes Refs.
  \cite{PhysRevA.82.033412,SIBALIC2017319,PhysRevApplied.16.034013,PhysRevApplied.22.024073,annurev-conmatphys-070909-104059,PhysRevLett.116.235301,doi:10.1126/science.aad9041,doi:10.1126/science.1250057,Bluvstein2022,Bluvstein2024,PhysRevB.101.125126,Tenpy,10.21468/SciPostPhys.2.1.003,10.21468/SciPostPhys.7.2.020}.}\BibitemShut
  {Stop}%
\bibitem [{\citenamefont {Chin}\ \emph {et~al.}(2010)\citenamefont {Chin},
  \citenamefont {Grimm}, \citenamefont {Julienne},\ and\ \citenamefont
  {Tiesinga}}]{RevModPhys.82.1225}%
  \BibitemOpen
  \bibfield  {author} {\bibinfo {author} {\bibfnamefont {C.}~\bibnamefont
  {Chin}}, \bibinfo {author} {\bibfnamefont {R.}~\bibnamefont {Grimm}},
  \bibinfo {author} {\bibfnamefont {P.}~\bibnamefont {Julienne}},\ and\
  \bibinfo {author} {\bibfnamefont {E.}~\bibnamefont {Tiesinga}},\ }\bibfield
  {title} {\emph {\bibinfo {title} {Feshbach resonances in ultracold gases}},\
  }\href {https://doi.org/10.1103/RevModPhys.82.1225} {\bibfield  {journal}
  {\bibinfo  {journal} {Rev. Mod. Phys.}\ }\textbf {\bibinfo {volume} {82}},\
  \bibinfo {pages} {1225} (\bibinfo {year} {2010})}\BibitemShut {NoStop}%
\bibitem [{\citenamefont {Regal}\ \emph {et~al.}(2004)\citenamefont {Regal},
	\citenamefont {Greiner},\ and\ \citenamefont {Jin}}]{PhysRevLett.92.083201}%
\BibitemOpen
\bibfield  {author} {\bibinfo {author} {\bibfnamefont {C.~A.}~\bibnamefont
		{Regal}}, \bibinfo {author} {\bibfnamefont {M.}~\bibnamefont {Greiner}},\ and\
	\bibinfo {author} {\bibfnamefont {D.~S.}~\bibnamefont {Jin}},\ }\bibfield
{title} {\emph {\bibinfo {title} {Lifetime of Molecule-Atom Mixtures near a Feshbach Resonance in \({}^{40}\mathrm{K}\)}},\ }\href
{https://doi.org/10.1103/PhysRevLett.92.083201} {\bibfield  {journal}
	{\bibinfo  {journal} {Phys. Rev. Lett.}\ }\textbf {\bibinfo {volume} {92}},\
	\bibinfo {pages} {083201} (\bibinfo {year} {2004})}\BibitemShut {NoStop}%
\bibitem [{foo()}]{footnote2}%
\BibitemOpen
\href@noop {} {}\bibinfo {note} {The typical bare tunneling coefficient $t_0$ between adjacent tweezers typically ranges from 100 to 300 Hz \cite{doi:10.1126/science.1250057,PhysRevLett.128.223202,PhysRevLett.129.123201}. Then an on-site interaction strength over several kHz is sufficient to eliminate the double occupacy, and can be well achieved according to the experiment studies of the Fermi-Hubbard model in optical lattices~\cite{PhysRevLett.104.080401,Hart2015,Mazurenko2017,Shao2024}. 	
Moreover, the lifetime for fermions confined in lattice configurations can be significantly enhanced by suppressing the three-body loss. For $^{40}$K atoms in optical lattices in strongly repulsive Mott-insulating regime, the lifetime can be up to $1$s~\cite{PhysRevLett.104.080401}, much longer than the typical lifetimes in optical dipole traps~\cite{PhysRevLett.92.083201}. The similar long lifetime is expected for the current simulation and is sufficient.}

\BibitemShut {Stop}%
\bibitem [{\citenamefont {Hines}\ \emph {et~al.}(2023)\citenamefont
	{Hines}, \citenamefont {Rajagopal}, \citenamefont {Moreau}, \citenamefont
	{Wahrman}, \citenamefont {Lewis}, \citenamefont {Marković},\ and\
	\citenamefont {Schleier-Smith}}]{Hines2023}%
\BibitemOpen
\bibfield  {author} {\bibinfo {author} {\bibfnamefont {J.~A.}\ \bibnamefont
		{Hines}}, \bibinfo {author} {\bibfnamefont {S.~V.}\ \bibnamefont
		{Rajagopal}}, \bibinfo {author} {\bibfnamefont {G.~L.}\ \bibnamefont
		{Moreau}}, \bibinfo {author} {\bibfnamefont {M.~D.}\ \bibnamefont {Wahrman}},
	\bibinfo {author} {\bibfnamefont {N.~A.}\ \bibnamefont {Lewis}}, \bibinfo
	{author} {\bibfnamefont {O.}~\bibnamefont {Marković}},\ and\ \bibinfo {author}
	{\bibfnamefont {M.}~\bibnamefont {Schleier-Smith}},\ }\href
{https://link.aps.org/doi/10.1103/PhysRevLett.131.063401} {\emph {\bibinfo
		{title} {Spin Squeezing by Rydberg Dressing in an Array of Atomic Ensembles}}},\
\bibinfo  {journal} {Phys. Rev. Lett.}\ \textbf {\bibinfo {volume} {131}},\
\bibinfo {pages} {063401} (\bibinfo {year} {2023})\BibitemShut {NoStop}%
\bibitem [{foo()}]{footnote3}%
\BibitemOpen
\href@noop {} {}\bibinfo {note} {Unlike more demanding Rydberg dressing in optical lattices, which has also demonstrated experimentally in \cite{PhysRevX.11.021036,Weckesser2024}, our present scheme temporally separates the engineering of hoppings and interactions, and confines atoms at longer distance in spatially separated tweezers when dressing is applied. This setting eliminates interatomic distance fluctuations and avoids the complex behavior of DDI at short range, thereby ensuring greater stability and control. The employed control techniques in our scheme are standard and widely used, closely following the established experiments that couple the ground states to Rydberg states to generate many-body correlations \cite{Labuhn2016,Scholl2021,Bernien2017,Ebadi2021,PhysRevLett.130.243001,Bluvstein2022,Bluvstein2024}. Moreover, the stroboscopic dressing sequences method \cite{Hines2023,Weckesser2024}, in which the dressing light is applied periodically in short pulses rather than continuously, as applied in our scheme, can further enhance the lifetime of the dressed atoms.}\BibitemShut {Stop}%

\bibitem [{\citenamefont {Suzuki}(1991)}]{10.1063/1.529425}%
  \BibitemOpen
  \bibfield  {author} {\bibinfo {author} {\bibfnamefont {M.}~\bibnamefont
  {Suzuki}},\ }\bibfield  {title} {\emph {\bibinfo {title} {General theory of
  fractal path integrals with applications to many‐body theories and
  statistical physics}},\ }\href {https://doi.org/10.1063/1.529425} {\bibfield
  {journal} {\bibinfo  {journal} {Journal of Mathematical Physics}\ }\textbf
  {\bibinfo {volume} {32}},\ \bibinfo {pages} {400} (\bibinfo {year}
  {1991})}\BibitemShut {NoStop}%
\bibitem [{\citenamefont {Berry}\ \emph {et~al.}(2007)\citenamefont {Berry},
  \citenamefont {Ahokas}, \citenamefont {Cleve},\ and\ \citenamefont
  {Sanders}}]{Berry2007}%
  \BibitemOpen
  \bibfield  {author} {\bibinfo {author} {\bibfnamefont {D.~W.}\ \bibnamefont
  {Berry}}, \bibinfo {author} {\bibfnamefont {G.}~\bibnamefont {Ahokas}},
  \bibinfo {author} {\bibfnamefont {R.}~\bibnamefont {Cleve}},\ and\ \bibinfo
  {author} {\bibfnamefont {B.~C.}\ \bibnamefont {Sanders}},\ }\bibfield
  {title} {\emph {\bibinfo {title} {{Efficient Quantum Algorithms for
  Simulating Sparse Hamiltonians}}},\ }\href
  {https://doi.org/10.1007/s00220-006-0150-x} {\bibfield  {journal} {\bibinfo
  {journal} {Communications in Mathematical Physics}\ }\textbf {\bibinfo
  {volume} {270}},\ \bibinfo {pages} {359} (\bibinfo {year}
  {2007})}\BibitemShut {NoStop}%
\bibitem [{\citenamefont {Zhao}\ \emph {et~al.}(2023)\citenamefont {Zhao},
  \citenamefont {Bukov}, \citenamefont {Heyl},\ and\ \citenamefont
  {Moessner}}]{PRXQuantum.4.030319}%
  \BibitemOpen
  \bibfield  {author} {\bibinfo {author} {\bibfnamefont {H.}~\bibnamefont
  {Zhao}}, \bibinfo {author} {\bibfnamefont {M.}~\bibnamefont {Bukov}},
  \bibinfo {author} {\bibfnamefont {M.}~\bibnamefont {Heyl}},\ and\ \bibinfo
  {author} {\bibfnamefont {R.}~\bibnamefont {Moessner}},\ }\bibfield  {title}
  {\emph {\bibinfo {title} {{Making Trotterization Adaptive and
  Energy-Self-Correcting for NISQ Devices and Beyond}}},\ }\href
  {https://doi.org/10.1103/PRXQuantum.4.030319} {\bibfield  {journal} {\bibinfo
   {journal} {PRX Quantum}\ }\textbf {\bibinfo {volume} {4}},\ \bibinfo {pages}
  {030319} (\bibinfo {year} {2023})}\BibitemShut {NoStop}%
\bibitem [{\citenamefont {Schymik}\ \emph {et~al.}(2021)\citenamefont
  {Schymik}, \citenamefont {Pancaldi}, \citenamefont {Nogrette}, \citenamefont
  {Barredo}, \citenamefont {Paris}, \citenamefont {Browaeys},\ and\
  \citenamefont {Lahaye}}]{PhysRevApplied.16.034013}%
  \BibitemOpen
  \bibfield  {author} {\bibinfo {author} {\bibfnamefont {K.-N.}\ \bibnamefont
  {Schymik}}, \bibinfo {author} {\bibfnamefont {S.}~\bibnamefont {Pancaldi}},
  \bibinfo {author} {\bibfnamefont {F.}~\bibnamefont {Nogrette}}, \bibinfo
  {author} {\bibfnamefont {D.}~\bibnamefont {Barredo}}, \bibinfo {author}
  {\bibfnamefont {J.}~\bibnamefont {Paris}}, \bibinfo {author} {\bibfnamefont
  {A.}~\bibnamefont {Browaeys}},\ and\ \bibinfo {author} {\bibfnamefont
  {T.}~\bibnamefont {Lahaye}},\ }\bibfield  {title} {\emph {\bibinfo {title}
  {{Single Atoms with 6000-Second Trapping Lifetimes in Optical-Tweezer Arrays
  at Cryogenic Temperatures}}},\ }\href
  {https://doi.org/10.1103/PhysRevApplied.16.034013} {\bibfield  {journal}
  {\bibinfo  {journal} {Phys. Rev. Appl.}\ }\textbf {\bibinfo {volume} {16}},\
  \bibinfo {pages} {034013} (\bibinfo {year} {2021})}\BibitemShut {NoStop}%
\bibitem [{\citenamefont {Pichard}\ \emph {et~al.}(2024)\citenamefont
  {Pichard}, \citenamefont {Lim}, \citenamefont {Bloch}, \citenamefont
  {Vaneecloo}, \citenamefont {Bourachot}, \citenamefont {Both}, \citenamefont
  {M\'eriaux}, \citenamefont {Dutartre}, \citenamefont {Hostein}, \citenamefont
  {Paris}, \citenamefont {Ximenez}, \citenamefont {Signoles}, \citenamefont
  {Browaeys}, \citenamefont {Lahaye},\ and\ \citenamefont
  {Dreon}}]{PhysRevApplied.22.024073}%
  \BibitemOpen
  \bibfield  {author} {\bibinfo {author} {\bibfnamefont {G.}~\bibnamefont
  {Pichard}}, \bibinfo {author} {\bibfnamefont {D.}~\bibnamefont {Lim}},
  \bibinfo {author} {\bibfnamefont {E.}~\bibnamefont {Bloch}}, \bibinfo
  {author} {\bibfnamefont {J.}~\bibnamefont {Vaneecloo}}, \bibinfo {author}
  {\bibfnamefont {L.}~\bibnamefont {Bourachot}}, \bibinfo {author}
  {\bibfnamefont {G.-J.}\ \bibnamefont {Both}}, \bibinfo {author}
  {\bibfnamefont {G.}~\bibnamefont {M\'eriaux}}, \bibinfo {author}
  {\bibfnamefont {S.}~\bibnamefont {Dutartre}}, \bibinfo {author}
  {\bibfnamefont {R.}~\bibnamefont {Hostein}}, \bibinfo {author} {\bibfnamefont
  {J.}~\bibnamefont {Paris}}, \bibinfo {author} {\bibfnamefont
  {B.}~\bibnamefont {Ximenez}}, \bibinfo {author} {\bibfnamefont
  {A.}~\bibnamefont {Signoles}}, \bibinfo {author} {\bibfnamefont
  {A.}~\bibnamefont {Browaeys}}, \bibinfo {author} {\bibfnamefont
  {T.}~\bibnamefont {Lahaye}},\ and\ \bibinfo {author} {\bibfnamefont
  {D.}~\bibnamefont {Dreon}},\ }\bibfield  {title} {\emph {\bibinfo {title}
  {Rearrangement of individual atoms in a 2000-site optical-tweezer array at
  cryogenic temperatures}},\ }\href
  {https://doi.org/10.1103/PhysRevApplied.22.024073} {\bibfield  {journal}
  {\bibinfo  {journal} {Phys. Rev. Appl.}\ }\textbf {\bibinfo {volume} {22}},\
  \bibinfo {pages} {024073} (\bibinfo {year} {2024})}\BibitemShut {NoStop}%
\bibitem [{foo()}]{footnote1}%
  \BibitemOpen
  \href@noop {} {}\bibinfo {note} {Here, these clusters refer to any
  spatial arrangements of the characteristic spin bound states, with at least
  one vacuum state separating these states. For instance, $\lvert
  \mathrm{B}0\mathrm{C}\rangle$, $\lvert \mathrm{C}0\mathrm{B}\rangle$ and
  $\lvert \mathrm{B}00\mathrm{C}\rangle$ are clusters comprising one
  B and one C states. Other multiparticle spin bound states, such as four-body and
  five-body states, lie out of the Krylov subspace considered in the main
  text. This is because examination of
  	entangled spin states up to seven particles reveals no similar energy degeneracy relation with configurations
  	constructed from $\lvert\mathrm{A}\rangle$, $\lvert\mathrm{B}\rangle$, and $\lvert\mathrm{C}\rangle$. Further, local spin bound states with more particles have negligible impact on the dynamics, as transitions to these states from configurations built by $\lvert\mathrm{A}\rangle$, $\lvert\mathrm{B}\rangle$, and $\lvert\mathrm{C}\rangle$ states occur only via very high-order processes. Nonetheless, they constitute local bases for other fragmented subspaces, and
  these states serve as the intermediate states in the transition process
  between $\lvert\mathrm{A}\rangle$, $\lvert\mathrm{B}\rangle$, and
  $\lvert\mathrm{C}\rangle$ spin bound states~\cite{supplement}.}\BibitemShut {Stop}%

\bibitem [{\citenamefont {Vidal}(2004)}]{PhysRevLett.93.040502}%
  \BibitemOpen
  \bibfield  {author} {\bibinfo {author} {\bibfnamefont {G.}~\bibnamefont
  {Vidal}},\ }\bibfield  {title} {\emph {\bibinfo {title} {{Efficient
  Simulation of One-Dimensional Quantum Many-Body Systems}}},\ }\href
  {https://doi.org/10.1103/PhysRevLett.93.040502} {\bibfield  {journal}
  {\bibinfo  {journal} {Phys. Rev. Lett.}\ }\textbf {\bibinfo {volume} {93}},\
  \bibinfo {pages} {040502} (\bibinfo {year} {2004})}\BibitemShut {NoStop}%
\bibitem [{\citenamefont {Schollwöck}(2011)}]{SCHOLLWOCK201196}%
  \BibitemOpen
  \bibfield  {author} {\bibinfo {author} {\bibfnamefont {U.}~\bibnamefont
  {Schollwöck}},\ }\bibfield  {title} {\emph {\bibinfo {title} {The
  density-matrix renormalization group in the age of matrix product states}},\
  }\href {https://doi.org/https://doi.org/10.1016/j.aop.2010.09.012} {\bibfield
   {journal} {\bibinfo  {journal} {Annals of Physics}\ }\textbf {\bibinfo
  {volume} {326}},\ \bibinfo {pages} {96} (\bibinfo {year} {2011})}\BibitemShut
  {NoStop}%
 \bibitem [{foo()}]{footnotetjz}%
 \BibitemOpen
 \href@noop {} {}\bibinfo {note} {Here, we consider the 1D $t$-$J_z$ model whose Hamiltonian is $H_{t\mathrm{-}J_z}=\sum_{i,\mathrm{\sigma}} (-t P_{i}c_{i,\mathrm{\sigma}}^{\dagger}c_{i+1,\mathrm{\sigma}}P_{i+1}+\mathrm{H.c.})+\sum_{i} J_z S_{z,i}S_{z,i+1}$, where $P_i=1-n_{i,\uparrow}n_{i,\downarrow}$ is the projection operator excluding double occupancy on site $i$.}\BibitemShut {Stop}%
\bibitem [{\citenamefont {Rakovszky}\ \emph {et~al.}(2020)\citenamefont
  {Rakovszky}, \citenamefont {Sala}, \citenamefont {Verresen}, \citenamefont
  {Knap},\ and\ \citenamefont {Pollmann}}]{PhysRevB.101.125126}%
  \BibitemOpen
  \bibfield  {author} {\bibinfo {author} {\bibfnamefont {T.}~\bibnamefont
  {Rakovszky}}, \bibinfo {author} {\bibfnamefont {P.}~\bibnamefont {Sala}},
  \bibinfo {author} {\bibfnamefont {R.}~\bibnamefont {Verresen}}, \bibinfo
  {author} {\bibfnamefont {M.}~\bibnamefont {Knap}},\ and\ \bibinfo {author}
  {\bibfnamefont {F.}~\bibnamefont {Pollmann}},\ }\bibfield  {title} {\emph
  {\bibinfo {title} {{Statistical localization: From strong fragmentation to
  strong edge modes}}},\ }\href {https://doi.org/10.1103/PhysRevB.101.125126}
  {\bibfield  {journal} {\bibinfo  {journal} {Phys. Rev. B}\ }\textbf {\bibinfo
  {volume} {101}},\ \bibinfo {pages} {125126} (\bibinfo {year}
  {2020})}\BibitemShut {NoStop}%
\bibitem [{\citenamefont {Deutsch}(1991)}]{PhysRevA.43.2046}%
  \BibitemOpen
  \bibfield  {author} {\bibinfo {author} {\bibfnamefont {J.~M.}\ \bibnamefont
  {Deutsch}},\ }\bibfield  {title} {\emph {\bibinfo {title} {Quantum
  statistical mechanics in a closed system}},\ }\href
  {https://doi.org/10.1103/PhysRevA.43.2046} {\bibfield  {journal} {\bibinfo
  {journal} {Phys. Rev. A}\ }\textbf {\bibinfo {volume} {43}},\ \bibinfo
  {pages} {2046} (\bibinfo {year} {1991})}\BibitemShut {NoStop}%
\bibitem [{\citenamefont {Srednicki}(1994)}]{PhysRevE.50.888}%
  \BibitemOpen
  \bibfield  {author} {\bibinfo {author} {\bibfnamefont {M.}~\bibnamefont
  {Srednicki}},\ }\bibfield  {title} {\emph {\bibinfo {title} {Chaos and
  quantum thermalization}},\ }\href {https://doi.org/10.1103/PhysRevE.50.888}
  {\bibfield  {journal} {\bibinfo  {journal} {Phys. Rev. E}\ }\textbf {\bibinfo
  {volume} {50}},\ \bibinfo {pages} {888} (\bibinfo {year} {1994})}\BibitemShut
  {NoStop}%
\bibitem [{\citenamefont {Rigol}\ \emph {et~al.}(2008)\citenamefont {Rigol},
  \citenamefont {Dunjko},\ and\ \citenamefont {Olshanii}}]{Rigol2008}%
  \BibitemOpen
  \bibfield  {author} {\bibinfo {author} {\bibfnamefont {M.}~\bibnamefont
  {Rigol}}, \bibinfo {author} {\bibfnamefont {V.}~\bibnamefont {Dunjko}},\ and\
  \bibinfo {author} {\bibfnamefont {M.}~\bibnamefont {Olshanii}},\ }\bibfield
  {title} {\emph {\bibinfo {title} {Thermalization and its mechanism for
  generic isolated quantum systems}},\ }\href
  {https://doi.org/10.1038/nature06838} {\bibfield  {journal} {\bibinfo
  {journal} {Nature}\ }\textbf {\bibinfo {volume} {452}},\ \bibinfo {pages}
  {854} (\bibinfo {year} {2008})}\BibitemShut {NoStop}%
\bibitem [{\citenamefont {Deutsch}(2018)}]{Deutsch_2018}%
  \BibitemOpen
  \bibfield  {author} {\bibinfo {author} {\bibfnamefont {J.~M.}\ \bibnamefont
  {Deutsch}},\ }\bibfield  {title} {\emph {\bibinfo {title} {Eigenstate
  thermalization hypothesis}},\ }\href
  {https://doi.org/10.1088/1361-6633/aac9f1} {\bibfield  {journal} {\bibinfo
  {journal} {Reports on Progress in Physics}\ }\textbf {\bibinfo {volume}
  {81}},\ \bibinfo {pages} {082001} (\bibinfo {year} {2018})}\BibitemShut
  {NoStop}%
\bibitem [{\citenamefont {Hauschild}\ and\ \citenamefont
  {Pollmann}(2018)}]{Tenpy}%
  \BibitemOpen
  \bibfield  {author} {\bibinfo {author} {\bibfnamefont {J.}~\bibnamefont
  {Hauschild}}\ and\ \bibinfo {author} {\bibfnamefont {F.}~\bibnamefont
  {Pollmann}},\ }\bibfield  {title} {\emph {\bibinfo {title} {{E}fficient
  numerical simulations with {T}ensor {N}etworks: {T}ensor {N}etwork {P}ython
  ({T}e{NP}y)}},\ }\href {https://doi.org/10.21468/SciPostPhysLectNotes.5}
  {\bibfield  {journal} {\bibinfo  {journal} {SciPost Phys. Lect. Notes}\ ,\
  \bibinfo {pages} {5}} (\bibinfo {year} {2018})}\BibitemShut {NoStop}%
\bibitem [{\citenamefont {Šibalić}\ \emph {et~al.}(2017)\citenamefont
  {Šibalić}, \citenamefont {Pritchard}, \citenamefont {Adams},\ and\
  \citenamefont {Weatherill}}]{SIBALIC2017319}%
  \BibitemOpen
  \bibfield  {author} {\bibinfo {author} {\bibfnamefont {N.}~\bibnamefont
  {Šibalić}}, \bibinfo {author} {\bibfnamefont {J.}~\bibnamefont
  {Pritchard}}, \bibinfo {author} {\bibfnamefont {C.}~\bibnamefont {Adams}},\
  and\ \bibinfo {author} {\bibfnamefont {K.}~\bibnamefont {Weatherill}},\
  }\bibfield  {title} {\emph {\bibinfo {title} {{ARC: An open-source library
  for calculating properties of alkali Rydberg atoms}}},\ }\href
  {https://doi.org/https://doi.org/10.1016/j.cpc.2017.06.015} {\bibfield
  {journal} {\bibinfo  {journal} {Computer Physics Communications}\ }\textbf
  {\bibinfo {volume} {220}},\ \bibinfo {pages} {319} (\bibinfo {year}
  {2017})}\BibitemShut {NoStop}%
\bibitem [{\citenamefont {Cheuk}\ \emph
  {et~al.}(2016{\natexlab{b}})\citenamefont {Cheuk}, \citenamefont {Nichols},
  \citenamefont {Lawrence}, \citenamefont {Okan}, \citenamefont {Zhang},\ and\
  \citenamefont {Zwierlein}}]{PhysRevLett.116.235301}%
  \BibitemOpen
  \bibfield  {author} {\bibinfo {author} {\bibfnamefont {L.~W.}\ \bibnamefont
  {Cheuk}}, \bibinfo {author} {\bibfnamefont {M.~A.}\ \bibnamefont {Nichols}},
  \bibinfo {author} {\bibfnamefont {K.~R.}\ \bibnamefont {Lawrence}}, \bibinfo
  {author} {\bibfnamefont {M.}~\bibnamefont {Okan}}, \bibinfo {author}
  {\bibfnamefont {H.}~\bibnamefont {Zhang}},\ and\ \bibinfo {author}
  {\bibfnamefont {M.~W.}\ \bibnamefont {Zwierlein}},\ }\bibfield  {title}
  {\emph {\bibinfo {title} {{Observation of 2D Fermionic Mott Insulators of
  $^{40}\mathrm{K}$ with Single-Site Resolution}}},\ }\href
  {https://doi.org/10.1103/PhysRevLett.116.235301} {\bibfield  {journal}
  {\bibinfo  {journal} {Phys. Rev. Lett.}\ }\textbf {\bibinfo {volume} {116}},\
  \bibinfo {pages} {235301} (\bibinfo {year} {2016}{\natexlab{b}})}\BibitemShut
  {NoStop}%
\bibitem [{\citenamefont {Greif}\ \emph {et~al.}(2016)\citenamefont {Greif},
  \citenamefont {Parsons}, \citenamefont {Mazurenko}, \citenamefont {Chiu},
  \citenamefont {Blatt}, \citenamefont {Huber}, \citenamefont {Ji},\ and\
  \citenamefont {Greiner}}]{doi:10.1126/science.aad9041}%
  \BibitemOpen
  \bibfield  {author} {\bibinfo {author} {\bibfnamefont {D.}~\bibnamefont
  {Greif}}, \bibinfo {author} {\bibfnamefont {M.~F.}\ \bibnamefont {Parsons}},
  \bibinfo {author} {\bibfnamefont {A.}~\bibnamefont {Mazurenko}}, \bibinfo
  {author} {\bibfnamefont {C.~S.}\ \bibnamefont {Chiu}}, \bibinfo {author}
  {\bibfnamefont {S.}~\bibnamefont {Blatt}}, \bibinfo {author} {\bibfnamefont
  {F.}~\bibnamefont {Huber}}, \bibinfo {author} {\bibfnamefont
  {G.}~\bibnamefont {Ji}},\ and\ \bibinfo {author} {\bibfnamefont
  {M.}~\bibnamefont {Greiner}},\ }\bibfield  {title} {\emph {\bibinfo {title}
  {{Site-resolved imaging of a fermionic Mott insulator}}},\ }\href
  {https://doi.org/10.1126/science.aad9041} {\bibfield  {journal} {\bibinfo
  {journal} {Science}\ }\textbf {\bibinfo {volume} {351}},\ \bibinfo {pages}
  {953} (\bibinfo {year} {2016})}\BibitemShut {NoStop}%
\bibitem [{\citenamefont {Weinberg}\ and\ \citenamefont
  {Bukov}(2017)}]{10.21468/SciPostPhys.2.1.003}%
  \BibitemOpen
  \bibfield  {author} {\bibinfo {author} {\bibfnamefont {P.}~\bibnamefont
  {Weinberg}}\ and\ \bibinfo {author} {\bibfnamefont {M.}~\bibnamefont
  {Bukov}},\ }\bibfield  {title} {\emph {\bibinfo {title} {{QuSpin: a Python
  package for dynamics and exact diagonalisation of quantum many body systems
  part I: spin chains}}},\ }\href
  {https://doi.org/10.21468/SciPostPhys.2.1.003} {\bibfield  {journal}
  {\bibinfo  {journal} {SciPost Phys.}\ }\textbf {\bibinfo {volume} {2}},\
  \bibinfo {pages} {003} (\bibinfo {year} {2017})}\BibitemShut {NoStop}%
\bibitem [{\citenamefont {Weinberg}\ and\ \citenamefont
  {Bukov}(2019)}]{10.21468/SciPostPhys.7.2.020}%
  \BibitemOpen
  \bibfield  {author} {\bibinfo {author} {\bibfnamefont {P.}~\bibnamefont
  {Weinberg}}\ and\ \bibinfo {author} {\bibfnamefont {M.}~\bibnamefont
  {Bukov}},\ }\bibfield  {title} {\emph {\bibinfo {title} {{QuSpin: a Python
  package for dynamics and exact diagonalisation of quantum many body systems.
  Part II: bosons, fermions and higher spins}}},\ }\href
  {https://doi.org/10.21468/SciPostPhys.7.2.020} {\bibfield  {journal}
  {\bibinfo  {journal} {SciPost Phys.}\ }\textbf {\bibinfo {volume} {7}},\
  \bibinfo {pages} {020} (\bibinfo {year} {2019})}\BibitemShut {NoStop}%
\end{thebibliography}
\end{document}